\documentclass[fleqn,usenatbib]{mnras}

\usepackage[T1]{fontenc}
\usepackage{ae,aecompl}
\usepackage{floatflt}

\usepackage{graphicx}	% Including figure files
\usepackage{amsmath}	% Advanced maths commands
\usepackage{amssymb}	% Extra maths symbols
\usepackage{rotating}
\usepackage{tabularx}
\usepackage{multicol}
\newcommand\textlcsc[1]{\textsc{\MakeLowercase{#1}}}

\def\lae{\mathrel{<\kern-1.0em\lower0.9ex\hbox{$\sim$}}}
\def\gae{\mathrel{>\kern-1.0em\lower0.9ex\hbox{$\sim$}}}

\title[Colour gradients in cluster ellipticals at z $\sim$ 1.4]{Colour gradients in cluster ellipticals at z $\sim$ 1.4:\\
the hidden content of the galaxy central regions}

\author[F. Ciocca et al.]{
F. Ciocca$^{1,2}$\thanks{E-mail: federica.ciocca@brera.inaf.it},
P. Saracco$^{1}$,
A. Gargiulo$^{3}$,
R. De Propris$^{4}$
\\
$^{1}$INAF- Osservatorio Astronomico di Brera, Via Brera 28, 20121 Milano, Italy\\
$^{2}$Universit\`a degli Studi dell'Insubria, Via Valleggio 11, 22100 Como, Italy \\
$^{3}$INAF - Istituto di Astrofisica Spaziale e Fisica Cosmica (IASF), Via E. Bassini 15, 20133 Milano, Italy\\
$^{4}$Finnish Centre for Astronomy with ESO, University of Turku, Finland
}

\date{Accepted 2017 January 3. Received 2016 December 22; in original form 2016 July 15}

\pubyear{2016}

\begin{document}
\label{firstpage}
\pagerange{\pageref{firstpage}--\pageref{lastpage}}
\maketitle

\begin{abstract}
We present F775W-F850LP (rest-frame UV-U) and F850LP-F160W (rest-frame U-R) 
colour gradients for a sample of 17 elliptical galaxies morphologically selected 
in the cluster XMMU J2235.3-2557 at $z$=1.39. 
We detected significant negative (redder inwards) U-R colour gradients in 
$\sim$70 per cent of the galaxies and flat gradients for the remaining ones. 
On the other hand, the UV-U gradients are significant positive (bluer inwards) 
for $\sim$80 per cent of the galaxies and flat for the remaining ones. 
Using stellar population synthesis models, we found that the behaviour of the 
two colour gradients cannot be simultaneously explained by a radial variation 
of age, metallicity and/or dust. 
The observed  U-R gradients are consistent with a metallicity gradient 
(mean value $\nabla_{Z} =-0.4$) in agreement with the one observed 
in the local elliptical galaxies. 
The positive UV-U gradients cannot be explained with age or metallicity 
variations and imply an excess of UV emission towards the galaxies' central 
regions. 
 This excess calls into question mechanisms able to efficiently produce UV emission.
 The data require either steady weak star formation  
($\lae 1$ M$_\odot$ yr$^{-1}$) or an He-rich population in the cores of these 
galaxies in order to simultaneously reproduce both the colour gradients. 
On the contrary, the presence of a QSO cannot account for the observed UV excess on its own.
We discuss these hypotheses on the basis of current observations and available models.

\end{abstract}


%%%%%%%%%%%%%%%%%%%%%%%%%%%%%%%%%%%%%%%%%%%%%%%%%%

%%%%%%%%%%%%%%%%% BODY OF PAPER %%%%%%%%%%%%%%%%%%

\section{Introduction}
The mass assembly and the shaping of the elliptical morphology of early-type galaxies 
(ellipticals and lenticulars) are not yet completely understood and they 
represent an important challenge for observational cosmology. 
Colour gradients represent a powerful tool to constrain the possible mechanisms 
ruling these two processes.  
The presence of colour gradients indicates that the properties of
the stellar population within galaxy are not homogeneous. 
Stars may not have been formed in a single burst of star formation, 
or may have been formed in regions characterized by different physical 
conditions of the interstellar medium (ISM), or be the result of later episodes of
accretion of small satellites.
The different mechanisms of mass assembly that may have taken place affect in different way the internal distribution of stellar populations probed by the colour gradients.

If elliptical galaxies are formed in dissipational scenarios, such as gas-rich major mergers in which  most ($\sim$80 per cent) of the 
stars are formed in a dissipative central starburst 
\citep[e.g.][]{khochfar06,naab07,sommer10}, or from primordial mergers of
lumps of dark and baryonic matter occurring at very early epochs ($z>4$)
\citep[e.g.][]{chiosi02,pipino04, merlin06,merlin12}, the radial variation
of their stellar populations 
resembles the one expected in the classical monolithic scenario \citep{eggen62}.
In this scenario, gas is more efficiently retained in the centre of the 
galaxy and thus fuels both more prolonged star formation and chemical 
evolution, resulting in a negative metallicity gradient 
(galaxies are more metal-rich in the centre) and a mild positive age 
gradient (the mean age of stellar populations is younger in the centre) 
\citep{kobayashi04}.

However, this early phase dominated by \textit{in situ} star formation could be 
followed by an extended phase of \textit{ex situ} accretion of stars 
from dry minor mergers with smaller satellite stellar systems,
as recently proposed by e.g. \cite{oser10}.
In this case, the above gradients could be strongly affected.
In this two-phase process, early-type galaxies continue to grow from the 
inside-out redistributing the stellar content of the satellites in the outer 
regions, without mixing with the pre-existing  central bulge.
However, given the  stochastic nature of merger, the resulting galaxies will 
not show systematics in their colour variation:
if age and metallicity of the satellite stellar systems are similar to the 
central early-type galaxy, the gradients may flatten or, opposite, steepen if the 
satellites are younger.

In the classical hierarchical merging scenario 
\citep[e.g.][]{kauffmann96,delucia06},
early-type galaxies form by the wet (gas-rich) major mergers of disk galaxies. 
During these mergers, star formation \textit{in situ} is triggered by gas cooling 
and shock heating \citep[e.g.][]{dekel06,cattaneo08} and provides only
a minor fraction of the final stellar mass of the remnant,
whose stars come principally from the disc progenitors. 
These major mergers will tend to mix the 
stellar content of the progenitors producing nearly flat gradients in the remnant.   
Hereafter, the remnant may later experience minor dry mergers whose effects are those discussed above.  

The environment also plays an important role in the evolution of elliptical galaxies.
Some theoretical studies suggest that ellipticals in dense environment are
formed through bursts of star formation at $z > 3$, unlike field ellipticals 
that are formed through major mergers of disk galaxies at lower redshift
(e.g. $z$ < 2)  \citep[e.g.][]{peacock99}.
These different formation patterns would produce different colour gradients
as discussed above.
Field and cluster galaxies also experience different 
 evolutionary processes due to the different environment. 
 Cluster galaxies, unlike field galaxies, are subject to two main classes of 
 processes: first, the gravitational ones, including the tidal interactions 
 (galaxy--galaxy, galaxy--cluster, harassment) 
 \citep[e.g.][]{fujita98,merritt83,merritt84,mihos95} and, secondly, the interactions 
 taking place between the galaxy's ISM and the hot intergalactic medium (ram pressure, 
 thermal evaporation, etc.) \citep[e.g.][]{gunn72,abadi99,cowie77,dressler83}. 
 Processes like ram pressure stripping or thermal evaporation, removing the gas 
 reservoir from the galaxy, can lead to a rapid quenching of star formation 
 \citep[e.g.][]{larson80,drake00}. 
 These processes can either act at global galaxy scale or involve only a region 
 of the galaxy, introducing a gradient of the stellar population properties (mainly age),
  whose steepness depends on the fraction of gas removed. 

All these different predictions show how spatially resolving the properties of the 
stellar populations within the galaxies may therefore constrain the history of mass assembly and evolution of elliptical galaxies. 

Colour gradients in local early-type galaxies  have been extensively
studied during the last 30 years.
Local and intermediate redshift elliptical galaxies 
show negative optical or optical-NIR rest-frame colour 
gradients (e.g., B-R,  U-R, V-K), i.e. they are redder towards the centre. 
Almost all these works agree on the fact that a metallicity gradient is the main driver of the 
observed colour variation 
\citep[e.g.][]{peletier90,tamura00,saglia00,wu05,labarbera05,tortora10}: 
the cores are metal-rich compared to the outer region. 
A few studies have also found that slightly positive 
age gradients, in addition to the metallicity gradient, are necessary to explain the  observed radial colour variation, i.e.  early-type galaxies tend to have younger and 
more metal-rich cores \citep[e.g.][]{clemens09,rawle10,labarbera09}.
However, the contribution 
played by radial age variation is still controversial. 

The presence or otherwise of an age gradient is a key piece in understanding 
of the mass assembly of elliptical galaxies. 
However, its detection in low redshift galaxies is a challenging 
task: 
an age variation of 1--2 Gyr in a stellar population with a mean age of about 8--9 Gyr
reflects a variation of $\sim$0.05 mag in the U-R colour, 
much smaller than the one produced by the observed metallicity gradient and
comparable to typical photometric errors.

On the contrary, a variation of just 1 Gyr in a  stellar population 
$\sim$2--3 Gyr old, as in ellipticals at $z$ $\sim$ 1.5, would produce a 
colour variation of $\sim$0.2 mag, comparable to the one produced by
a metallicity variation (see Section 5).
Hence, an age variation, if present, would be detectable in high redshift 
early-type galaxies.
Hence, age gradients, if present, would be more easily detectable in high redshift galaxies.

Only a few studies of colour gradients in high-redshift elliptical galaxies have been carried out so far.
\cite{gargiulo12} studied the rest-frame optical U-R and UV-U 
\citep{gargiulo11} colour gradients in a sample of field ellipticals at 1 < $z$ < 2 
and they found that only for half of the sample an age/metallicity gradient 
can account for the observed gradients. 
For the remaining half of the sample, the variation of more than one parameter
 is required to account for the observed colour variation.
Recently, \cite{chan16} found that the rest-frame optical U-R colour gradients in 
passive cluster galaxies at $z$=1.39 are mostly negative,  
with a median value twice the value in the local Universe \citep{wu05}. 
To reproduce the colour gradients locally observed, the presence of both age and 
metallicity gradient is needed. It is a conclusion also reached by
 \cite{depropris15,depropris16} in cluster early-type galaxies at $z>1$, where age gradients were associated with the presence of disc-like
morphologies.

In this paper, we jointly studied the rest-frame UV-U and U-R 
colour gradients for a sample of 17 elliptical galaxies morphologically selected in 
the cluster XMMU J2235.3-2557 at $z$=1.39. 
Here, we focus on the study of the colour gradients in cluster ellipticals at 
$z$ $\sim 1.4$, while we refer to a forthcoming paper for their evolution to $z$=0.
Our colour gradients are derived from optical and near-IR data for this 
cluster in the Hubble Space Telescope (HST) archive (see the next section). 
 
This paper is organized as follows. 
In Section 2 we describe the data and the sample selection. 
In Section 3 we derive the physical (stellar mass, age, absolute magnitudes) and 
structural (effective radius, index of concentration $n$) parameters of the galaxies 
of our sample. In Section 4 we derive the colour gradients with different methods. 
In Sections 5 and 6 we investigate the origin of the observed colour gradients
showing the key role played by the UV-U colour variation in the reconstruction
of the history of the mass assembly of elliptical galaxies. 
In Section 7 we compare our results with previous works. 
In Section 8 we summarize our results and present our conclusions.

Throughout this paper magnitudes are in the Vega system and we used a standard 
cosmology with $H_0 = 70$ km s$^{-1}$ Mpc$^{-1}$, 
$\Omega_{0}=0.3$ and $\Omega_{\Lambda}=0.7$. 
With these parameters, the age of the Universe at z = 1.39 is $\sim$ 4.5 Gyr.

\section{Data description}
\subsection{Optical and near-IR data}
The analysis presented in this paper is based on data obtained from the
Hubble Space Telescope (HST), Spitzer Space Telescope data and  
 the Very Large Telescope (VLT).

The HST data are composed of archival images of cluster XMMUJ2235-2557  \citep[]{mullis05} at $z$ = 1.39
obtained with the 
Advanced Camera for Surveys (ACS) in the F775W (5060 s) and 
F850LP (6240 s) filters ($i_{775}$ and $z_{850}$ hereafter)
and with the Wide Field Camera 3 (WFC3) in the F160W ($\sim$1200 s) filter.
The ACS images have a pixel scale of 0.05 arcsec pixel$^{-1}$ and a resolution of FWHM$_{850}$ $\simeq$ 0.11 arcsec.
ACS observations cover a field of about 11 arcmin$^2$
surrounding the cluster, while WFC3 observations a field of about 
5 arcmin$^2$.
At $z\simeq1.4$, the ACS F775W and F850LP bands probe the younger stellar 
populations in the near UV, while the WFC3 F160W closely corresponds to the 
rest-frame R band and traces the older stellar
populations in these galaxies.
The original WFC3 images have a pixel scale of  0.123 arcsec pixel$^{-1}$ and a 
resolution FWHM$_{160}\simeq0.2$ arcsec.
We have re-reduced these images with
 the software \textlcsc{ multidrizzle}  \citep{koekemoer02} to match the pixel 
 scale of the ACS images (0.05 arcsec pixel$^{-1}$), checking  
that the procedure had not introduced any spurious effects and conserved
the flux.

VLT High Acuity Wide field K-band Imager (HAWK-I) observations covering
a 13$\times$13 arcmin$^2$ region centred on the cluster were obtained
in the J and Ks filters ($\sim10500$ s each) under excellent seeing 
conditions 
(FWHM(J)$\simeq$ 0.5 arcsec and FWHM(Ks)$\simeq$ 0.35 arcsec).
These data are described in \cite{lidman08} and
\cite{lidman13}. 

U-band data were obtained from VIMOS at VLT, for a total exposure of 
21000 s and FWHM of 0.7 arcsec \cite[see][for details]{nonino09}.

Spitzer data in all four Infrared Array Camera (IRAC) bandpasses (3.6, 4.5, 5.8 and 8.0 $\mu$m) 
were retrieved as fully reduced images from the Spitzer Science Archive, 
with exposure times of $\sim$ 2000 s in all bands. 

\begin{table}
\caption{ 
Sample of 17 ETGs selected as cluster members at $z_{850}<24$, within 1 Mpc 
radius from the cluster centre and having a colour
$i_{775}-z_{850}=1.08\pm0.12$, i.e. within 2$\sigma$ from the mean colour
of the five galaxies spectroscopically confirmed cluster members \citep{rosati09}.
}
\centerline{
\begin{tabular}{rrrr}
\hline
\hline
 ID &    RA	   &	    Dec	  &   $z_{spec}$\\ 
\hline
358	&22:35:27.003	& -25:58:14.11 & ----   \\ 
595	&22:35:26.220	& -25:56:45.54 & ----   \\ 
684	&22:35:25.804	& -25:56:46.00 & ----   \\ 
692	&22:35:25.680	& -25:56:58.48 & ----  \\ 
837	&22:35:24.895	& -25:56:37.03 & ----   \\ 
1284	&22:35:22.814	& -25:56:24.92 & ----   \\ 
1539	&22:35:22.472	& -25:56:15.17 & ----  \\ 
1740	&22:35:20.839	& -25:57:39.76 & 1.39  \\ 
1747	&22:35:20.588	& -25:58:20.68 & ----  \\ 
1758	&22:35:20.920	& -25:57:35.90 & 1.39   \\ 
1782	&22:35:20.707	& -25:57:44.43 & 1.39   \\ 
1790	&22:35:20.707	& -25:57:37.70 & ----   \\ 
2054	&22:35:19.078	& -25:58:27.32 & ----   \\ 
2147	&22:35:19.046	& -25:57:51.42 & ----   \\ 
2166	&22:35:18.168	& -25:59:05.89 & ----  \\ 
2429	&22:35:17.867	& -25:56:13.06 & ----   \\ 
2809	&22:35:18.251	& -25:56:06.17 & ----   \\ 
 \hline									 
\hline									 
\end{tabular}								 
}
\label{tab:1}									 
\end{table}

\begin{figure}
\begin{center}
\includegraphics[width=9cm]{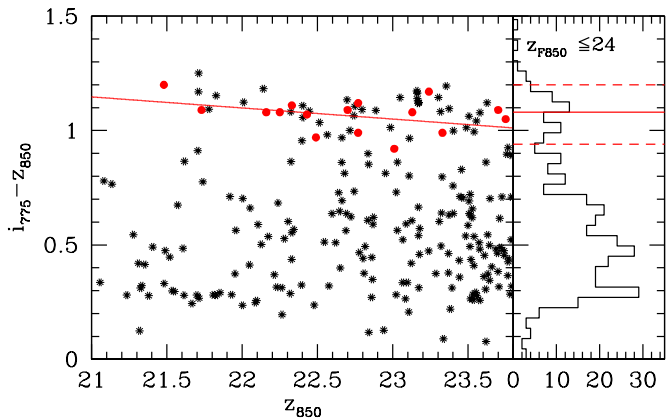}
\vskip -0.5truecm
\caption{\textit{Left panel}: colour--magnitude relation. The $i_{775}$  - $z_{850}$ colours of the 17 selected ellipticals (red points) as a function of $z_{850}$ are shown together with those of the 352 galaxies with $z_{850}<24$ (black asterisks). The solid red line is the colour-magnitude relation ($i_{775}$ - $z_{850}$) = 2.16($\pm$0.58) - 0.048($\pm$0.026)$z_{850}$ best-fitting the 17 ellipticals. \textit{Right panel}: the $i_{775}-z_{850}$ colour distribution of the 352 galaxies
brighter than $z_{850}<24$ falling within 1 Mpc from the cluster centre 
is shown. 
The red solid line marks the mean colour 
$\langle i_{775}-z_{850}\rangle=1.08\pm0.06$ of the 5 cluster members
spectroscopically confirmed \citep{rosati09}.
The dashed lines represent $\pm2\sigma$. A second peak in the distribution centered at the
mean colour of the five cluster galaxies is evident. 
}
\label{fig:ist}
\end{center}
\end{figure}

\subsection{Sample selection}
\begin{table*}
\caption{
For each galaxy of the sample we report the photometry in the 10 photometric bands.
Optical U-band  and near-IR J- and Ks-band magnitudes come from VIMOS and HAWKI 
VLT observations respectively; 
$i_{775}$, $z_{850}$ and  $H_{160}$ from ACS and WFC3 HST observations
in the F775W, F850LP and F160W filters respectively; m3.6, m4.5, m5.8 and m8.0 
from the Spitzer archival images in the corresponding filters.
We adopted the \textlcsc{MAG\_BEST} provided by SE\textlcsc{xtractor} as total magnitude.
A detailed description of the photometric measurements is given in Section 2.3.
}
\centerline{
\begin{tabular}{rrrrrrrrrrr}
\hline
\hline
 ID& U&   $i_{775}$&      $z_{850}$&   J&  $H_{160}$&	 Ks&  m3.6       &      m4.5	 &	m5.8	  &	m8.0	     \\ 
\hline
358  &  $>$28	 99  & 24.85$\pm$0.05& 23.91$\pm$0.04 & 22.14$\pm$0.1  &  ---             & 20.41$\pm$0.05&  19.84$\pm$0.1  &  19.69$\pm$0.1   &  19.53$\pm$0.4 & 19.33$\pm$0.4 \\  
595  &  27.8$\pm$0.3 & 23.17$\pm$0.02& 22.16$\pm$0.01 & 20.23$\pm$0.03 & 19.35$\pm$0.03 & 18.35$\pm$0.01&  16.70$\pm$0.1  &  16.23$\pm$0.1   &  15.99$\pm$0.1 & 15.94$\pm$0.2 \\	
684  &  25.9$\pm$0.1 & 24.19$\pm$0.03& 23.33$\pm$0.03 & 21.92$\pm$0.07 & 21.10$\pm$0.06 & 20.11$\pm$0.03&  18.15$\pm$0.02 &  17.79$\pm$0.03  &  17.31$\pm$0.1 & 17.41$\pm$0.2 \\  
692  &  $>$28	 99  & 24.80$\pm$0.06& 23.75$\pm$0.04 & 21.90$\pm$0.08 & 21.12$\pm$0.06 & 19.97$\pm$0.03&  18.69$\pm$0.1  &  18.46$\pm$0.1   &  18.29$\pm$0.3 & 18.03$\pm$0.3 \\  
837  &  $>$28	 99  & 23.76$\pm$0.03& 22.77$\pm$0.03 & 20.59$\pm$0.05 & 19.83$\pm$0.04 & 18.61$\pm$0.02&  16.99$\pm$0.01 &  16.49$\pm$0.02  &  16.38$\pm$0.08& 16.10$\pm$0.1 \\  
1284 &  $>$28	 99  & 23.65$\pm$0.02& 22.70$\pm$0.02 & 20.91$\pm$0.04 & 20.22$\pm$0.04 & 19.16$\pm$0.02&  17.56$\pm$0.01 &  17.09$\pm$0.02  &  16.77$\pm$0.09& 16.84$\pm$0.1 \\  
1539 &  26.02$\pm$0.1& 23.86$\pm$0.02& 23.01$\pm$0.02 & 21.28$\pm$0.05 &  ---             & 19.34$\pm$0.02&  17.63$\pm$0.01 &  17.20$\pm$0.03  &  17.14$\pm$0.1 & 16.87$\pm$0.1 \\  
1740 &  24.30	0.05 & 22.42$\pm$0.02& 21.48$\pm$0.02 & 18.55$\pm$0.02 & 18.68$\pm$0.02 & 17.44$\pm$0.01&  15.60$\pm$0.1  &  15.17$\pm$0.1   &  15.17$\pm$0.1 & 14.82$\pm$0.2 \\  
1747 &  $>$28	 99  & 24.79$\pm$0.05& 23.70$\pm$0.04 & 22.06$\pm$0.09 & 21.25$\pm$0.06 & 20.08$\pm$0.03&  18.73$\pm$0.02 &  18.41$\pm$0.04  &  18.56$\pm$0.3 & 99.00$\pm$99.0\\  
1758 &  $>$28	 99  & 24.16$\pm$0.03& 23.13$\pm$0.02 & 20.46$\pm$0.04 & 20.33$\pm$0.04 & 19.09$\pm$0.01&  17.58$\pm$0.1  &  17.14$\pm$0.1   &  16.70$\pm$0.1 & 16.89$\pm$0.2 \\  
1782 &  24.6$\pm$0.5 & 23.45$\pm$0.02& 22.49$\pm$0.02 & 19.97$\pm$0.03 & 19.60$\pm$0.04 & 18.40$\pm$0.01&  16.99$\pm$0.1  &  16.58$\pm$0.1   &  16.55$\pm$0.1 & 16.49$\pm$0.2 \\  
1790 &  $>$28	 99  & 23.22$\pm$0.02& 22.25$\pm$0.02 & 19.70$\pm$0.02 & 19.35$\pm$0.03 & 18.18$\pm$0.01&  16.34$\pm$0.1  &  15.91$\pm$0.1   &  15.91$\pm$0.1 & 15.56$\pm$0.2 \\  
2054 &  $>$28	 99  & 23.43$\pm$0.02& 22.43$\pm$0.02 & 20.67$\pm$0.03 & 19.86$\pm$0.04 & 18.83$\pm$0.01&  16.96$\pm$0.01 &  16.61$\pm$0.03  &  16.27$\pm$0.1 & 15.79$\pm$0.1 \\  
2147 &  $>$28	 99  & 23.78$\pm$0.03& 22.77$\pm$0.03 & 20.90$\pm$0.04 & 20.15$\pm$0.04 & 19.02$\pm$0.02&  16.82$\pm$0.02 &  16.36$\pm$0.03  &  16.37$\pm$0.1 & 15.42$\pm$0.1 \\  
2166 &  26.50$\pm$0.1& 22.80$\pm$0.01& 21.73$\pm$0.01 & 20.53$\pm$0.03 &  ---             & 18.73$\pm$0.01&  17.33$\pm$0.01 &  17.13$\pm$0.03  &  17.02$\pm$0.1 & 16.68$\pm$0.1 \\  
2429 &  $>$28	 99  & 24.41$\pm$0.03& 23.24$\pm$0.02 & 21.96$\pm$0.08 & 21.04$\pm$0.06 & 20.05$\pm$0.03&  18.15$\pm$0.02 &  17.98$\pm$0.04  &  17.91$\pm$0.2 & 17.36$\pm$0.2 \\  
2809 &  $>$28	 99  & 23.41$\pm$0.01& 22.33$\pm$0.01 & 20.92$\pm$0.03 &  ---             & 19.28$\pm$0.01&  18.36$\pm$0.1  &  17.99$\pm$0.1   &  17.96$\pm$0.2 & 16.71$\pm$0.2 \\  
 \hline									 
\hline									 
\end{tabular}								 
}	
\label{tab:sex}								 
\end{table*}

Our sample consists of 17 galaxies that we have
selected to belong to the cluster XMMUJ2235-2557 at $z=1.39$ \citep{mullis05,rosati09}.
We selected our final sample
following the criteria described in \cite{saracco14}. 
Briefly, we first
detected all the sources ($\sim3000$) in the F850LP band image up to the 
magnitude limit $z_{850}\simeq27$
in the $\sim11$ arcmin$^2$ region surrounding the cluster. 
We ran SE\textlcsc{xtractor} \citep{bertin96} in double image mode on the F850LP and F775W band
images, using the F850LP band image as the reference 
image and adopted the
\textlcsc{MAG\_BEST} as our fiducial estimate of total magnitude.
In order to perform a reliable and robust visual morphological classification,
we selected galaxies with magnitudes $z_{850}\le24$.
At this magnitude limit, the sample 
is 100 per cent complete.
From this flux limited sample we removed stars using SE\textlcsc{xtractor}'s stellarity index (\textlcsc{CLASS\_STAR > 0.9}) and we only considered galaxies 
within $\le1$ Mpc ($\sim120$ arcsec) from the cluster centre. This leaves 352 galaxies, 219 of which also have IR data from WFC3. 
This sample contains the 5 central cluster member galaxies 
spectroscopically confirmed by \cite{rosati09}, three out of which,
1740, 1758 and 1782 of Table \ref{tab:1}, are
ellipticals
(see below for the morphological classification).
The mean colour of these five cluster member ETGs is 
$\langle i_{775}-z_{850}\rangle=1.08$
with a dispersion $\sigma_{iz}=0.06$. 
This colour well traces the redshift of the galaxies.
In particular, for z < 0.8 the $ i_{775}-z_{850}$ colour is always < 0.8 mag, while for z > 0.8 - 0.9 its value rapidly increases and remains always > 0.8 mag, independently of the age of the stellar population considered \citep[see e.g.][]{saracco14}.
In Fig. \ref{fig:ist} (right panel) we show the $i_{775}-z_{850}$ colour distribution
of the 352 galaxies with $z_{850}\le24$, which reflects the behaviour described above. Two peaks are present.
We selected all the galaxies in the colour range 
$0.96<i_{775}-z_{850}<1.2$ mag defining the second peak of the colour distribution centred at the mean colour (marked by red solid line in the figure) of the five red sequence member galaxies. 
The dashed red lines represent $\pm 2 \sigma$ from the mean colour of the five cluster members.

According to these criteria, 50 galaxies have been selected as cluster 
member candidates. 
Thus, in order to identify the elliptical galaxies, we have performed a 
morphological classification based on the
visual inspection of the galaxies
on the ACS-F850LP image and on the fitting of their surface
brightness profile as described in Section 3.2.
We regard
galaxies as ellipticals if they have regular shapes, no signs of a disc and smooth residuals after profile fitting and model subtraction.
On this basis, 
17 galaxies out of the 50 
turned out to be ellipticals.
The sample is summarized in Table \ref{tab:1}.

Fig. \ref{fig:ist} (left panel) shows the $i_{775}$  - $z_{850}$ colours of the 17 selected ellipticals (red points) as a function of $z_{850}$ together with those of the 352 galaxies with $z_{850}<24$ (black asterisks). The selected ellipticals define a red sequence, as expected. The solid red line is the colour-magnitude relation ($i_{775}$ - $z_{850}$) = 2.16($\pm$0.58) - 0.048($\pm$0.026)$z_{850}$ best fitting the 17 ellipticals.

\subsection{Multiwavelength photometry}

We measured the magnitudes for the 17 galaxies selected in all the available
bands  using SE\textlcsc{xtractor}. We adopted the \textlcsc{MAG\_BEST} as the best
estimator of the magnitude.

As mentioned in Section 2.1, the field of view of the WFC3 is smaller than the ACS
field (5 arcmin$^2$ instead of 11 arcmin$^2$); hence, only 13 ellipticals out of
the 17 are covered by WFC3 observations, with galaxies 358, 1539, 2166 and
2809 falling outside. 

VIMOS U-band observations cover the whole sample to a limiting depth of
U $\simeq28$. Thank to this depth, we detected six galaxies at this wavelength.

HAWKI J- and Ks-band images cover the whole sample. Thanks to the excellent
resolution of the images all the galaxies are resolved in both the filters.

\begin{figure}
\includegraphics[width=2.7cm]{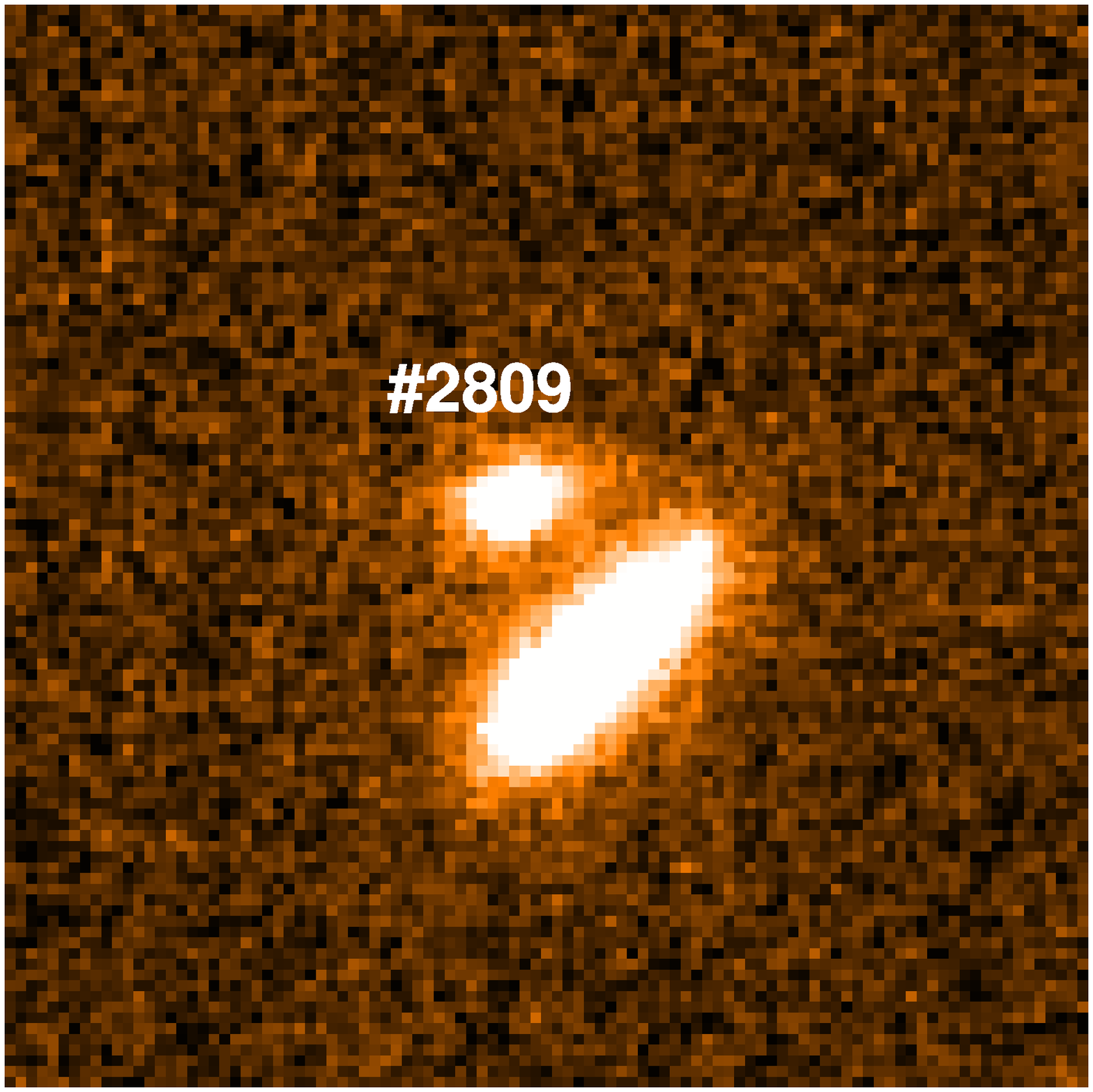}
\includegraphics[width=2.7cm]{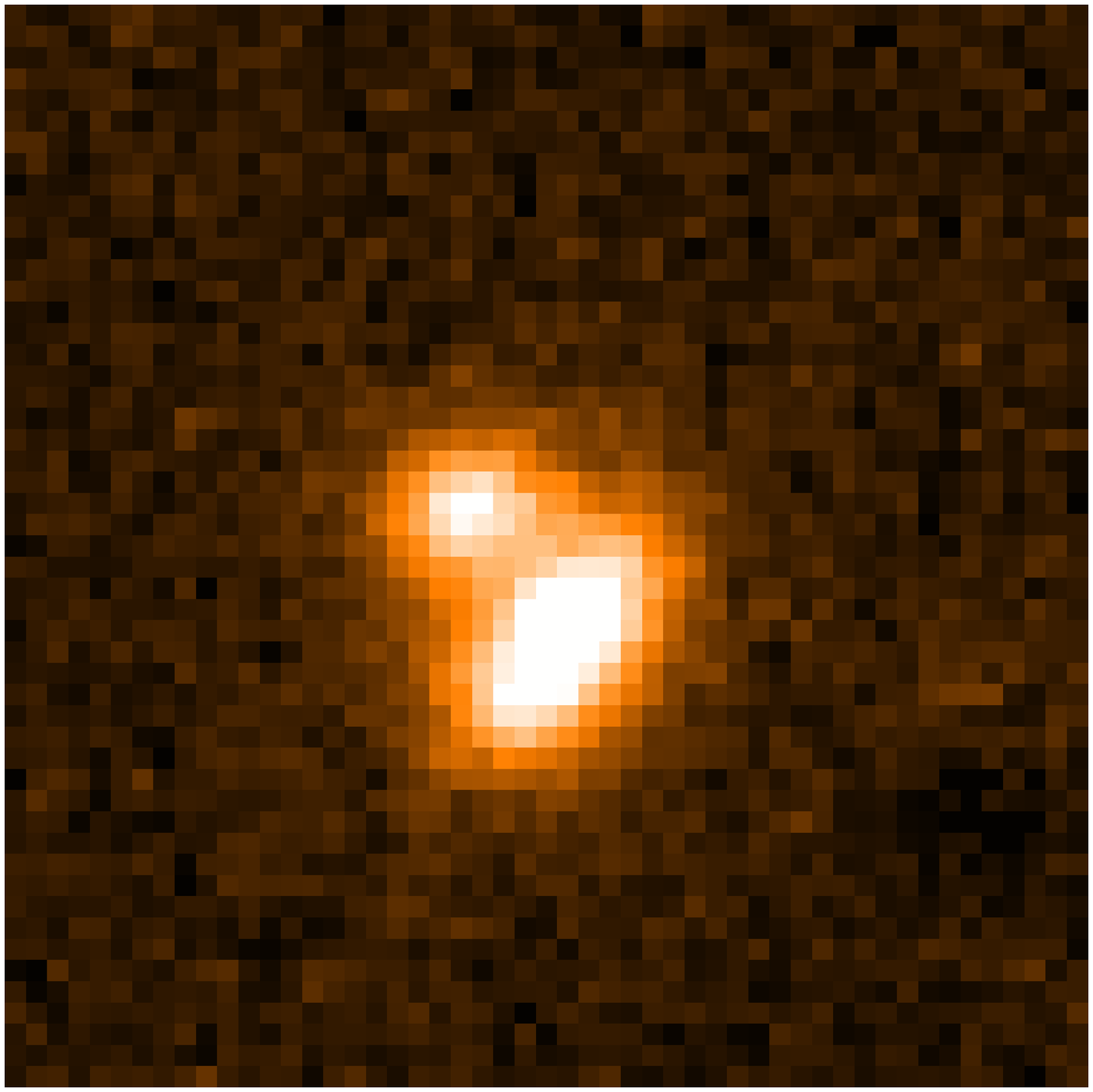}
\includegraphics[width=2.7cm]{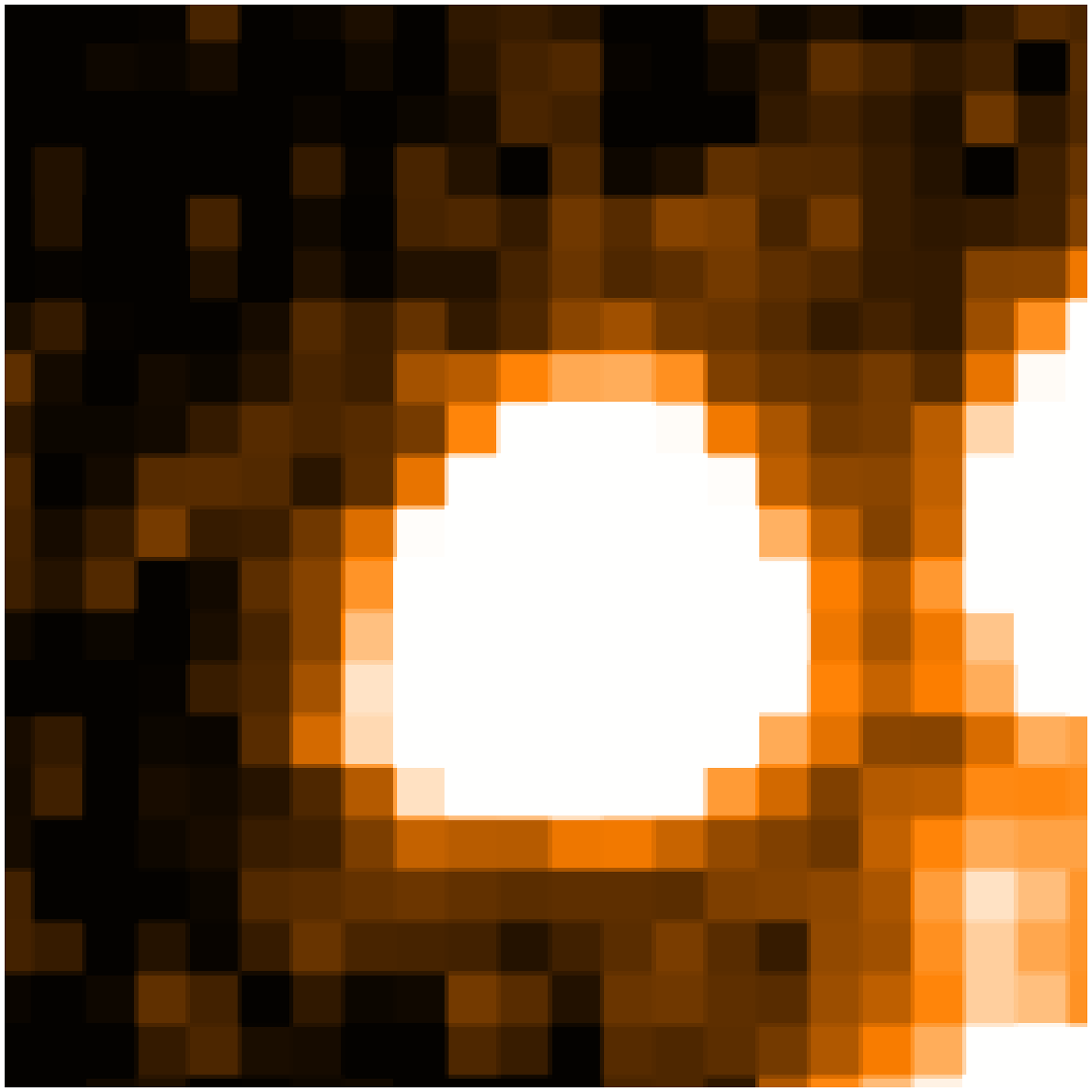}
\caption{We show galaxy 2809, which is closed to another galaxy. 
The two galaxies are resolved both in the F850LP band image (\textit{left panel}) 
(FWHM $\sim$ 0.12 arcsec) and in the HAWKI J band image (FWHM $\sim$ 0.5 arcsec) 
(\textit{central panel}), whereas they are completely blended in the Spitzer - IRAC 3.6
 $\mu$m band image due to the lower resolution (FWHM > 2.7 arcsec) and bigger pixel size 
 (0.6 arcsec pixel$^{-1}$) (\textit{right panel}). The F850LP and J images are 5 $\times$ 5 arcsec$^2$, and the IRAC 3.6 
 $\mu$m image is 10 $\times$ 10 arcsec$^2$.}
\label{fig:ex}
\end{figure}

Also Spitzer-IRAC observations cover the whole sample. Magnitudes have been
estimated in the four IRAC bands using SE\textlcsc{xtractor} in double image mode adopting
the 3.6 $\mu$m image as reference. We have tested for the reliability of the
flux measurement by comparing the flux of some stars in the field measured with Sextractor with the flux obtained using the IRAF task phot.  Due to the low
resolution (FWHM > 2.7 arcsec), eight galaxies (2809, 595, 692, 358, 1740,
1790, 1782 and 1758) are not resolved but appear fully blended with other
close sources. For these galaxies, we have derived their magnitudes in the four
IRAC bands  by redistributing the total IRAC flux of the blended sources
according the fluxes measured for each of them in the F160W and Ks filters. 
In Fig. \ref{fig:ex} we show as example galaxy 2809, which is closed to another
galaxy, in the ACS-F850LP ({left panel}), HAWKI J ({central panel}) and Spitzer-IRAC 
3.6 $\mu$m ({right panel}) images. 
The two galaxies are clearly resolved in the F850LP and in the HAWKI J band images, while they are 
fully blended in the Spitzer-IRAC 3.6 $\mu$m image.

In Table \ref{tab:sex} we report the photometry in the 10 photometric bands for
the 17 ellipticals of the sample.

\section{Physical and structural parameters}
\subsection{Global physical properties of galaxies}
\label{SED}
For each galaxy we derived the mean global stellar parameters (stellar mass, age, 
absolute magnitudes) 
by fitting the spectral energy distribution (SED) defined by the 10 available 
photometric points. 
The fit has been performed with the software \textlcsc{hyperzmass} \citep{bolzonella00} 
using a 
set of templates characterized by the composite stellar population models of 
\citet[hereafter BC03]{bruzual03}, the Chabrier {initial mass function}  
(IMF) \citep{chabrier03}, an exponentially declining star formation history 
(SFH $\propto exp^{-t/\tau}$). 
In all the cases, we considered five star formation time-scale 
($\tau = [0.1,0.3,0.4,0.6,1]$ Gyr) and solar metallicity Z$_{\odot}$. 
For the extinction curve we adopted the Calzetti law \citep{calzetti00}, with $A_V$ left as a free parameter and allowed to vary in the
range 0.0 to 0.6 mag.

The F775W, F850LP and F160W bands sample $\lambda_{rest} \sim 3200$ \AA, 
$\lambda_{rest} \sim 3800$ \AA \, and $\lambda_{rest} \sim 6400$ \AA, respectively, 
at the cluster redshift.
Hence, we derived 
absolute magnitudes in UV, U and R bands (M$_{UV}$, M$_{U}$ and M$_{R}$, respectively), with a $k-$correction derived from the best fitting
template. 
In Table \ref{tab:2} we report the best-fitting values for the age, the logarithm of 
the stellar mass, M$_{UV}$, M$_{U}$ and M$_{R}$ for the 17 galaxies of the sample.

For 14 galaxies out of the 17, the best fitting template is defined by a SFH with 
$\tau = 0.1$ Gyr, while the remaining three galaxies are characterized by  
SFHs with $\tau = 0.3$ Gyr (837 and 1539) and with $\tau = 0.6$ Gyr (1740). 
The 17 ellipticals have stellar masses in the range 
$1.2\times 10^{10}$ M$_{\odot}< M_{\ast}<8.5\times 10^{11}$ M$_{\odot}$ 
with a mean value $M_{\ast} \simeq 3\times10^{10}$ M$_{\odot}$ 
and ages in the range 0.7 Gyr < age $\le$ 4.3 Gyr with a mean value of $\sim 1.7$ Gyr.

\begin{table}
\caption{ 
For each of the 17 galaxies of the sample we report the age, the logarithm of the stellar mass and the absolute magnitudes M$_{UV}$, M$_{U}$ and M$_{R}$ derived from the best fitting template.
}
\centerline{
\begin{tabular}{cccccc}
\hline
\hline
 ID &    Age	   &	    log $M_{\ast}$	  &   $ M_{UV} $ &$M_{U}$ & $ M_{R}$\\ 
         &  [Gyr]        &        [M$_{\odot}$]                & [mag]& [mag]& [mag]\\
\hline
358	& 0.72	& 10.08 &-20.14&-20.19  &-21.18\\ 
595	& 3.00	& 11.56 &-21.83 &-21.95  &-23.62\\ 
684	& 1.70	& 10.56 &-20.71&-20.79  &-21.91\\ 
692	& 1.14	& 10.56 &-20.23&-20.35  &-21.87\\ 
837	& 2.50	& 11.47 &-21.22&-21.36  &-23.22\\ 
1284	& 1.01	& 11.05 &-21.29 &-21.41  &-22.88\\ 
1539	& 1.80	& 11.01 &-21.07&-21.15  &-22.67\\ 
1740	& 3.50      & 11.93 &-22.56&-22.65  &-24.39\\ 
1747	& 1.02	& 10.55 &-20.22&-20.34  &-21.77\\ 
1758	& 1.80	& 11.14 &-20.85 &-20.99&-22.73  \\ 
1782	& 1.90	& 11.57 &-21.52&-21.65 &-23.43 \\ 
1790	& 4.25	& 11.60 &-21.76&-21.89  &-23.71\\ 
2054	& 1.70	& 11.22 &-21.55&-21.66  &-23.21\\ 
2147	& 1.14	& 11.01 &-21.20&-21.31  &-22.90\\ 
2166	& 0.90	& 10.99 &-22.19&-22.27  &-23.39\\ 
2429	& 0.72	& 10.58 &-20.66&-20.76  &-21.98\\ 
2809	& 0.65	& 10.73 &-21.59&-21.67  &-22.74\\
\hline
Mean&1.73 &10.45 &-21.21 &-21.32 &-22.80 \\
 \hline									 
\hline									 
\end{tabular}								 
}
\label{tab:2}									 
\end{table}

\subsection{Structural parameters} 
\label{Structural parameters}

\begin{table*}
\caption{For each galaxy of the sample we report the ID number, the S\'{e}rsic index $n$, the
total magnitude and the effective radius $R_{e}$ [kpc] as derived from
the fitting of the surface brightness profile in the F775W, F850LP
and F160W images, taking as reference the position angle and the axis ratio $(b/a)_{850}$
obtained in the F850LP image. For each galaxy we report, in the second row, the values
obtained for a de Vaucouleur's profile ($n=4$). }
\centerline{
\begin{tabular}{rccccccccccc}
\hline
\hline
  ID &$n_{775}$& $i_{775}^{fit}$&R$_e^{775}$ & $n_{850}$&$z_{850}^{fit}$& R$_e^{F850}$&(b/a)$_{850}$&$n_{160}$ & $H_{160}^{fit}$ &R$_e^{F160}$  \\
     &     &[mag]  & [kpc]                 &     & [mag]    &[kpc]   &  &           &   [mag]     & [kpc] \\
\hline
    358&  6.0$\pm$0.4 &  24.5$\pm$0.2&  1.1$\pm$0.1 &  6.0$\pm$0.4   &  23.5$\pm$0.1 & 1.5$\pm$0.2 & 0.6$\pm$0.1&  ---         &   ---            &  ---\\ 
    "&  --- &  24.8$\pm$0.2&  0.8$\pm$0.1 &  4.0   &  23.7$\pm$0.1 & 1.0$\pm$0.2&  ---       &   ---  &---	  &  ---\\ 
    595&  5.7$\pm$0.3 &  22.9$\pm$0.1&  2.2$\pm$0.3 &  5.6$\pm$0.3   &  21.6$\pm$0.1 & 3.5$\pm$0.5&0.9$\pm$0.1& 4.5$\pm$0.1 &   19.2$\pm$0.1 &  2.5$\pm$0.1\\ 
    "&  --- &  23.1$\pm$0.1&  1.5$\pm$0.3 &  4.0   &  21.9$\pm$0.1 & 2.2$\pm$0.5&  ---   &  ---     &   19.2$\pm$0.1 &  2.3$\pm$0.1\\ 
    684&  6.0$\pm$0.4 &  24.0$\pm$0.2&  0.7$\pm$0.1 &  6.0$\pm$0.4   &  22.9$\pm$0.1 & 1.4$\pm$0.2 & 0.6$\pm$0.1& 4.1$\pm$0.1 &   20.9$\pm$0.1 &  0.8$\pm$0.1\\ 
    "&  --- &  24.2$\pm$0.2&  0.5$\pm$0.1 &  4.0   &  23.1$\pm$0.1 & 1.0$\pm$0.2&  ---   &  ---     &   20.9$\pm$0.1 &  0.8$\pm$0.1\\ 
    692&  4.5$\pm$0.2 &  24.3$\pm$0.2&  1.3$\pm$0.1 &  4.1$\pm$0.2   &  23.5$\pm$0.2 & 1.5$\pm$0.2 & 0.7$\pm$0.1 &3.0$\pm$0.1 &   21.0$\pm$0.1 &  0.9$\pm$0.1\\ 
    "&  --- &  24.7$\pm$0.2&  1.3$\pm$0.1 &  4.0   &  23.5$\pm$0.2 & 1.5$\pm$0.2&  ---   &  ---     &   20.9$\pm$0.1 &  1.1$\pm$0.1\\ 
    837&  6.0$\pm$0.2 &  23.3$\pm$0.1&  5.2$\pm$0.6 &  6.0$\pm$0.4   &  21.9$\pm$0.1 & 7.8$\pm$1.1 & 0.7$\pm$0.1 &6.0$\pm$0.2 &   19.4$\pm$0.1 &  5.2$\pm$0.2\\
    "&  --- &  23.6$\pm$0.1&  2.3$\pm$0.6 &  4.0   &  22.3$\pm$0.1 & 8.0$\pm$1.1&  ---   &  ---     &   19.7$\pm$0.1 &  7.2$\pm$0.2\\
  1284&  4.3$\pm$0.3 &  23.4$\pm$0.1&  1.8$\pm$0.3 &  4.3$\pm$0.2   &  22.3$\pm$0.1 & 2.4$\pm$0.3 & 0.9$\pm$0.1&4.0$\pm$0.1 &   20.0$\pm$0.1 &  1.7$\pm$0.1\\ 
   "&  --- &  23.4$\pm$0.1&  1.9$\pm$0.3 &  4.0   &  22.3$\pm$0.1 & 2.3$\pm$0.3&  ---   &  ---     &   20.0$\pm$0.1 &  1.7$\pm$0.1\\ 
   1539&  4.8$\pm$0.3 &  23.7$\pm$0.2&  1.1$\pm$0.1 &  3.5$\pm$0.2   &  22.7$\pm$0.1 & 1.5$\pm$0.2 & 0.6$\pm$0.1& ---         &   ---            &  ---         \\ 
   "&  --- &  23.8$\pm$0.2&  1.0$\pm$0.1 &  4.0   &  22.6$\pm$0.1 & 1.7$\pm$0.2&  ---   &  ---     &   ---		 &  --- 	\\ 
   1740&  3.3$\pm$0.2 &  21.8$\pm$0.1& 12.7$\pm$1.9 &  3.3$\pm$0.2  &  20.8$\pm$0.1 &12.8$\pm$1.9& 0.6$\pm$0.1& 4.6$\pm$0.1 &   17.8$\pm$0.1 &  20.7$\pm$0.9\\ 
   "&  --- &  21.5$\pm$0.1& 20.1$\pm$2.0 &  4.0   &  20.5$\pm$0.1 &19.5$\pm$1.9&  ---   &  ---     &   18.0$\pm$0.1 &  15.4$\pm$0.9\\ 
   1747&  4.3$\pm$0.2 &  24.6$\pm$0.2&  0.8$\pm$0.1 &  4.8$\pm$0.3   &  23.3$\pm$0.1 & 1.7$\pm$0.2 & 0.6$\pm$0.1& 4.0$\pm$0.2 &   21.1$\pm$0.1 &  0.9$\pm$0.1\\ 
   "&  --- &  24.7$\pm$0.2&  0.7$\pm$0.1 &  4.0   &  23.4$\pm$0.1 & 1.5$\pm$0.2&  ---   &  ---     &   21.1$\pm$0.1 &  0.9$\pm$0.1\\ 
   1758&  2.9$\pm$0.2 &  23.9$\pm$0.2&  1.9$\pm$0.2 &  2.9$\pm$0.2   &  22.6$\pm$0.1 & 2.5$\pm$0.3 & 0.8$\pm$0.1& 2.90$\pm$0.04&   20.3$\pm$0.1 &  1.5$\pm$0.1\\ 
   "&  --- &  23.7$\pm$0.2&  1.9$\pm$0.2 &  4.0   &  22.4$\pm$0.1 & 3.8$\pm$0.3&  ---   &  ---     &   20.1$\pm$0.1 &  1.0$\pm$0.1\\ 
   1782&  2.5$\pm$0.2 &  23.5$\pm$0.1&  1.7$\pm$0.1 &  3.6$\pm$0.2   &  22.5$\pm$0.1 & 3.4$\pm$0.5 &0.6$\pm$0.1& 3.11$\pm$0.04&  19.4$\pm$0.1 &  2.2$\pm$0.1\\ 
   "&  --- &  22.9$\pm$0.1&  3.4$\pm$0.1 &  4.0   &  22.0$\pm$0.1 & 3.5$\pm$0.5&  ---   &  ---     &   19.4$\pm$0.1 &  2.0$\pm$0.1\\ 
   1790&4.4$\pm$0.3 &  23.1$\pm$0.1&  1.9$\pm$0.2 &  4.4$\pm$0.3   &  21.9$\pm$0.1 & 2.4$\pm$0.3 & 0.6$\pm$0.1&4.1$\pm$0.1 &   19.3$\pm$0.1 &  2.3$\pm$0.1\\ 
   "&  --- &  23.2$\pm$0.1&  1.8$\pm$0.2 &  4.0   &  22.0$\pm$0.1 & 2.2$\pm$0.3&  ---   &  ---     &   19.3$\pm$0.1 &  2.2$\pm$0.1\\ 
   2054&  5.1$\pm$0.3 &  23.2$\pm$0.1&  2.9$\pm$0.4 &  4.4$\pm$0.3   &  21.9$\pm$0.1 & 4.2$\pm$0.6 & 0.7$\pm$0.1& 5.1$\pm$0.1 &   19.5$\pm$0.1 &  3.7$\pm$0.2\\
   "& --- &  23.3$\pm$0.1&  2.2$\pm$0.4 &  4.0   &  22.0$\pm$0.1 & 3.7$\pm$0.6&  ---   &  ---     &   19.6$\pm$0.1 &  2.9$\pm$0.2\\
   2147&   5.4$\pm$0.3 &  23.5$\pm$0.1&  3.6$\pm$0.5 &  5.3$\pm$0.3   &  22.2$\pm$0.1 & 5.3$\pm$0.8 & 0.7$\pm$0.1& 4.1$\pm$0.1 &   20.0$\pm$0.1 &  2.3$\pm$0.1\\
   "&    --- &  23.7$\pm$0.1&  2.3$\pm$0.5 &  4.0   &  22.4$\pm$0.1 & 3.4$\pm$0.8&  ---   &  ---     &   20.0$\pm$0.1 &  2.3$\pm$0.1\\
   2166&  3.6$\pm$0.2 &  22.6$\pm$0.1&  1.2$\pm$0.1 &  3.0$\pm$0.2   &  21.5$\pm$0.1 & 1.4$\pm$0.1 & 0.6$\pm$0.1& ---         &   ---            &  ---         \\
   "&  --- &  22.6$\pm$0.1&  1.3$\pm$0.1 &  4.0   &  21.4$\pm$0.1 & 1.7$\pm$0.1&  ---   &  ---     &   ---		 &  --- 	\\
   2429&  2.1$\pm$0.4 &  24.3$\pm$0.2&  0.7$\pm$0.1 &  2.1$\pm$0.2   &  23.0$\pm$0.1 & 0.9$\pm$0.1& 0.6$\pm$0.1& 2.1$\pm$0.1&   21.0$\pm$0.1 &  0.9$\pm$0.1\\
   "&  --- &  24.3$\pm$0.2&  0.7$\pm$0.1 &  4.0   &  22.8$\pm$0.1 & 1.2$\pm$0.1&  ---   &  ---     &   20.9$\pm$0.1 &  1.0$\pm$0.1\\
   2809&  4.8$\pm$0.3 &  23.3$\pm$0.1&  1.2$\pm$0.1 &  3.2$\pm$0.2   & 22.7$\pm$0.1 & 1.4$\pm$0.2 & 0.6$\pm$0.1& --- & ---    & ---  &\\
   "&  --- &  23.3$\pm$0.1&  1.1$\pm$0.1 &  4.0   &  22.0$\pm$0.1 & 1.6$\pm$0.2&  ---   &  ---     &   ---	       & ---  &\\
\hline
\end{tabular}
}
\label{tab:tab_gal_fix}
\end{table*}

We derived the structural parameters of the ellipticals of our sample by fitting their 
surface brightness profile with a single S\'ersic law

\begin{equation}
\mu(r)=\mu_e+\frac{2.5b_n}{ln(10)}[(r/r_e)^{1/n}-1]
\label{eq:sersic}
\end{equation}
in the F775W, F850LP and F160W bands.
The two-dimensional fitting  has been performed 
using \textlcsc{Galfit} software \citep{peng02},
which returns as best-fitting set of structural parameters
the one that minimizes the residuals between the model convolved with the PSF
and the observed image.
A detailed description of this procedure may be found in Appendix \ref{galfit}. 
The fitting was performed both assuming $n$ as a free parameter and assuming $n$ = 4 (appropriate for ellipticals that have a de Vaucouleurs profile). 
In order to test for the dependence  of our results on the PSF used, 
for each galaxy we have convolved the S\'ersic profile 
with two different PSFs, derived empirically (as recommended in the \textlcsc{Galfit} manual) from two bright, unsaturated and uncontaminated
stars in the field.
The residual maps lacked of any structure in both the cases and the returned 
$\chi^2$ values were not statistically different, showing that the results do not 
depend on the PSF used. 
Thus, we chose the mean values of the fitting results as best fitting value 
for the structural parameters. 
The structural parameters obtained through the fit are shown in Table \ref{tab:tab_gal}.

In order to measure colour gradients (see Section \ref{Color gradients}), 
we need to measure the radial surface brightness profiles in ellipses having the same orientation
and ellipticity in all bands in order to prevent any artificial 
gradients.
Hence, we re-run \textlcsc{Galfit} on the F775W and the F160W images fixing the position angle $PA$ and the axial ratio $b/a$  to those
derived for the F850LP image.
We choose the F850LP image as reference band because it has better resolution 
(FWHM $\sim 0.11$ arcsec) and the smallest pixel scale (0.05 arcsec pixel$^{-1}$).
Moreover, given the shallower surface brightness, it determines the maximum radius at which
we can reliably estimate radial gradients (see discussion in the next section).
In Table \ref{tab:tab_gal_fix} we show for each galaxy the best-fitting $n$, $R_{e}$ 
[kpc] and $m_{tot}$  in the F775W, F850LP and F160W bands obtained taking as reference 
the position angle and the axis ratio derived in the F850LP image. 
In the same table, we also show in the second row the parameters obtained assuming
$n=4$, i.e. de Vaucouleur's profile. 
The effective radii $R_e$, as derived by fitting the S\'ersic law in the F850LP image, 
are in the range 0.9-7.8 kpc, with the exception of the dominant central galaxy of 
the cluster, which has an effective radius R$_e$ = 12.8 kpc. 
The S\'ersic index varies in the range 2.1 - 6, while the total magnitude $m_{tot}$ 
varies in the range 20.78 - 23.47 mag. 

We have tested the reliability of the surface brightness profile fitting and we have estimated the uncertainties on the structural parameters by using simulated galaxies,
as described in Appendix \ref{simulations}.

\section{Colour gradients}
\label{Color gradients}

\begin{table*}
\caption{ 
For each of the 17 galaxies of the sample we report the F775W - F850LP and 
F850LP - F160W colour gradients derived as the logarithmic slope of the colour 
profile within 1R$_e$ and 2R$_e$ and via the ratio of the effective radii as 
measured in the F775W, F850LP and F160W filters.
}
\centerline{
\begin{tabular}{ccccccc}
\hline
\hline
 ID &    $\nabla_{F775W-F850LP}$   &	$(\nabla_{F775W-F850LP})_{2r_e}$  &  $\nabla_{F850LP-F160W}$& $(\nabla_{F850LP-F160W})_{2r_e}$ &$log(R_{775}/R_{850})$  & $log(R_{850}/R_{160})$\\ 
\hline
358	& 0.2$\pm$0.1	&0.3$\pm$0.2&   --&--& -0.2& --\\ 
595	& 0.4$\pm$0.2	&0.4$\pm$0.2& -0.1$\pm$0.1&-0.2$\pm$0.2& -0.2 & 0.1\\ 
684	& 0.6$\pm$0.1	&0.6$\pm$0.2& -0.3$\pm$0.1&-0.5$\pm$0.2& -0.4 & 0.2\\ 
692	& 0.2$\pm$0.2	&0.2$\pm$0.2& -0.3$\pm$0.1&-0.5$\pm$0.3& -0.1 & 0.2\\ 
837	& 0.4$\pm$0.3	&0.4$\pm$0.3&-0.3$\pm$0.2&-0.3$\pm$0.2& -0.3 & 0.2\\ 
1284	& 0.2$\pm$0.1	&0.2$\pm$0.2& -0.3$\pm$0.1&-0.3$\pm$0.2& -0.1 & 0.2 \\ 
1539	& 0.5$\pm$0.2	&0.4$\pm$0.2& --&--& -0.1 & --\\ 
1740	& 0.0$\pm$0.2  &0.0$\pm$0.2& 0.05$\pm$0.10&0.06$\pm$0.2& 0.02 & -0.2\\ 
1747	& 0.7$\pm$0.2	&0.8$\pm$0.3& -0.4$\pm$0.1&-0.6$\pm$0.2& -0.4 & 0.3\\ 
1758	& 0.4$\pm$0.2	&0.5$\pm$0.2& -0.7$\pm$0.2&-0.7$\pm$0.3& -0.1 & 0.2\\ 
1782	& 0.6$\pm$0.1	&1.0$\pm$0.3& -0.3$\pm$0.2&-0.4$\pm$0.2& -0.3 & 0.2\\ 
1790	& 0.2$\pm$0.1	&0.2$\pm$0.2&  0.0$\pm$0.1&-0.04$\pm$0.02& -0.1 & 0.2\\ 
2054	& 0.5$\pm$0.1	&0.4$\pm$0.2& -0.2$\pm$0.2&-0.2$\pm$0.2& -0.2 & 0.1\\ 
2147	& 0.3$\pm$0.2	&0.3$\pm$0.3& -0.6$\pm$0.3&-0.7$\pm$0.3& -0.1 & 0.1\\ 
2166	& 0.3$\pm$0.1	&0.3$\pm$0.2& --&--& -0.2& --\\ 
2429	& 0.5$\pm$0.1	&0.6$\pm$0.1& 0.00$\pm$0.04&0.0$\pm$0.1& -0.1 & 0.4\\ 
2809	& 0.5$\pm$0.1	&0.4$\pm$0.2& --&--&-0.2 & --\\
 \hline									 
\hline									 
\end{tabular}					 
}			
\label{tab:grad}									 
\end{table*}

\begin{figure*}
\begin{center}
\includegraphics[width=6cm]{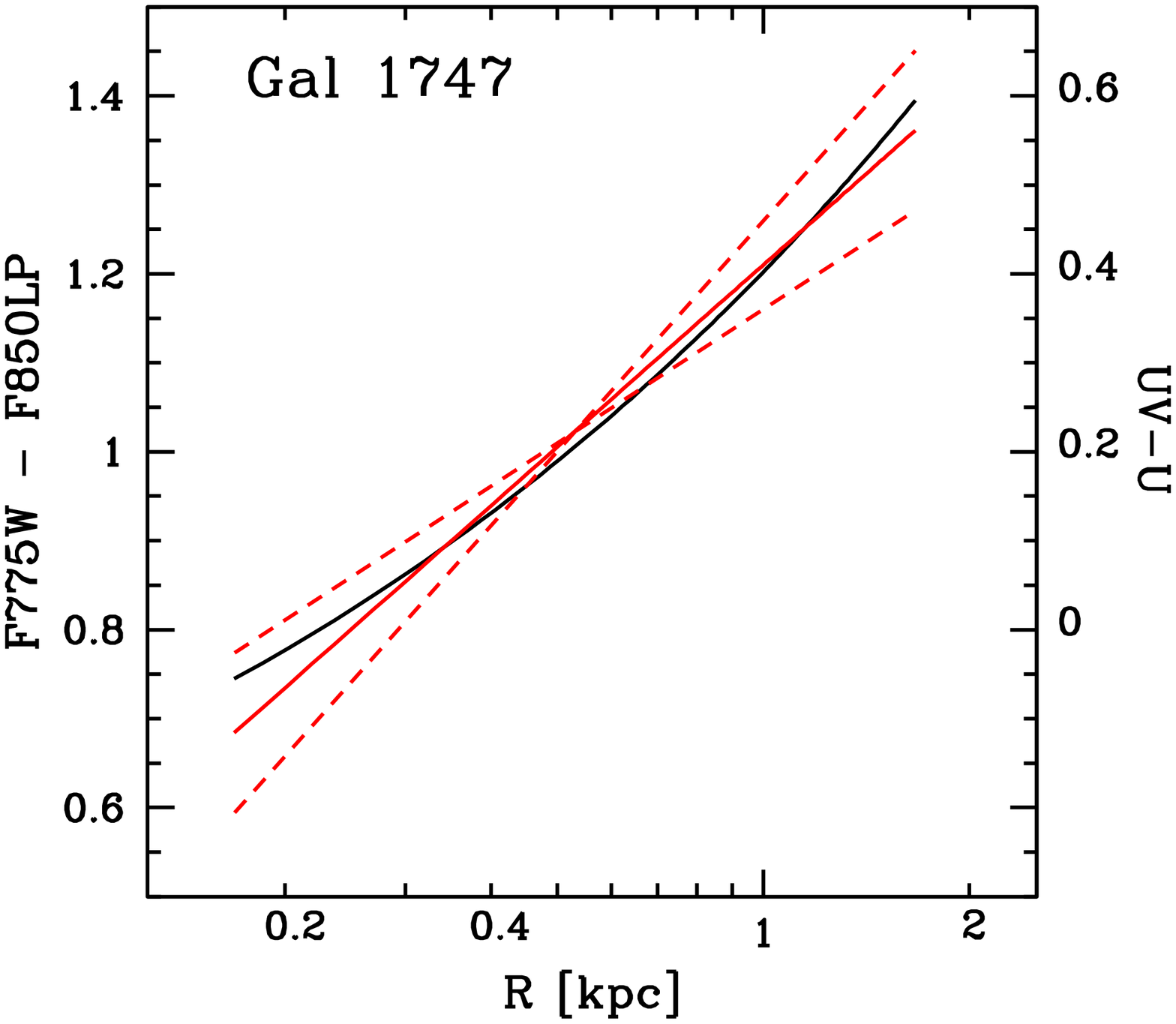}
\includegraphics[width=6cm]{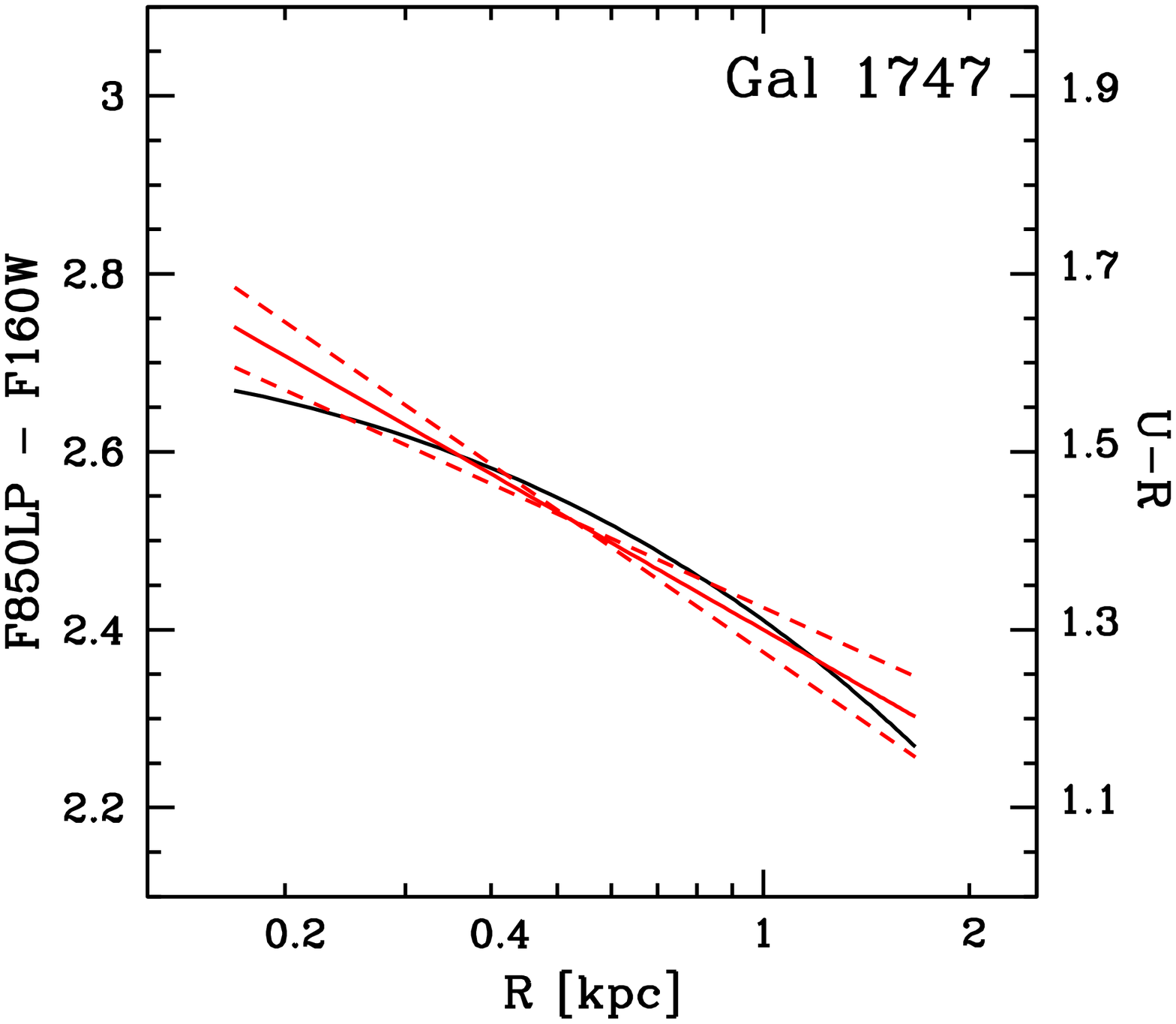}
\includegraphics[width=6cm]{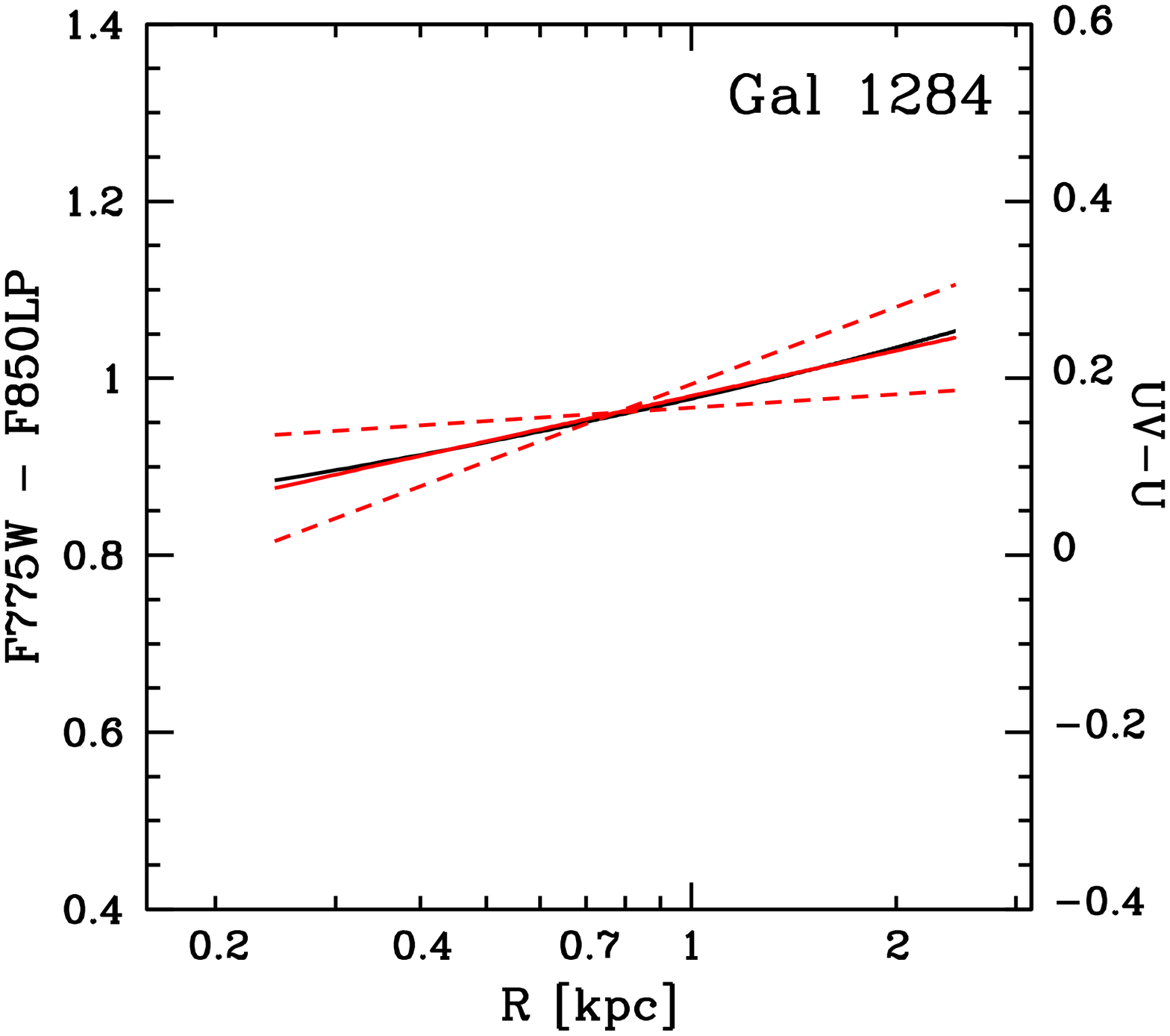}
\includegraphics[width=6cm]{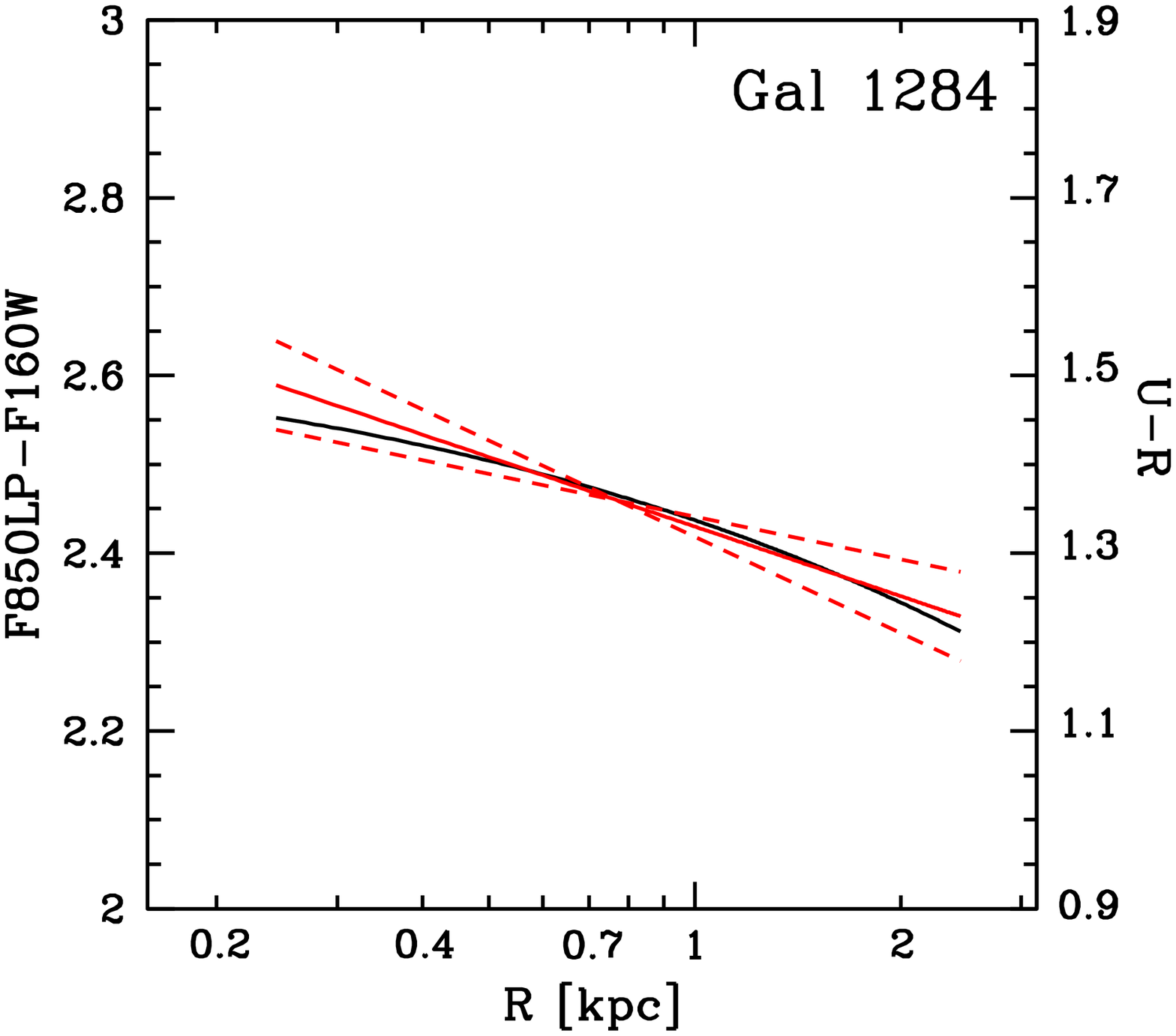}
\vskip 0.1truecm
\caption{The F775W - F850LP (left panels) and F850LP-F160W (right panels) colour 
gradients of two galaxies of our sample (1747 and 1284). 
Black curves correspond to the intrinsic colour profiles of the galaxies fitted between 
0.1$R_e$ and 1$R_e$ and red lines represent the best fitted lines to the colour profile. 
The dashed red lines correspond to 1$\sigma$ errors on colour gradients. 
The y axis on the right shows the rest-frame UV-U and U-R colours.
The width of the colour interval is the same ($\Delta$colour=1 mag) for both the colours to facilitate comparison.}
\label{fig:grad}
\end{center}
\end{figure*}

\begin{figure}
\begin{center}
\includegraphics[width=8cm]{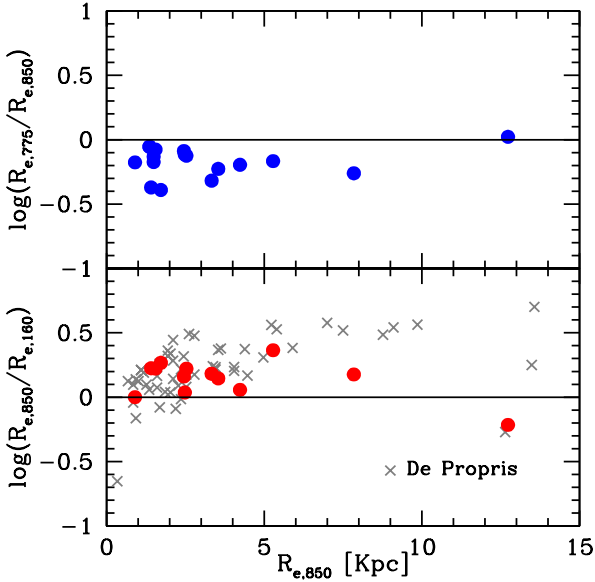}
\caption{Colour gradients derived as $log(R_e(775)/R_e(850))$ (upper panel) and 
as $log(R_e(850)/R_e(160))$ (lower panel) versus the effective radius in the 
F850LP band (R$_{e,850}$). 
We plotted with grey crosses the cluster galaxies at <z> $\sim$ 1.25 from \citet{depropris15} sample with $n$ consistent with our ellipticals. }
\label{fig:rapp}
\end{center}
\end{figure}

\begin{figure*}
\begin{center}
\includegraphics[width=6.cm]{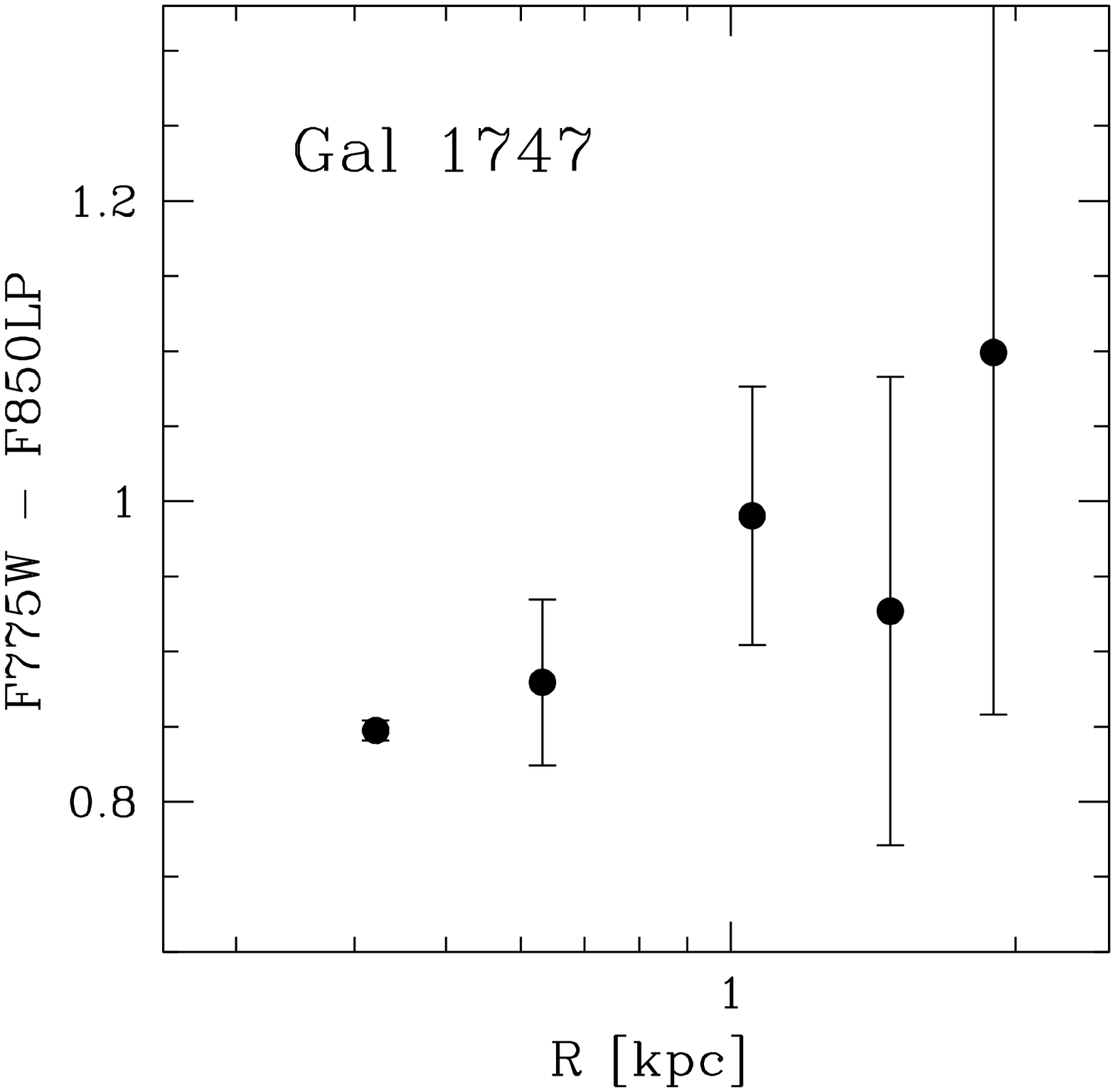} 
\includegraphics[width=6.cm]{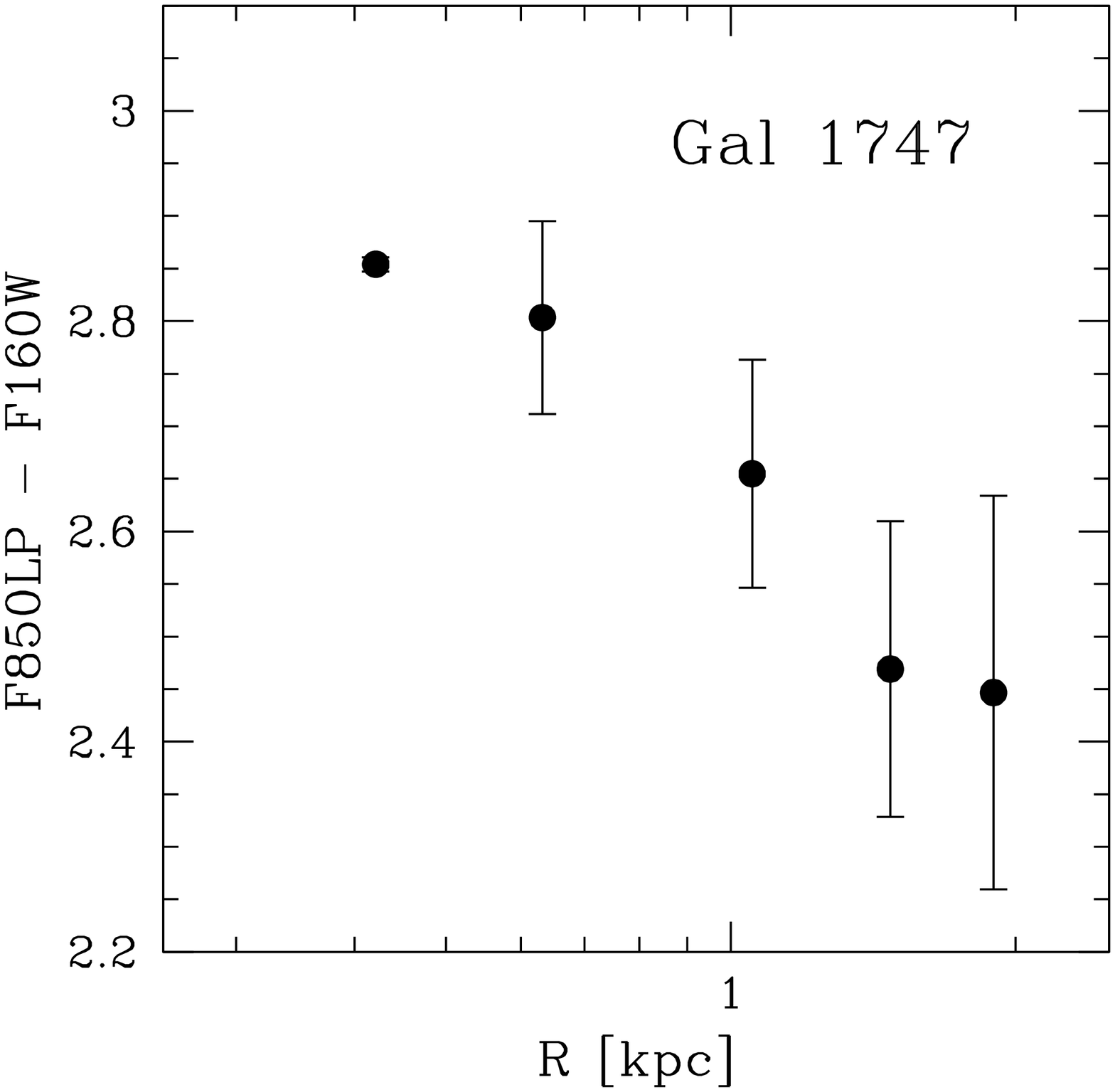}
\includegraphics[width=6.cm]{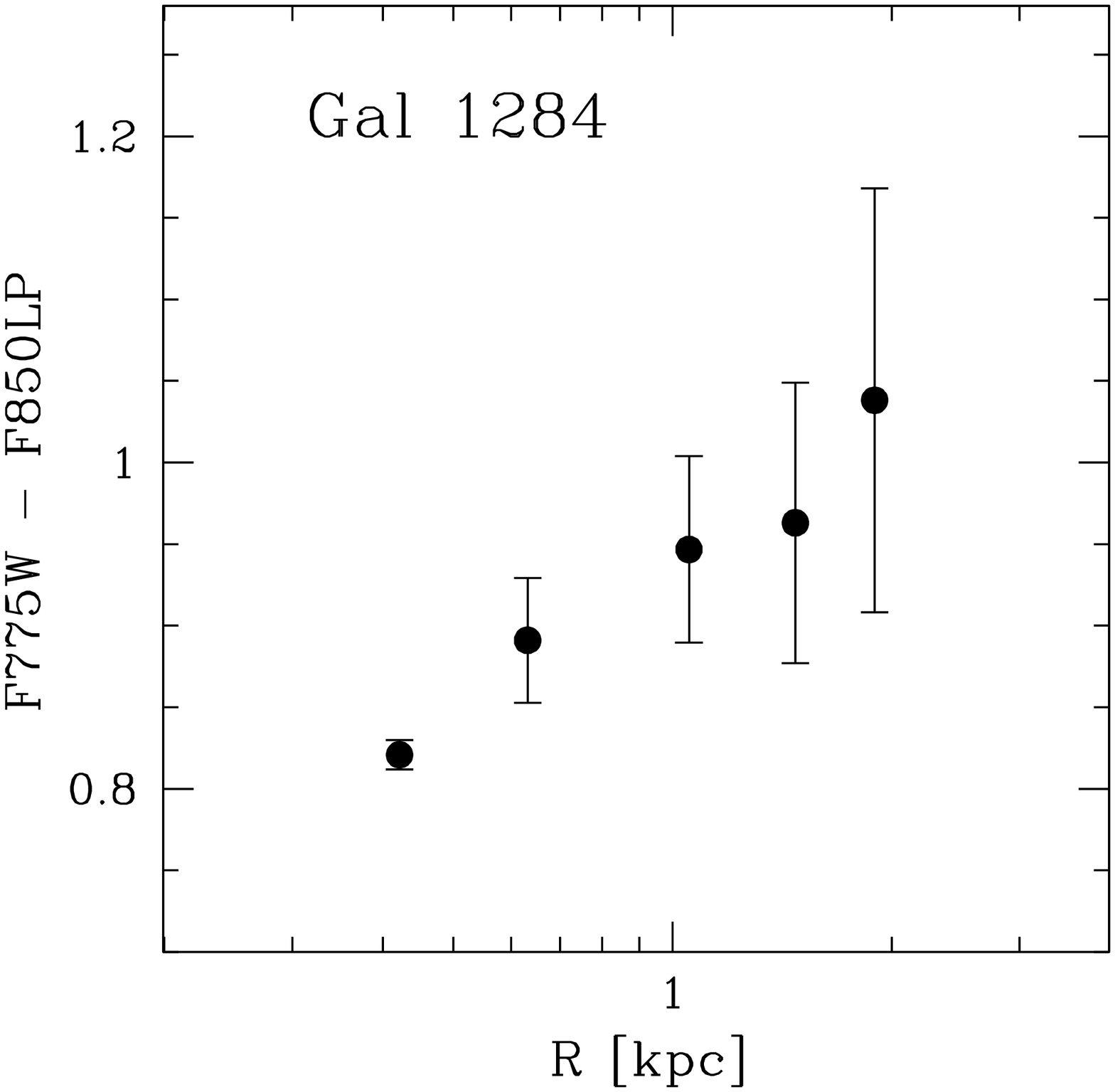} 
\includegraphics[width=6.cm]{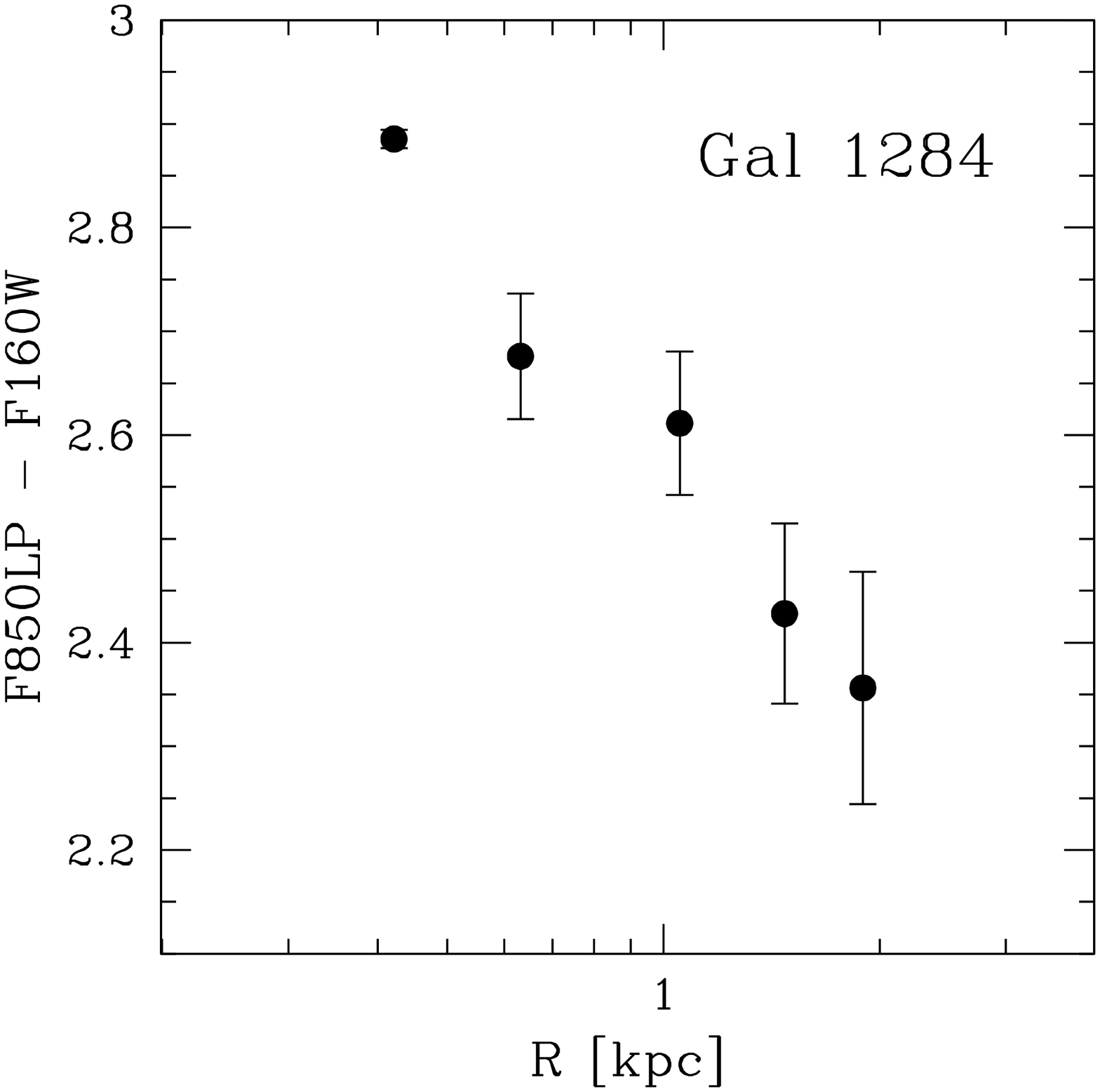}
\caption{Observed F775W - F850LP (upper panels) and F850LP - F160W (lower panels) 
colour profiles derived by measuring the fluxes within circular apertures centred on 
each galaxy. Errors have been estimated by propagating the uncertainties on the observed
 brightness profiles.  }
\label{fig:oss}
\end{center}
\end{figure*}

\begin{figure}
\begin{center}
\includegraphics[width=8cm]{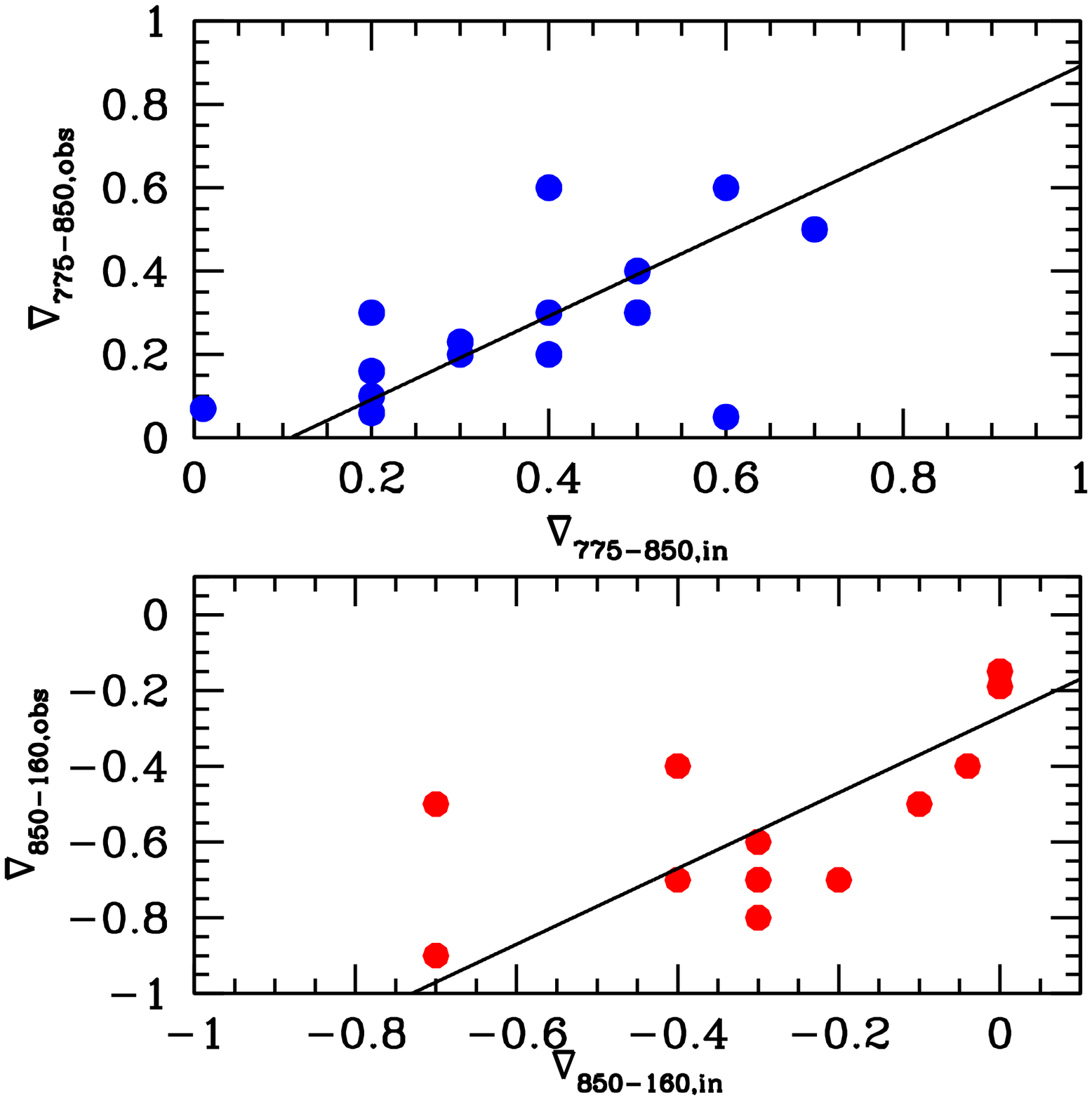}
\caption{Comparison between the F775W-F850LP (upper panel) and the F850LP-F160W (lower panel) colour gradients derived from the intrinsic surface brightness profiles and the observed profiles. }
\label{fig:obs_in}
\end{center}
\end{figure}

For each of the 17 galaxies of the sample we have derived F775W - F850LP and F850LP - F160W colour gradients, 
corresponding to  $\sim$(UV-U) and $\sim$(U-R) in the rest-frame. 
The transformations to obtain the rest-frame colours are: 
UV-U= (F775W-F850LP) - 0.8 mag and U-R = (F850LP-F160W) - 1.1 mag. 

We have used three different methods to estimate the colour
gradients: the logarithmic slope of the deconvoluted colour profiles, the ratio of the effective radii as measured in the three different filters and through the observed colour profiles.
Our results are broadly independent of the method used and 
are in good agreement
with previous work (see Section \ref{data}).

\subsection{Intrinsic colour gradients}
We first derived the colour gradient as the logarithmic slope of the colour 
profile \citep[e.g.][]{peletier90}

\begin{equation}
\nabla_{X-Y} = \frac{\Delta (\mu_{X}(R)-\mu_{Y}(R))}{\Delta \log R}
\label{sb}
\end{equation}

where $\mu_{X}$(R) and $\mu_{Y}$(R) are the surface brightness
profiles of the galaxy in the generic X and Y bands, respectively. 

We fitted the slope of the colour profile between 0.1R$_{e,}$ and 1R$_{e}$ to be 
consistent with previous works.
In Fig. \ref{fig:grad} we show as an example the F775W - F850LP and F850LP-F160W 
colour gradients for two galaxies of our sample (1747 and 1284). 
The colour gradients of all other galaxies are shown in Appendix \ref{colour gradients}.
The black curves represent the deconvolved colour profiles, whereas the red lines are 
the best fitting slopes. 
The 1$\sigma$ error around the fit is shown as red dashed
lines. 
The values of the colour  gradients and their relative errors are reported in Table 
\ref{tab:grad}. 
 We assigned as error on the colour gradient the error on the observed colour estimated within a thin circular annulus of width 0.1 arcsec and centred  
 at 1R$_e$ (if the colour gradients are fitted between 0.1R$_e$ and 1R$_e$) 
 or at 2R$_e$ (if the colour gradients are fitted between 0.1R$_e$ and 2R$_e$). 
 This represents an upper limit to the error on the colour gradient. 
 Indeed, we verified that, due to the lower signal-to-noise ratio (S/N), 
 the error at 1R$_e$ is, on average, approximately three times the error at 0.1R$_e$.
 
Ellipticals in our sample have positive or null rest-frame
 UV-U gradients. 
 Indeed, three galaxies have a colour gradient comparable with zero at 
 1$\sigma$, whereas the remaining galaxies present a significant positive gradient 
 showing the presence of a bluer stellar population towards the centre in this colour. 
 On the contrary, these galaxies have negative or null rest-frame U-R colour gradients. 
  In particular, four galaxies have a U-R colour gradient comparable with zero at 1$\sigma$, 
  whereas the remaining ones present a significant negative gradient showing that the 
  stellar populations are redder towards the centre in this colour. 
  These results do not depend on the region over which the fit is carried out.
  The intrinsic profile from \textlcsc{Galfit} well
reproduces our observed profile up to 2R$_e$ for the majority ($\sim$ 80 per cent) 
of the sample (see Appendix \ref{galfit}).
  Hence, we have estimated the colour gradients at most up to this 
  distance from the centre confirming the UV-U and  U-R gradients for all the 
  galaxies of our sample. 
 At larger radii, 
we cannot reliably fit the galaxy profile as the surface brightness of galaxies falls below the level of the sky noise (see below). 
 
 We also derived the colour gradients from the ratio of the effective radii as measured 
in the F775W, F850LP and F160W: $\Delta logR_{e}=log(R_e(X)/R_e(Y))$ 
\citep{labarbera03,depropris15}, where $X$ 
stands for the bluer band and $Y$ stands for the redder band. 
As shown in Fig. \ref{fig:rapp}
 most of the high-redshift cluster galaxies show significant negative 
 $log(R_e(F775W)/R_e(F850LP))$ and simultaneously significant positive
  $log(R_e(F850LP)/R_e(F160W))$.
  This implies positive UV-U and negative U-R colour gradients for most of the 
  galaxies in agreement with the previous method. 
The ratios $\Delta$log(R$_e$) are also shown in Table \ref{tab:grad}. 
 
 \subsection{Observed colour gradients}
As a test for the robustness of our results derived by 
  fitting the brightness profiles with \textlcsc{Galfit}, we also derived the 
  observed surface brightness profiles, and hence the colour profiles,
  by measuring the fluxes in each HST band within circular apertures centered on the 
  galaxy. 
  This method is not based on any fit of the surface brightness profile; hence, 
  no dependence either on any assumption on the analytic function of the profile 
  or on the fitting procedure is present. 
  Since ACS F850LP images and WFC3 F160W images have a different 
  resolution, in order not to introduce spurious gradients, we have degraded the 
  F850LP image to the same PSF of the F160W image before deriving the F850LP - F160W 
  observed colour profiles. 
   In Fig. \ref{fig:oss} we report as an example the observed colour profiles 
   for the same galaxies 1747 and 1284 of Fig. \ref{fig:grad} 
up to 1R$_e$.
 
These gradients cannot be easily 
compared to the results in the previous subsection, as they depend on the PSF.
However, both the methods provide the same systematics for all the galaxies as shown in Fig. \ref{fig:obs_in},
where we compare the `intrinsic' and `observed' F775W-F850LP (upper panel) and F850LP-F160W (lower panel) colour gradients out to 1R$_e$.
The `observed' F775W-F850LP gradients are, on average, slightly less steep than 
the `intrinsic'  gradients (with an offset < 0.1 mag) due to the PSF that 
smooths both the profiles. 
The `observed' F850LP-F160W colour gradients, instead, are, on average, steeper with an 
offset of $\sim$0.3 mag. 
  This is mainly due to the PSF-match procedure. 
  Indeed, degrading the F850LP image to the same PSF of the F160W image, the 
  same amount of flux of the F850LP image has been redistributed over a larger 
  area. 
  This implies a decrease of the flux in the inner region in favour to
  the outer region and, consequently, the gradients become steeper.  

We have also checked the effect of estimating colour gradients beyond 2R$_e$.
Fig. \ref{fig:comp_gal} shows as example the observed F850LP-F160W colour 
profiles up to 3R$_e$ for galaxies 595 and 837, with effective radius  
3.5 and 7.8 kpc, respectively. 
We can see that at R $\gae$ 1.5-2R$_e$ the colour profiles reverse because the sky noise dominates the surface brightness profiles in the F850LP band image (see Fig.
\ref{fig:prof}), introducing a spurious colour gradient.
This artificially flattens the colour gradient from the steeper one measured within 2R$_e$ (and with \textlcsc{Galfit}).
In agreement with what we previously found with the \textlcsc{Galfit} fitting, 
estimates of the colour gradients for R $\gae$ 2R$_e$ are not reliable
for most of the galaxies ($\sim75$ per cent of the sample)
since they are dominated by the background of the ACS images.

These three different methods used to derive colour gradients are mutually consistent,
implying that our results are independent of the 
method used.
We conclude that the UV-U colour gradients are systematically positive 
($\sim$80 per cent) or null ($\sim$20 per cent), never negative. On the contrary, the U-R 
colour gradients are systematically negative ($\sim$70 per cent) or null ($\sim$30 per cent), 
never positive.

\section{Investigating the nature of the observed colour gradients}

\begin{figure*}
\includegraphics[width=5.55cm]{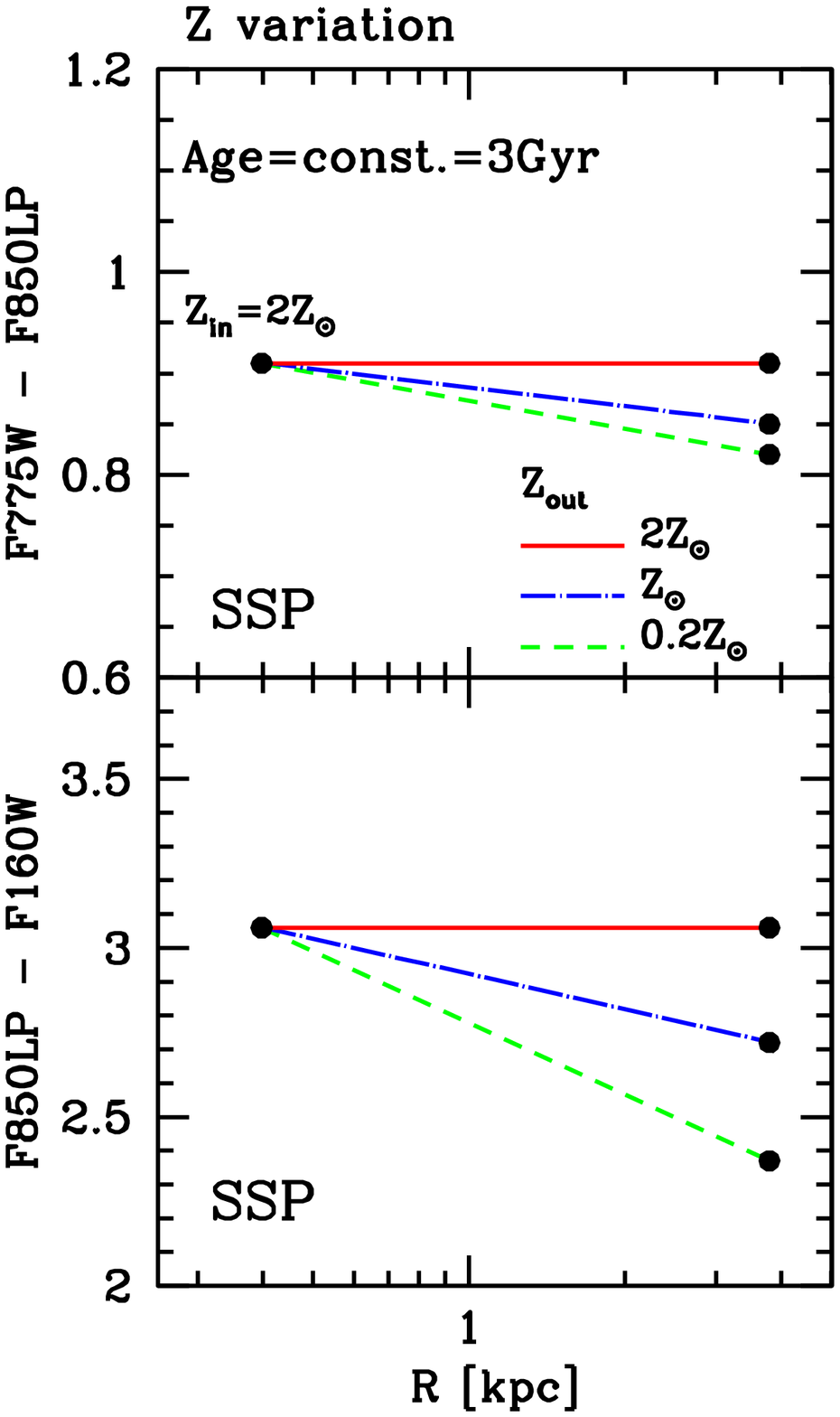}
\includegraphics[width=5.1cm]{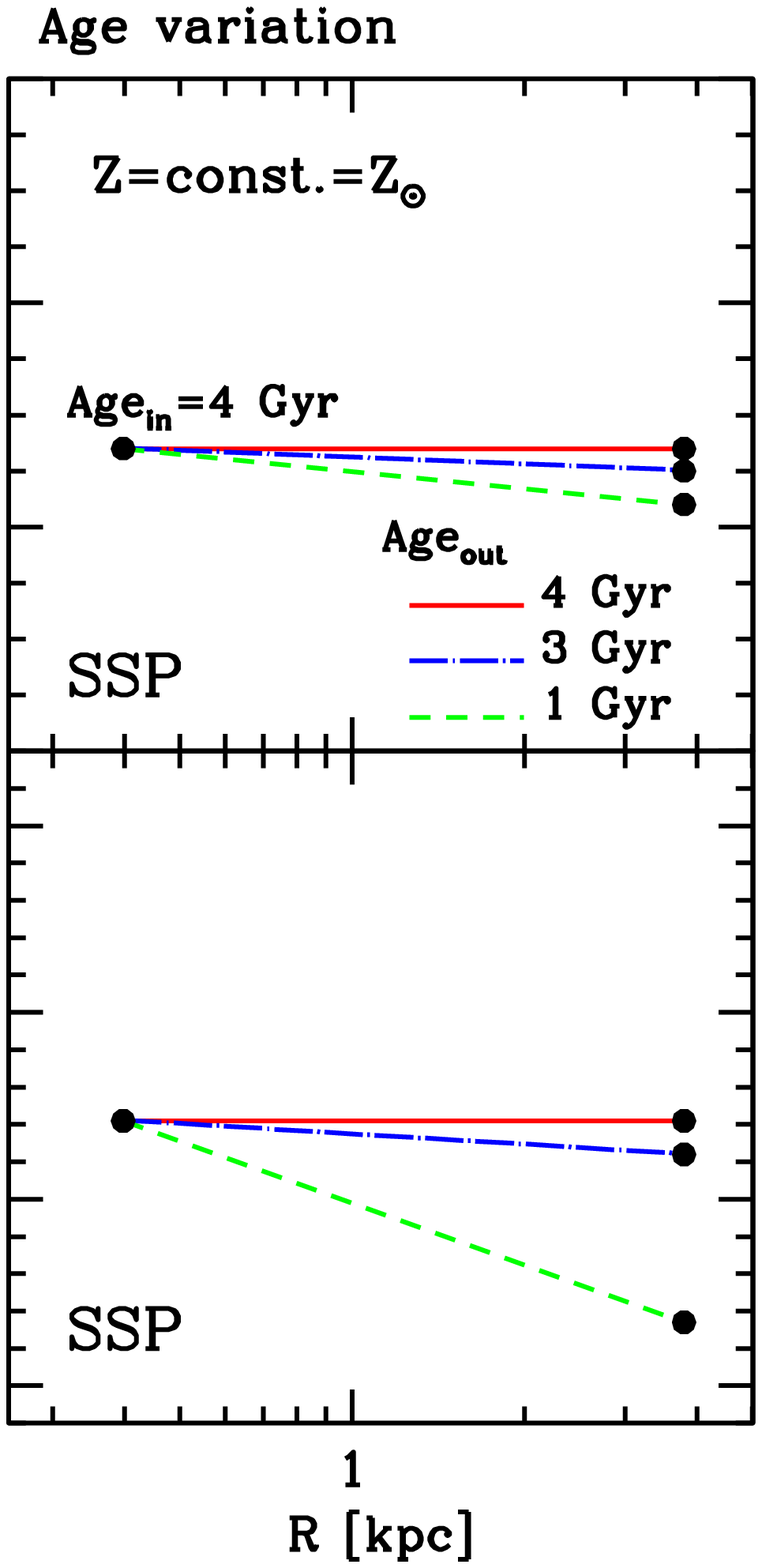}
\includegraphics[width=6.2cm]{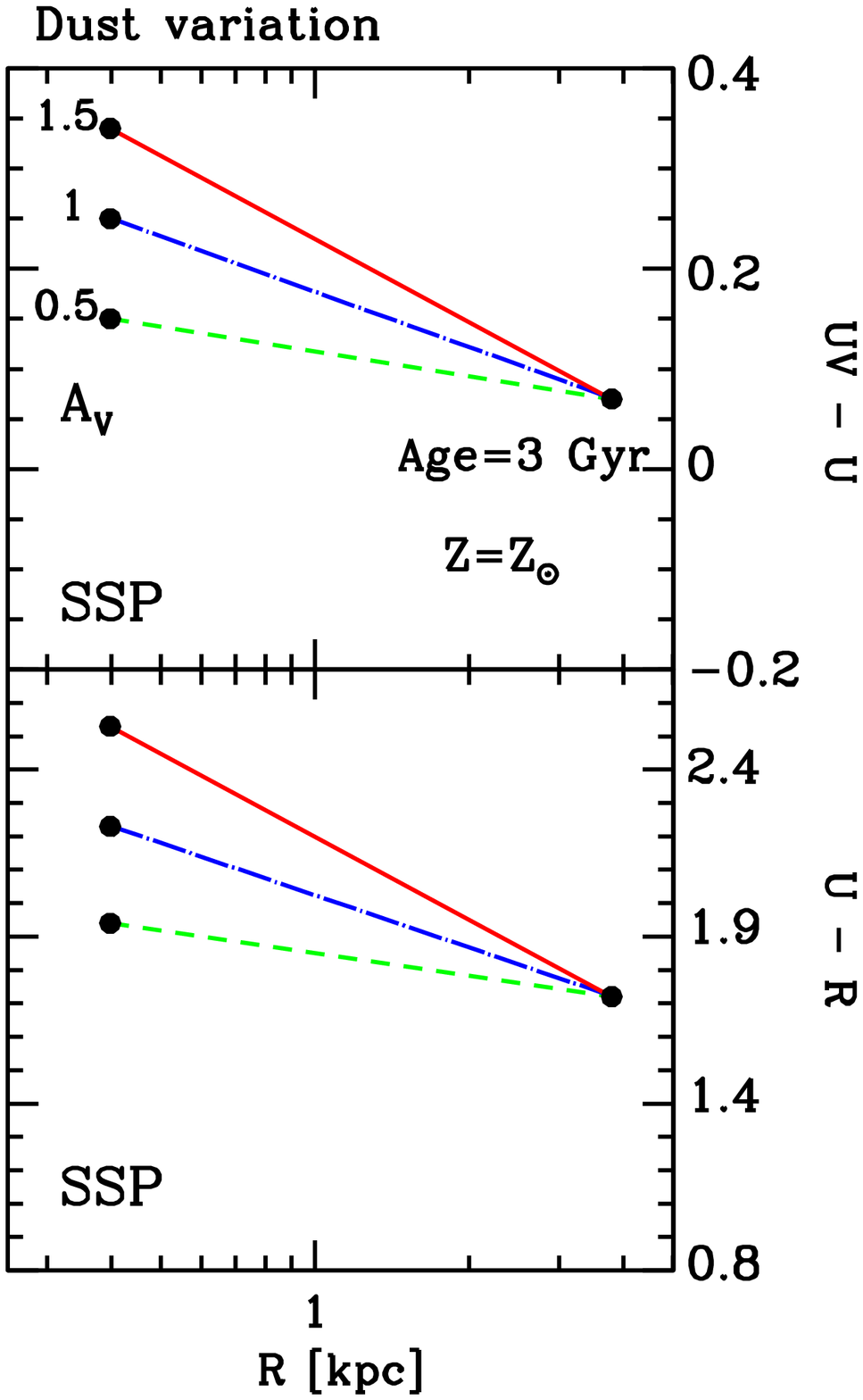}
\caption{F775W-F850LP and F850LP-F160W colour gradients variation as a 
function of metallicity, age and dust.  
 \textit{Left-hand panel}: we fixed the age of the stellar population at 3 Gyr
and varied the metallicity from the central region (2 Z$_{\odot}$) to the 
outer region [2 Z$_{\odot}$ (red solid line), Z$_{\odot}$ 
(blue dot-long dashed line) and 0.2 Z$_{\odot}$ (green short dashed line)]. 
\textit{Central panel}: we fixed the metallicity at 
Z$_{\odot}$ and varied the age from the central region (4 Gyr) to 
the outer region [4 Gyr (red solid line), 3 Gyr 
(blue dot-long dashed line) and 1 Gyr 
(green short dashed line)].   
 \textit{Right-hand panel}: the gradients variation as a function of the 
dust extinction $A_V$ at fixed age and metallicity for three different values 
of $A_V$: 0.5 (green short dashed line), 1 (blue dot-long dashed line) and 1.5 (red solid line) is shown. 
}
\label{fig:grad_model}
\end{figure*}

\begin{figure}
\begin{center}
\includegraphics[width=8cm]{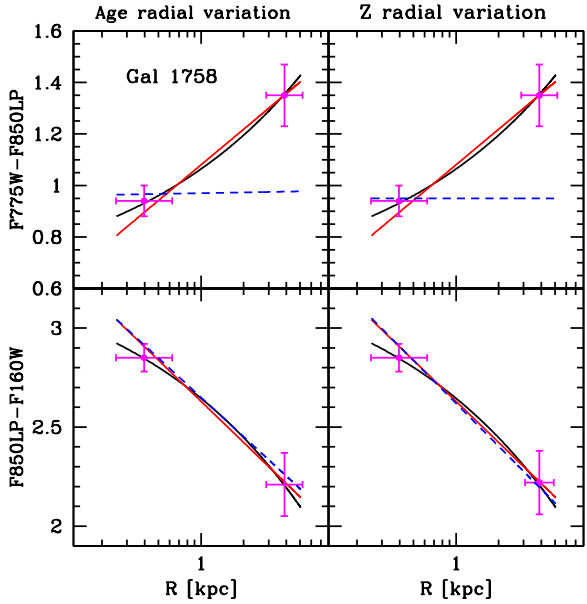}
\caption{\textit{Left}: radial age variation as origin of the observed colour gradients. The upper
panel shows the F775W-F850LP colour profile (black line) and best fitting slope (red line); the lower panel on this
side shows the same information but for the F850LP-F160W gradient. In both panels, the blue dashed lines show the effect of
a pure age variation as a function of radius on colour. The magenta points show the mean colour of the
stellar populations in the inner and outer regions. \textit{Right}: same as above but for metallicity. }
\label{fig:real_data1}
\end{center}
\end{figure}

\begin{figure}
\includegraphics[width=8cm]{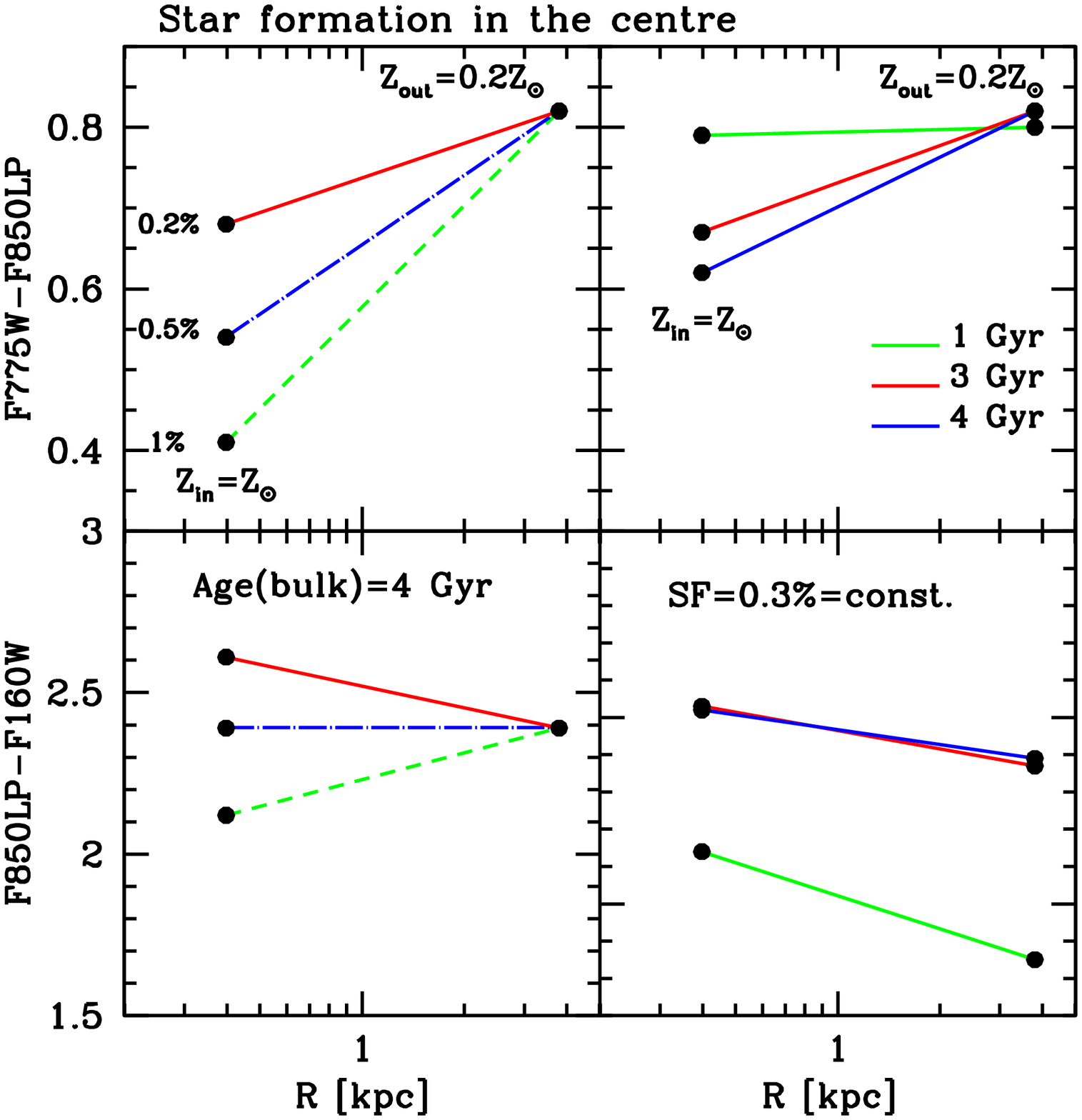}
\caption{We show how the F775W-F850LP and F850LP-F160W 
colour gradients change by adding a weak star formation towards the centre. 
\textit{Left-hand panel}: we fixed the age of the main stellar population at 4 Gyr and varied the metallicity from the centre (Z$_{\odot}$) to the 
outer 
region (0.2 Z$_{\odot}$), adding a weak star formation towards the centre. 
The percentages indicate the contributes in term of stellar mass of the stellar population, 
which is producing stars. \textit{Right-hand panel}: we show how both colours change if we add a
star-forming component equivalent to 0.3 per cent in stellar mass for three different ages of the main stellar population 
[1 Gyr (green line), 3 Gyr (red line) and 4 Gyr (blue line)] and with the same metallicity gradient.}
\label{fig:grad_model2}
\end{figure}

The presence of colour gradients implies that one or more properties of the stellar 
population vary radially.
In this section we test whether a radial gradient of a single parameter (age, metallicity or dust) can account for the opposite slopes of the observed
UV-U and U-R colour gradients.

\subsection{Age, metallicity and dust variation}

In the local Universe, colour gradients are usually interpreted as metallicity gradients, with the possible role
of age gradients being still controversial. 
On this basis, we first tested whether the colour gradients we measure at high 
redshift can be reproduced by a pure metallicity gradient. 
 
In the left-hand panel of Fig. \ref{fig:grad_model} the predicted UV-U and U-R colour gradients 
are shown as a function of metallicity. 
We modelled the central region of the galaxy with a 3 Gyr old simple
stellar population (SSP) (the results are independent on this assumption) 
with supersolar metallicity (2 Z$_\odot$), while we modelled the outer region 
 with the same SSP (i.e., we assume no radial age variation), but with metallicity 
 2 Z$_\odot$, Z$_\odot$ and 
 0.2 Z$_\odot$. 
 It can be seen that a metallicity gradient affects the derived  U-R colour 
 gradient more strongly than the UV-U gradient, but the slope is always negative
 (e.g for the extremes of the range, where the central region has Z$_{in}$=2 Z$_\odot$ 
and the outer region Z$_{out}$=0.2 Z$_\odot$, we find $\nabla_{UV-U} \sim -0.1$ mag and $\nabla_{U-R} 
\sim -0.7$ mag).
Therefore, we cannot explain both gradients at z=1.4 with a
simple metal abundance gradient as in the local Universe.
 
We repeated the same exercise in the case of a pure age
 (Fig. \ref{fig:grad_model}, central panel) and dust gradient
(Fig. \ref{fig:grad_model}, right-hand panel). 
To investigate the effect of a pure age variation,
we have modelled the central region of the galaxy with a SSP with solar
metallicity and age 4 Gyr, and the external region with a SSP with the same
metallicity and age 4 Gyr, 3 Gyr and 1 Gyr. 
Also in this case we observe that the U-R colour is more sensitive to age 
variation 
and that the variation of a single parameter is not able to reproduce 
both the negative U-R gradient and the positive
UV-U gradient
(e.g., for a galaxy whose centre has Age$_{in}$=4 Gyr and Age$_{out}$=1 Gyr in the outskirts we obtain $\nabla_{UV-U}
 \sim -0.05$ mag and $\nabla_{U-R} \sim -0.6$ mag).

Finally, we considered a radial change in extinction.
To model the dust variation, we fixed both the metallicity and the age and we
assumed  a more dusty
core.
The dust extinction produces an attenuation of the intrinsic flux of the galaxy
that depends on the wavelength: 
$f_{\rm obs}(\lambda)=f_{\rm int}(\lambda) 10^{-0.4 A_{\lambda}}$,
where $f_{\rm obs}$ and $f_{\rm int}$ are the observed and the intrinsic fluxes, 
respectively, and $A_{\lambda} = {k(\lambda) A_V / R_V}$ is the extinction at a 
wavelength $\lambda$,  $k(\lambda)$  is the Calzetti law and $R_V$ = 4.05$\pm$0.80 
\citep{calzetti00}. 
In the right-hand panel of Fig. \ref{fig:grad_model} it is shown how the two colour 
gradients vary as a function of the dust extinction for three different values of 
$A_V$: 0.5, 1 and 1.5. 
Also in this case, the radial variation of dust produces concordant UV-U and U-R 
gradients, which become steeper for increasing values of $A_V$. 

Therefore, simple radial trends in age, metallicity or dust in their own cannot reproduce
the observed UV-U and U-R gradients at the same time.
Radial trends in metallicity, age
or dust (with the metal abundance, age or extinction decreasing outwards) always produce
negative UV-U and U-R gradients (and opposite if the metallicity, age or extinction 
decrease inwards).
Furthermore, Fig. \ref{fig:grad_model} shows also that
the UV-U colour is much less sensitive to metallicity and age variation than the U-R colour. 
In fact, we find that while the U-R colour variations induced by the age/metallicity 
variations are comparable to those observed in our galaxies, 
the same age/metallicity variations do not produce UV-U colour variations
as large as those observed (see Fig. \ref{fig:col1}).  

To verify whether this result is just related to our choice of the input values 
of metallicity/age in the toy model of Fig. \ref{fig:grad_model}, 
in the next section we check whether an age or a metallicity variation 
is able to reproduce the amplitude of the observed gradients or it is easier to 
reproduce a certain colour gradient than the other one.

\subsection{Modelling of the real data}
Following the analysis technique proposed by \cite{gargiulo12}, 
we model the stellar content of a galaxy as composed of two main
stellar populations, one population dominating towards the centre and the
other one dominating towards the outskirts. 
We model the two populations using BC03 composite stellar populations.
To test for the radial age (metallicity, Z) variation as origin of the observed colour
gradients, we fixed Z (age), A$_V$ and $\tau$ of the two populations to those
derived from the fit to the global SED and leave age (Z) as free parameter of
the fit.
The best fitting age of the central (external) stellar population is
the one that better reproduces both the UV-U and U-R colours observed in the inner (outer) 
region of the galaxy.
We also
required that the sum of the stellar populations matches the global SED.

On the basis of the colour profiles (see Appendix \ref{fig:prof}), 
we have chosen as inner region the one enclosed within the FWHM, that is within a 
radius of $\sim0.06$ arcsec ($\sim$0.5 kpc) and, as outer region, a circular annulus 
$\sim$ 1 kpc width centred at 1.5R$_e$.
We verified that for each galaxy 1.5R$_e$ 
corresponds to at least approximately three times the radius of the PSF. 

In Fig. \ref{fig:real_data1}, we show as an example the result obtained for the 
galaxy 1758, the one showing the most extreme colour gradients. 
In the left-hand panels we show the colour gradients from a pure age variation, 
whereas in the right-hand panels we show those obtained by considering a pure metallicity 
gradient, where we have adopted a grid of metallicity values as
follows: 0.2 Z$_{\odot}$, 0.4 Z$_{\odot}$, Z$_{\odot}$ and 2 Z$_{\odot}$.
The black curves represent the colour profiles of the galaxy, whereas the red lines are 
the  observed colour gradients. 
The blue dashed lines represent the best fitting colour gradients as resulting
from the two-component model. 

What we found is that the best fitting procedure finds a solution to the observed U-R colour gradients ($-0.7<\nabla<-0.1$ mag; 
see Table \ref{tab:grad}) both in the case of an age gradient 
and a metallicity gradient.
The resulting metallicity gradients (at fixed age)
are in the range $\nabla_{Z}=d$log(Z)/$d$log(R)<-0.8, with a median value of $\sim$ -0.4, consistent with the metallicity gradients 
derived in local and intermediate redshift ellipticals, where $\nabla_{Z}$ ranges from -0.2 to -0.4  \citep[e.g.][]{saglia00,wu05,labarbera09}.
The resulting age gradients (at fixed metallicity), instead, are in the range  $\nabla_{age}=d$log(age)/$d$log(R) $\lae$ -0.4. 
Such age gradients would be barely detectable locally. 
 For instance, $\nabla_{age}$= -0.3 due to an age variation of 
 2 Gyr corresponds at $z$=0 to a colour gradient of -0.07 mag, smaller than 
 the typical errors on colour gradients.
On the contrary, the procedure does not find a solution for the observed large 
UV-U colour gradients and for the opposite slopes shown by the two gradients.

It is worth noting that, in the case of dust variation, to reproduce even the mean 
amplitude (-0.3 mag) of the U-R gradients an $A_V >> 1$ mag would be needed in the inner 
regions, values never obtained from the fitting to the global SED of our galaxies.
Hence, dust can play only a marginal role, if it has one, in the origin of 
the observed colour gradients.

Hence, it is easier to reproduce the negative U-R colour gradients rather than the
positive UV-U colour gradients. 
UV-U colour gradients can not be accounted for by age and/or by metallicity variations.
Actually, what the data show is that galaxies are bluer in the centre when the UV-U colour 
is considered. Hence, there is an UV excess towards the galaxy central regions 
with respect to the external ones.
Starting from this evidence, in the next section, we try to constrain the scenario 
that best reproduces simultaneously both the colour gradients, focusing on the nature 
of the bluer UV-U colour in the galaxy central regions.

\section{The key information stored in the galaxy central regions}
While an age gradient or a metallicity gradient, as the one observed in local 
elliptical galaxies \citep[e.g.][]{saglia00,wu05,labarbera05,tortora10} can account for the U-R colour gradients, the same gradient cannot produce the opposite (positive) UV-U
gradients. 
To reproduce the positive UV-U colour gradient without affecting the observed negative U-R 
gradient, a mechanism able to efficiently enhance UV emission without altering the spectrum at
longer wavelengths 
has to be hypothesized in the galaxy central regions.
Possible mechanisms able to fulfil these requirements are the star formation, the presence of a QSO,
stars in the He-burning phase or He-rich stars. 
In the following, we discuss each of these mechanisms.  

\subsection{A steady mild star formation?}
\label{mild_SF}
One mechanism that produces UV emission is the star formation. 
In order to test whether weak star formation can produce positive UV-U gradients without affecting the negative U-R gradients,
we have added in the centre a young stellar component seen while it is forming 
most of its stars, i.e. characterized by $\tau$=0.1 Gyr and an age of 0.1 Gyr.
The results are shown in Fig. \ref{fig:grad_model2}.
In particular, in the left-hand panel, we fixed the age of the bulk of stellar population (4 Gyr) 
and varied the metallicity from the centre (Z$_{\odot}$) to the external region 
(0.2 Z$_{\odot}$) and added different percentages (in terms of stellar mass) of star formation 
towards the centre.
In the right-hand panel, instead, we show the same model of the left panel (metallicity 
varies from the central region to the outer region with different percentages 
of star formation in the centre) but for different constant age of the bulk 
of the stellar population.

We found that, although the radial decrease of metallicity (at constant age) would
reflect in negative UV-U colour gradient (e.g. see Fig. \ref{fig:grad_model}
),
 the presence of a very weak star formation in the centre is able to 
reverse the UV-U gradient turning it from negative to positive. 
Moreover, the older the mean age of the stellar population the 
easier to produce positive UV-U colour gradients. 
This is because the UV emission of the star-forming component tends to dominate the 
UV and U emission of the stellar population when it is old, enhancing the gradient 
with respect to the outer regions. 
For instance, the ratio between the UV emission of 0.3 per cent (in stellar mass) of 
star-forming component and the UV emission of 1 Gyr old component is $\sim$ 0.08. 
This ratio increases to $\sim$ 0.7 in the case of a 4 Gyr old component.

In order not to affect the negative U-R colour gradients, the percentage in 
stellar mass of
this star forming component must be less than $\sim$0.5 per cent of the total stellar
 mass.

We applied this model to our galaxies. 
We built a grid where the percentage of the star forming component 
ranges from 0.1 to 3 per cent with a step of 0.2.
At fixed percentage of the star forming component, we varied the age and the 
metallicity of the stellar population in the centre and in the outer region.
The ages considered range from 1 to 4 Gyr (the age of the Universe at $z$=1.39 is $\sim$ 4.5 Gyr) with a step of 0.5.
Metallicity variation has been modelled on a grid of subsolar and
supersolar metallicity values: 0.2 Z$_{\odot}$, 0.4 Z$_{\odot}$, Z$_{\odot}$ 
and 2 Z$_{\odot}$.
We found that 0.7 per cent represent an upper limit to the percentage of the star-forming component, since larger percentages produce flat or even positive U-R 
gradients.

Six galaxies (595, 1740, 1790, 2054, 2429, 837) present 
U-R colour gradients consistent with a null gradient at 1$\sigma$. 
However, these galaxies present positive UV-U gradients at 1$\sigma$.
The only addition of star formation towards the centre produces positive UV-U 
gradients, but also slightly positive U-R gradients.
The best solutions provide for these galaxies a contribution of 0.2 per cent in mass 
of star formation together with either a negative age gradient of $\sim$ -0.3 
(Age$_{in}$= 4 Gyr and Age$_{out}$= 2 Gyr) or a metallicity gradient 
of $\sim$ -0.4 (Z$_{in}$ = Z$_{\odot}$ and Z$_{out}$= 0.4 Z$_{\odot}$), consistent with the metallicity gradients found in local ellipticals.

Four galaxies (692, 1284, 1782, 2147) show opposite colour 
gradients at 1$\sigma$. 
For them, the best fitting model producing simultaneously both the observed 
colour gradients is composed of 0.2 per cent of star-forming component in the centre 
superimposed to an age gradient of $\sim$ -0.3 (Age$_{in}$= 4 Gyr and Age$_{out}$= 2 Gyr) and 
a metallicity gradient of $\sim$ -0.4 (Z$_{in}$ = Z$_{\odot}$ and Z$_{out}$= 0.4 Z$_{\odot}$).

\begin{figure}
\begin{center}
\includegraphics[width=8cm]{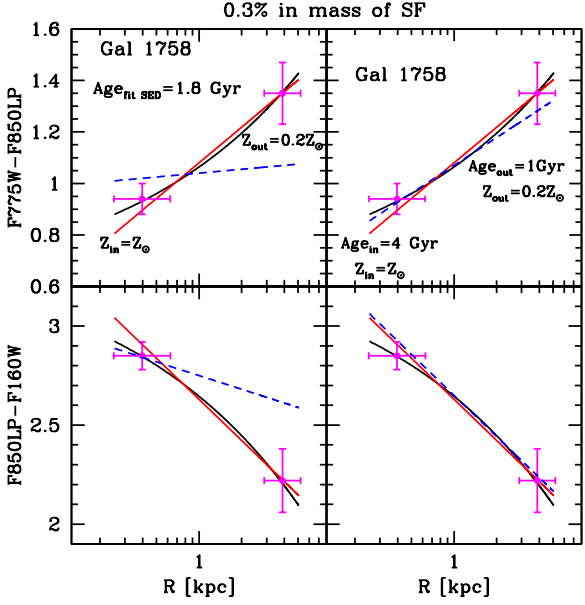}
\caption{The two panels show the F775W-F850LP (upper panel) 
colour profile (black line) with the best fitting slope (red line) and the F850LP-F160W (lower panel) 
colour profile (black line) with the best fitting slope (red lines). The blue dashed lines represent 
the colour gradients derived from the model described in Section \ref{mild_SF}. 
In the left panel, a metallicity gradient and a weak star formation in the centre have
been considered. In the right panel an age gradient has been also added. 
The magenta points represent the mean colour of the stellar population in the inner and outer regions.  
}
\label{fig:grad_SF}
\end{center}
\end{figure}

Galaxy 1758 presents an extreme value of the U-R gradient at 1$\sigma$ 
 ($\nabla$= -0.55 mag). 
 In order to produce such steep U-R gradient, it is needed to consider an even 
 lower metallicity towards the external region (Z$_{out}$= 0.2 Z$_{\odot}$). 
 This would imply a steeper metallicity gradient ($\nabla_{Z}$ $\sim$ -0.7)
 not observed in the local ellipticals.
For this galaxy, we simultaneously produce both the colour gradients by adding 
0.3 per cent of star formation  together with both an age gradient (Age$_{in}$= 4 Gyr and 
Age$_{out}
$= 2 Gyr) and a metallicity gradient (Z$_{in}$ = Z$_{\odot}$ and 
Z$_{out}$= 0.2 Z$_{\odot}$). 
This galaxy is shown in Fig. \ref{fig:grad_SF} as an example.
The upper panel shows the F775W-F850LP colour profile (black line) with the best 
fitting slope (red line), while the lower panel the F850LP-F160W colour profile. 
The blue dashed line is the colour gradient including star formation in the centre for an underlying metallicity gradient (left) 
and both a metallicity and an age gradient (right).

Finally, two galaxies (684 and 1747) present an extreme value of 
the UV-U gradients at 1$\sigma$ ($\nabla$= 0.47, 0.58 mag, respectively).
For these galaxies, we are not able to simultaneously produce both the colour 
gradients.
Indeed, in order to produce the steep observed UV gradients, $\gae$ 1 per cent in mass 
of star formation is needed, but this produces positive U-R gradients, even if 
we consider the lowest metallicity available in the external region 
(Z$_{out}$= 0.2 Z$_{\odot}$).
These galaxies are isolated; they are not the faintest in the sample and their 
structural parameters are not extreme. 
Hence, we have no reason to suppose that the colour gradients of these galaxies 
are overestimated.

In conclusion, by applying a model with star formation in the centre 
together with a metallicity and/or an age gradient, we are able to simultaneously 
produce at 1$\sigma$ both the colour gradients for 11 out of the 13 ellipticals 
($\sim$ 85 per cent of the sample).
On the basis of this model, these ellipticals would be composed by an older and more metal-rich
population in the centre, a younger and more metal-poor population in the outskirts and a small star-forming component in the core.

Taking as reference the mean value of the stellar mass 
(<M$_*$> $\sim$ 3 $\cdot$ 10$^{10}$ M$_\odot$) of the galaxies of our sample, this young 
star-forming component would contribute in stellar mass for just 
$\sim$ 2.7 $\cdot$ 10$^8$ M$_\odot$. 
Since the BC03 models are normalized at 1 M$_{\odot}$ and a stellar population 
of 0.1 Gyr with $\tau$=0.1 Gyr  is characterized by a 
SFR = 3.6 $\cdot$ 10$^{-9}$ M$_{\odot}$ yr$^{-1}$, 
the SFR associated with the young central star forming component would be 
SFR$\sim$1 M$_{\odot}$ yr$^{-1}$.
It is worth noting that this SFR is rather stable with respect to the assumptions made:
at fixed SFH (e.g. $\tau=0.1$ Gyr) a younger stellar population (<0.1 Gyr) would be 
characterized by a stronger UV emission and, consequently, the required mass percentage
would be lower (<0.5 per cent) leaving nearly unchanged the resulting SFR.

Following the relation 
\begin{equation}
SFR \ (M_{\odot} yr^{-1}) = (1.4 \pm 0.4) \cdot 10^{-41} \ L[OII] \ (erg \ s^{-1}),
\end{equation}
\citep{kennicutt98}, the flux $f$[0II]$\lambda$3727 associated with a SFR of $\sim$ 1 M$_{\odot}$ yr$^{-1}$ in a galaxy at $z$=1.39 would be $f$[0II] $\sim 6 \cdot 10^{-18}$ erg cm$^{-2}$ s$^{-1}$. 
Such faint $f[OII]$ fluxes are barely detectable even at a 10 m class telescope. 
For instance, \cite{cimatti08} report the detection of such faint fluxes in three 
passive galaxies in the GMASS survey, all of them observed for more than 30 hr. 
Thus, we could not directly confirm whether such a weak star formation activity is 
actually present in the galaxies of our sample, since it is too low to be detectable 
with the current instrumentations.
However, some works find indirect evidence of the presence of star formation in early-type 
galaxies both at high-$z$ and in the local universe. 
For instance, \cite{lonoce14}, measuring the spectral indices of a sample of early-type 
galaxies at  0.7 < z$_{spec}$ < 1.1, found that, in some cases, the measured H+K (Ca II) 
index is consistent with the presence of a small mass percentage (<1 per cent) of a young stellar 
component (< 1 Gyr) with recent weak star formation.
Also \cite{wagner15}, studying a sample of massive (M > $10^{10.1}$M$_\odot$)  cluster
early-type galaxies at 1.0 < z < 1.5, found that 12 per cent of them are likely
experiencing star formation activity. 
In the local Universe, \cite{kaviraj08}, studying $\sim$ 2100 early-type galaxies 
in the SDSS DR3, found that at least $\sim 30$ per cent of them
have UV to optical (NUV - r) colour consistent with recent star formation. 

If the star formation is the origin of the UV excess in the galaxy central 
regions of our sample, this scenario requires a certain degree of fine-tuning. 
The fact that we observe positive UV-U (or null) gradients in all the galaxies
implies that all of them are experiencing star formation in their centre,
otherwise we would have observed negative UV-U gradients in some of them at least. 
This implies, in turn, that the star formation cannot be either episodic or short, but 
that it protracts over time in a steady fashion at low levels to allow us to observe it  
as it is (low) in all the galaxies.
Being protracted over long time (say $\sim1$ Gyr, being at $z\sim1.4$), this SF even if weak, 
needs to be supplied with gas.  
The fuel cannot come from inflows of intracluster medium (ICM), since galaxies
are in thermal equilibrium with the ICM.
By the way, this kind of supplying would flatten the metallicity gradient and, consequently, 
the U-R gradient being the ICM less metal-rich. 
On the contrary, such star formation should be feeded by the residual processed gas 
in the galaxies (closed box).
This process would efficiently enhance the metallicity towards the centre
\citep[e.g.][]{peng15}
producing naturally the observed negative U-R gradients.

As already said, this scenario requires a fine-tuning in term of synchronism and duration 
of the star formation as possible origin of the observed UV excess in the centre of our 
galaxies.
However, apart from this, we have no evidence to exclude it.
Moreover, it is worth considering that this model gives rise to the observed colour 
gradients simultaneously reproducing both the UV-U and the U-R gradients (see Fig. \ref{fig:grad_SF}
)
and that it would naturally produce the metallicity gradient observed
also in the local ellipticals.

\subsection{A possible QSO contribution?}
\begin{figure}
\includegraphics[width=8.8truecm]{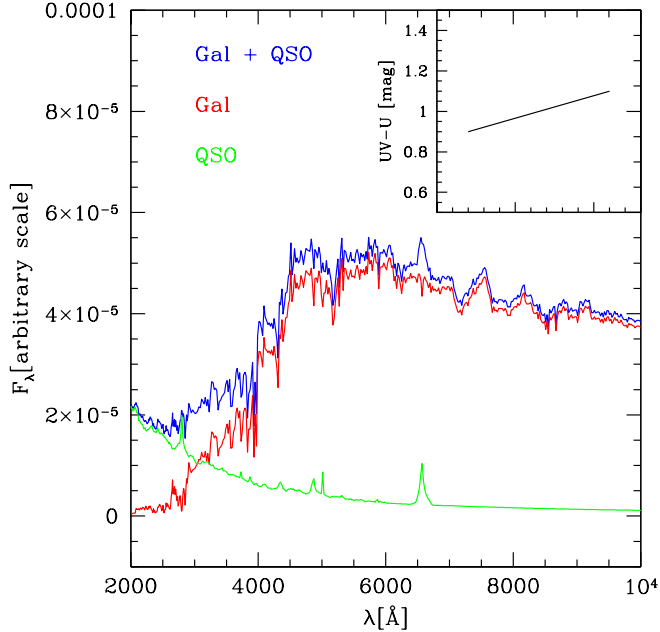}
\caption{QSO contribution to the UV excess in the central galaxy regions. Green line represents the composite spectrum of a QSO \citep[]{francis91}, the red line is the template of a galaxy (4 Gyr old) and the blue line is the sum of the two components.
The QSO is scaled to be 2 mag fainter than galaxy in the B band.
The addition of this QSO component together with an age and metallicity gradient reproduces simultaneously gradients in the range $\nabla(UV-U)\le0.25$ mag and $\nabla(U-R)>-0.3$ mag.
Brighter QSO component would flatten the U-R gradients and  would make the integrated UV-U and U-R colours of the galaxies bluer than those observed.
}
\label{fig:qso}
\end{figure}

Quasars are known to have a SED characterized by strong UV emission, peaking at $\sim$1500\; \AA \;
\citep[the so-called big blue bump;][]{shields78}.
To verify whether the UV excess in the galaxy central regions is
the sign of the blue bump, we have added in the centre 
a QSO component instead of the star-forming one.
We have considered the composite spectrum of \cite{francis91}.
In Fig. \ref{fig:qso} is shown, as example, the case of the QSO component  needed to 
produce a UV-U colour gradient of $\sim0.2$ mag.
In terms of absolute magnitude, this QSO is $\sim$ 2 mag fainter than 
the galaxy in the B band rest-frame.
In this case, the U-R gradient is not significantly affected and 
we can reproduce simultaneously gradients in the range 
$\nabla(UV-U)\le0.25$ mag and $\nabla(U-R)>-0.3$ mag, assuming a radial
decrease of metallicity and age.
For instance, for galaxies 692 and 1284, having gradients in the above ranges,
the QSO would have an absolute magnitudes of M$_B$ $\sim$ -18.7 mag and M$_B$ $\sim$ -19.7 mag (Seyfert like), respectively.
However, higher values of the UV-U gradient cannot be accounted for
by further increasing the QSO luminosity.
Indeed, in this latter case, the integrated UV-U and U-R colours of the
galaxy would be affected resulting in significantly bluer
(>0.3 mag in UV-U and in U-R) than those observed. 
Hence, a QSO could be present in the galaxies and contribute to 
the observed UV excess but cannot be the only cause.
It should be superimposed to, e.g. star formation to account
for the observed gradients without altering the colours of the galaxies.

\subsection{Or, rather, a population of He-rich stars?}
In the local Universe it has long been known that elliptical galaxies can present an UV excess
in their SED, stronger at wavelengths shortwards of 2100 \AA \; \citep[e.g.][]{code79,o'connel92,o'connel99}, with a tail extending up to $\sim$3200 \AA \;  \citep[see e.g.,][]{greggio90,ponder98,darman95}.
This was unexpected since ellipticals were supposed to contain mainly old and cold stellar 
populations. 
Since its discovery, the UV upturn phenomenon in elliptical galaxies has generated a big 
discussion about its origin. 
From the shape of the SED in the UV, it derives that the temperature of the stars, which give rise to the UV excess, is in the range 20000 $\pm$ 3000 K \citep{brown97}.
This effectively ruled out young stars as the main driver of the UV upturn.
The main mechanisms that are supposed to be responsible for the UV upturn are He-burning 
stars \citep[e.g.][]{han02,han03,han07,han08} and He-rich stars \citep[e.g.][]{yi11,carter11}.  
We have considered these scenarios  to test if they can explain 
the blue excess we observe towards the centre of our ellipticals at $z\sim1.4$.
Indeed, even though the emission peak of the UV upturn is at 
wavelengths < 2100 \AA, its tail extends up to $\sim$ 3200 \AA, affecting 
the F775W band (filter range 2700 \AA  \; $\lae \lambda_{rest} \lae$ 3600 \AA).

\subsubsection{He-burning stars}
He-burning stars are stars that have already stopped burning hydrogen in their cores and they 
are burning helium. 
Assuming for the stellar population a formation redshift $z_f\sim6$, at z=1.39 the
population would be $\sim$ 4 Gyr old. 
Hence, at z = 1.39, only stars with 1.2 M$_{\odot}\lae$ M $\lae$ 1.5 M$_{\odot}$ would be 
in the He-burning phase since their lifetime in the {Main Sequence} (MS) is 
$\sim$3-4 Gyr \citep{iben67a}. 
Stars with M<1.2 M$_{\odot}$ would still be in the MS (lifetime in MS phase $> 10^{10}$ yr), 
whereas stars with M>1.4M$_{\odot}$ have already evolved (lifetime in MS phase $< 10^{8}$ yr). 
The He-burning phase is characterized by a very short time-scale if compared
to the MS phase and, for stars with 1.2 M$_{\odot}$<M<1.5 M$_{\odot}$ it lasts just 
$\sim 10^8$ yr.   
Hence, this very short time makes highly unlikely that the UV excess seen 
in the galaxy central regions is due to an excess of intermediate mass stars in the 
He-burning phase. 

\subsubsection{He-rich stars}
One other possibility is that the cores of elliptical galaxies host a He-rich 
subpopulation of stars.
These stars are formed from the gas enriched by the evolution of very high mass 
stars. 
In the $\Lambda$cold dark matter model the first stars are predicted to be 
composed principally by hydrogen \citep[see e.g.][]{blumenthal84,komatsu09}.  
The absence of metals disadvantaged the formation of low and intermediate mass stars, since 
the cooling is less efficient. 
Results from recent numerical simulations of the collapse and fragmentation of primordial gas 
clouds suggest that the first stars were predominantly very massive, with typical masses 
M$_* >100$ M$_{\odot}$  \citep{bromm99,bromm02,nakamura11,abel00,abel02}.
These stars would evolve very quickly and the supernova (SN) explosions that ended their lives  
would enrich efficiently the ISM with heavy elements 
\citep[see e.g.][]{ostriker96,gnedin97,ferrara00,madau01}. 
These metals would fall towards the centre of the galaxies due to the potential well 
naturally explaining the observed negative metallicity gradients resulting from the subsequent
generations of stars.

We know that
these stars exist in local ellipticals, where they contribute to the classical
UV upturn \citep[e.g.][]{o'connel99}.
Moreover, the UV gradients observed 
in Coma galaxies by \cite{carter11} imply that there exists an He abundance gradient, established at very early times. 
He-enriched stars are
known to be bluer than their counterparts with standard metal abundance (see, e.g., \citealt{chantereau15}). This is observed 
in the bluer UV colours of He-enriched red giants in \cite{lardo12} and in the bluer secondary MSs of several globular 
clusters \citep[e.g.][]{piotto05}.

 Since
these stars will be bluer, they will naturally produce an excess in the UV in the
centre, but they will not strongly affect
optical colours, as also shown by the above observations in globular clusters. 
Thus, an He
abundance gradient, which
is known to exist in the local Universe, would reproduce the observed positive UV-U gradients
without fine-tuning. Unfortunately,
we lack population synthesis models for He-enriched populations to study this issue
in detail.

\section{Colour gradients and global properties of galaxies}

\begin{figure*}
\begin{center}
\includegraphics[width=14truecm]{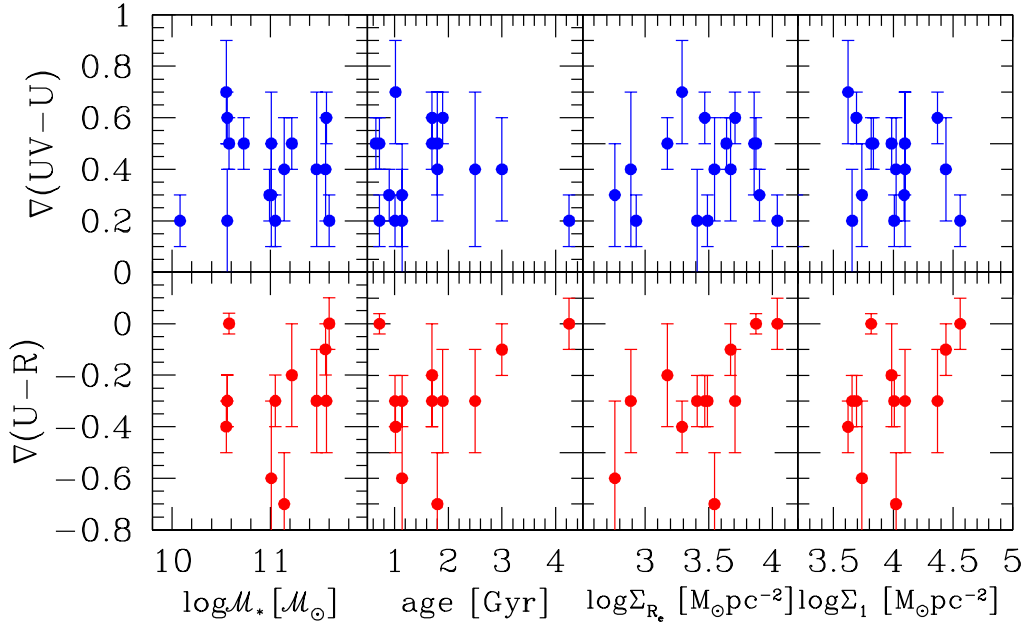}
\caption{The UV-U gradients (upper panels) and U-R gradients (lower panels) versus, from left to right, log$\mathcal{M}_*$, the mean age as derived by
the SED fitting, log$\Sigma_{R_e}$ and log$\Sigma_{1}$ within 1 kpc radius.
}
\label{fig:grad_glo}
\end{center}
\end{figure*}

To further investigate on the origin of the UV excess towards the
centre and on the mechanisms producing the observed colour gradients
in our galaxies, we have looked for the presence of correlations between
gradients and global properties of galaxies.
We considered the stellar mass $\mathcal{M}_*$, the mean age as derived by
the SED fitting, the effective stellar mass density
$\Sigma_{R_e}=\mathcal{M}_*/(2 \pi R_e^2)$ and the central stellar mass
density $\Sigma_{1kpc}$ within 1 kpc radius, as derived in \cite{saracco16}.
Fig. \ref{fig:grad_glo} shows the relation between colour gradients and global properties
of galaxies.
The U-R colour gradient seems to be weakly correlated with the effective
stellar mass density of galaxies in the sense that the larger the gradient
the lower the density. However, the Spearmen rank test provides a
probability
of just 90 per cent significance.
Actually, no correlation was detected with the global properties of
galaxies,
both considering the UV-U (upper panel) and the U-R (lower panel) colour
gradients.
This is not surprising since colour gradients trace principally the
variation
of the stellar component dominating the galaxy within 1-2R$_e$ while
the other properties are derived considering the whole galaxy mass.
Similar results are also found by \cite{depropris15,depropris16} for cluster red sequence galaxies at a comparable redshift and by \cite{gargiulo12} for a sample of field early-type galaxies at 0.9 < $z$ < 1.92.
We have also verified if the amplitude of the colour gradients depends on the position of the galaxy inside the cluster, namely on the radial distance from the cluster centre (i.e. on the local density).
Also in this case we did not detect a correlation.
The low statistics does not allow us to establish whether the local environment has affected the colour gradients of these galaxies.

\section{Comparison with previous works}
\label{data} 
In this section we compare our results with those present in literature. 
Although colour gradients are largely studied in 
the local universe, still few works are focused on gradients in cluster and field galaxies at 
$z > 1$. 

\subsection{Cluster environment}
 
 Our results are in agreement with those of \cite{depropris15}, who studied the 
 F850LP - F160W colour gradients  of red  sequence galaxies belonging to four clusters 
 (one of which in common with us) at $1 < z < 1.4$ using the ratio of their effective 
 radii measured in the two bands. 
  They found that the galaxies in their sample present significant positive 
 $log(R_e(z)/R_e(H)) $, hence negative colour gradients,  larger than those 
 observed locally (e.g. in Virgo cluster). 
 The authors divide their red sequence galaxies in red discs ($n$<2) and red spheroids ($n$>2). 
  Since 16 out to the 17 ellipticals of our sample have $n$>3, for a proper comparison, 
  in Fig. \ref{fig:rapp} we plotted with grey crosses the galaxies of \cite{depropris15}
  sample with $n$>3.  
  The median value of the $log(R_e(F850LP)/R_e(F160W))$ of their cluster 
  galaxies with $n$>3 is 0.18, consistent with the median value (0.21) of our 
  cluster ellipticals.
The authors interpret the 
negative F850LP - F160W gradients in term of metallicity gradients plus radial age gradients, with a younger stellar 
population towards the outskirts, due to the fact that star formation may have continued for a longer time in the outer 
regions of  the galaxies \citep[e.g.][]{tacchella16}.
    Our interpretation, as resulting from the analysis shown in Section \ref{mild_SF}, does not exclude
    a contribution of age gradient, specially to justify the most extreme U-R gradients.
    However,  it seems that the major role is played by the metallicity, as also seen in the 
    local ellipticals. 
     Anyway, the similarities in the colour gradients, independently found in the range 
     $1 < z < 1.4$, suggest that cluster galaxies have experienced 
     similar evolutionary processes in their past.

Colour gradients in cluster and field galaxies were also derived by 
\cite{allen15} at $z\sim2$.
The authors selected a mass-complete sample of 59 cluster galaxies at 
z = 2.095 and 478 field galaxies at $2 < z < 2.2$ with log$(M_*/M_{\odot}) \ge 9$. 
Galaxies in both samples are separated in quiescent and star-forming galaxies using the 
UVJ rest-frame colour - colour diagram \citep[e.g.][]{williams09,williams10,wuyts09,wild14}. The authors stacked galaxy images both for cluster and field sample to measure average F814W - F160W ($(U-V)_{rest frame}$) radial colour profiles as a function of the mass.
They found no colour gradients either in field or in cluster quiescent galaxies. 
Unlike \cite{allen15}, we find that most of our cluster elliptical galaxies presents 
significative colour gradients. 
This suggests that from $z \sim$ 2 to $z \sim$ 1.4 
cluster galaxies may have experienced some evolutionary processes, which have 
increased the amplitude of their colour gradients producing a significant radial 
variation of their stellar population properties.

We also compared our results with \cite{chan16}, who 
studied a sample of 36 passive galaxies selected in the same cluster 
considered in our analysis, XMMUJ2235-2557. 
They base their analysis only on the rest-frame U-R  gradient.
Since we selected elliptical galaxies, 
only 12 galaxies are in common (for a detailed comparison see Appendix  \ref{Chan}).
They derived U-R colour gradients from the observed colour profile up 
to 3.5$a_e$ ($a_e$ is the major axis), whereas we derived colour gradients from 
the intrinsic colour profile and up to 2 effective radii. 
For the 12 galaxies in common we found a median value of the F850LP-F160W colour gradients of -0.3 mag, 
consistent with the median value ($\sim$ -0.4 mag) found by \cite{chan16}. 
Anyway, for one (595) out of the 12 galaxies in common,  they derive a positive gradient, whereas 
we measured a negative gradient. 
We will discuss the possible origin of this difference in Appendix \ref{Chan}. 
 They explain the evolution of the colour gradients from $z$ = 1.39 to $z \sim$ 0 through 
 the presence of an age gradient ($\nabla_{age} \sim -0.33$) besides a metallicity gradient 
 ($\nabla_Z \sim -0.2$). 
 In agreement with \cite{chan16}, we also found that the rest-frame U-R gradients can be produced 
 by the radial variation of age and metallicity.
 However, our analysis shows that relevant information is stored in the UV-U 
 gradient.  
 As pointed out in Section 6, the rest-frame ultraviolet gradient is a 
 very important key to correctly interpret the properties of the stellar population  
 of galaxies and to constrain their history of mass accretion.
 
\subsection{Cluster versus field}
For field elliptical galaxies we first compare our results
with those of \cite{gargiulo12}. 
The authors have derived the F850LP - F160W colour gradients for 11 morphologically selected ellipticals  
 at $1 < z < 1.9$ in the GOODS-South region. 
Structural parameters and gradients have been derived as in this work.
They found significant negative $\nabla(F850LP - F160W)$ within 1R$_e$ in $\sim 70$ per cent
of the galaxies, a fraction that rises up to 100 per cent when
R>R$_e$ is considered, as for our sample. 
By considering the eight galaxies of their sample in redshift 
range $1.2 < z < 1.6$, to avoid evolutionary effects, we found that our cluster ellipticals
show U-R (F850LP - F160W) colour gradients of amplitudes ($<\nabla>=-0.3\pm$0.2 mag)
comparable to those of field ellipticals ($<\nabla>=-0.5\pm$0.3 mag).

In a previous work \citep{gargiulo11}, the authors studied the F606W - F850LP ($(UV-U)_{rest frame}$) for 
20 early-type galaxies at 0.9 < $z$ < 1.92, only four of these in common with the other study.
They detected significant radial UV-U colour variation in 10 out of the 20 galaxies,
five showing negative gradients and the remaining ones showing positive gradients.
Hence, contrary to cluster galaxies, field galaxies at similar redshift ($1.2 < z < 1.6$) 
seem to present both positive and negative ultraviolet gradients.  
Their analysis shows that, for a minor fraction, the observed gradients can be well reproduced by  
a pure radial age variation or by a pure metallicity variation while, for the remaining galaxies, 
more than one property of the stellar population must simultaneously 
vary to account for the observed gradients.
Hence, although the small statistic, this comparison may suggest that the environment could 
have already influenced the evolution of elliptical galaxies.

Finally, we compared our results with those of \cite{guo11}, who studied 
the F850LP - F160W colour gradients for six massive 
($M_* > 10^{10}$ M$_{\odot}$) early-type galaxies selected at redshift $1.3 < z < 2.5$
for their early-type morphology and low SSFRs (SSFR = SFR/$M_* \le 10^{-11}$ yr$^{-1}$).
By deriving colour gradients in concentric annular apertures up to $\approx $ 10 R$_e$,
they found, concordantly with our results, that the inner regions are redder than the outskirts. 
A radial variation of a single stellar parameter (age/metallicity) 
does not account for the observed colour gradients.
The authors found a correlation between the 
dust and the observed colour gradients: the redder the inner region the higher the central 
dust obscuration.
Like \cite{guo11} we also found that the radial variation of a single stellar 
 parameter can not account for the observed colour gradients, but we do not confirm 
 the correlation between the observed colour gradients and the dust content.
 However, their analysis considered only the $\sim$(U-R) colour gradients.  
 
 \section{Summary and conclusions}
 
In this paper, we investigated the radial variation of the stellar 
population properties in a sample of 17 ellipticals belonging to the 
cluster XMMU J2235.3-2557 at $z$=1.39.
The galaxies have been selected on the basis of their elliptical morphology 
that we assigned through a visual inspection of their ACS-F850LP images.

Making use of the ACS images in the F775W and F850LP bands and of the 
WFC3 images in the F160W band, we derived their rest-frame UV-U (F775W-F850LP) 
and U-R (F850LP-F160W) colour gradients using three different methods:
the logarithmic slope of the deconvoluted colour profiles,
the ratio of the effective radii as measured in the three different
filters and through the observed colour profiles. 
While the first two methods are both dependent on the best fitting 
procedure of the colour profiles, the third method does not 
involve any fit of the light profiles of the galaxies.
 
 Our main results are as follows.

\begin{itemize}

\item[(i)] The UV-U colour gradients are systematically positive ($\sim$80 per cent) or null ($\sim$20 per cent), never negative. On the contrary, the U-R colour gradients are systematically negative ($\sim$70 per cent) or null ($\sim$30 per cent), never positive. 
The mean value of the UV-U gradients is 0.4$\pm$0.2 mag, slightly larger than the 
mean value of the U-R gradients (-0.3$\pm$0.2 mag) in spite of the much narrower
wavelength interval of the UV-U gradient.

\item[(ii)]  The results do not depend on the method used to derive the colour gradients.

\end{itemize}

We then investigated the origin of the observed colour gradients 
using BC03 stellar population models. 
Our results are the following:

\begin{itemize}

\item[(i)]

The opposite slopes of the observed colour gradients cannot be accounted
for by the radial variation of a single parameter (age, metallicity or dust).
A variation of these parameters produces ever gradients with concordant slopes.
In particular, a radial decrease in metallicity/age/dust from the central to
external regions produces negative UV-U and U-R colour
gradients and vice versa.

\item[(ii)] The amplitude of the U-R colour variations observed in our galaxies can be 
easily accounted for by an age or by a metallicity radial decrease. 
On the contrary, we are not able to reproduce the amplitude of the observed UV-U colour 
gradients through age or metallicity variation.
\end{itemize}

The UV-U colour gradients we detected show 
the presence of an UV excess towards the centre of the galaxies with respect 
to the outer regions and call into question other mechanisms 
able to efficiently produce UV and U emission in the galaxy central
regions. 
We considered these possible UV and U emission sources: star formation, the presence of a QSO,
stars in the He-burning phase and He-rich stars.

We found that
a weak star formation ($\sim 1$ M$_\odot$ yr$^{-1}$) superimposed to an old 
stellar component towards the centre and to a negative metallicity gradient 
can reverse the UV-U colour gradient
simultaneously reproducing the amplitude and the opposite slopes of 
the UV-U and U-R colour gradients.
The required central star-forming component should not exceed $\sim$0.5 per cent 
in stellar mass not to affect the U-R gradient.
In this way, we are able to reproduce both the colour gradients for 10 out 
of the 13 ellipticals ($\sim$ 85 per cent of the sample), for which both colour gradients are available. 
However, if this is the cause of the bluer UV-U colour towards the centre, 
this implies that all the galaxies are currently experiencing SF in the
centre since all of them have positive or null UV-U gradients.
This in turn implies that the star formation should be steady and protracted
over a long time ($>1$ Gyr) and not episodic and/or erratic. 

We found that the presence of a QSO could contribute to the UV excess, but
it cannot justify alone the observed UV-U gradient unless flattening the U-R gradient and significantly altering the integrated colours of the galaxies.

We found that it is highly unlikely that the observed UV excess we observed is due to the abundance of He-burning stars towards the centre of the galaxies. Indeed, at $z$=1.39, only stars with 1.2 M$_{\odot}\lae$ M $\lae$ 1.5 M$_{\odot}$ would be 
in the He-burning phase. 
Since the He-burning time-scale for stars in this range of mass is very short ($\sim 10^8$ yr), it would require a very strong fine-tuning in term of synchronism to justify the UV excess observed in almost all the galaxies of our sample.

Finally, we also found that
an excess of He-rich stars towards the centre of the galaxies
would qualitatively explain the positive UV-U gradients and
the negative metallicity U-R gradients as well, without introducing any fine-tuning.
It is worth noting that the presence of a population of He-rich stars in the centre of elliptical galaxies would be in agreement with the  NUV-UV colour gradients observed in some ellipticals in the local universe. Unfortunately, we cannot test 
this scenario since no suited models are available.

 \section*{acknowledgments}
 
FC wishes to thank Paola Severgnini for the helpful comments and suggestions for the QSO discussion.
We thank the anonymous referee for the useful and constructive comments.
This work is based on observations collected at the European Organisation for Astronomical Research in the Southern Hemisphere under ESO programmes 60.A-9284 and 079.A-0758. This work is also based on observations made with the NASA/ESA Hubble Space Telescope, obtained from the data archive (ID 10698, 10496 and 12051) at the Space Telescope Science Institute, which is operated by the Association of Universities for Research in Astronomy, and with the Spitzer Space Telescope, which is operated by the Jet Propulsion Laboratory, California Institute of Technology under a contract with NASA.


\bibliographystyle{mnras}
\bibliography{paper_col_grad_fin} % if your bibtex file is called example.bib

%\begin{thebibliography}{99}
%\end{thebibliography}

%%%%%%%%%%%%%%%%%%%%%%%%%%%%%%%%%%%%%%%%%%%%%%%%%%

%%%%%%%%%%%%%%%%% APPENDICES %%%%%%%%%%%%%%%%%%%%%

\appendix

\section{Surface Brightness Profiles}
\label{galfit}

\begin{table*}
\caption{For each galaxy of the sample we report the ID number, the S\'{e}rsic index $n$, the
total magnitude and the effective radius $R_{e}$ [kpc] as derived from
the fitting of the surface brightness profile in the F775W, F850LP
and F160W images, respectively. For the the F850LP band we also report the axial ratio (b/a)$_{850}$. }
\centerline{
\begin{tabular}{rccccccccccc}
\hline
\hline
  ID &$n_{775}$& $i_{775}^{fit}$&R$_e^{775}$ & $n_{850}$&$z_{850}^{fit}$& R$_e^{F850}$&(b/a)$_{850}$&$n_{160}$ & $H_{160}^{fit}$ &R$_e^{F160}$  \\
     &     &[mag]  & [kpc]                 &     & [mag]    &[kpc]   &  &           &   [mag]     & [kpc] \\
\hline
    358&  6.0$\pm$0.4 &  24.5$\pm$0.2&  1.1$\pm$0.1 &  6.0$\pm$0.4   &  23.5$\pm$0.1 & 1.5$\pm$0.2 & 0.6$\pm$0.1&  ---         &   ---            &  ---\\ 
    595&  5.7$\pm$0.3 &  22.9$\pm$0.1&  2.1$\pm$0.3 &  5.6$\pm$0.3   &  21.6$\pm$0.1 & 3.5$\pm$0.5&0.9$\pm$0.1& 4.5$\pm$0.1 &   19.2$\pm$0.1 &  2.5$\pm$0.1\\ 
    684&  6.0$\pm$0.4 &  24.1$\pm$0.2&  0.6$\pm$0.1 &  6.0$\pm$0.4   &  22.9$\pm$0.1 & 1.4$\pm$0.2 & 0.6$\pm$0.1& 4.0$\pm$0.1 &   20.9$\pm$0.1 &  0.8$\pm$0.1\\ 
    692&  4.0$\pm$0.2 &  24.7$\pm$0.2&  1.3$\pm$0.1 &  4.1$\pm$0.2   &  23.5$\pm$0.2 & 1.5$\pm$0.2 & 0.7$\pm$0.1 &3.0$\pm$0.1 &   21.0$\pm$0.1 &  0.9$\pm$0.1\\ 
    837&  6.0$\pm$0.2 &  23.3$\pm$0.1&  4.3$\pm$0.6 &  6.0$\pm$0.4   &  21.9$\pm$0.1 & 7.8$\pm$1.1 & 0.7$\pm$0.1 &6.0$\pm$0.2 &   19.4$\pm$0.1 &  5.2$\pm$0.2\\
   1284&  4.3$\pm$0.3 &  23.4$\pm$0.1&  2.0$\pm$0.3 &  4.3$\pm$0.2   &  22.3$\pm$0.1 & 2.4$\pm$0.3 & 0.9$\pm$0.1&4.0$\pm$0.1 &   20.0$\pm$0.1 &  1.7$\pm$0.1\\ 
   1539&  4.8$\pm$0.3 &  23.7$\pm$0.2&  1.1$\pm$0.1 &  3.5$\pm$0.2   &  22.7$\pm$0.1 & 1.5$\pm$0.2 & 0.6$\pm$0.1& ---         &   ---            &  ---         \\ 
   1740&  3.3$\pm$0.2 &  21.8$\pm$0.1& 13.3$\pm$2.0 &  3.3$\pm$0.2  &  20.8$\pm$0.1 &12.8$\pm$1.9& 0.6$\pm$0.1& 4.6$\pm$0.1 &   17.8$\pm$0.1 &  20.7$\pm$0.9\\ 
   1747&  4.3$\pm$0.2 &  24.7$\pm$0.2&  0.7$\pm$0.1 &  4.8$\pm$0.3   &  23.3$\pm$0.1 & 1.7$\pm$0.2 & 0.6$\pm$0.1& 4.0$\pm$0.2 &   21.1$\pm$0.1 &  0.9$\pm$0.1\\ 
   1758&  2.9$\pm$0.2 &  23.9$\pm$0.2&  1.9$\pm$0.2 &  2.9$\pm$0.2   &  22.6$\pm$0.1 & 2.5$\pm$0.3 & 0.8$\pm$0.1& 2.90$\pm$0.04&   20.3$\pm$0.1 &  1.5$\pm$0.1\\ 
   1782&  2.6$\pm$0.2 &  23.5$\pm$0.1&  1.6$\pm$0.1 &  3.6$\pm$0.2   &  22.5$\pm$0.1 & 3.4$\pm$0.5 &0.6$\pm$0.1& 3.11$\pm$0.04&  19.5$\pm$0.1 &  2.2$\pm$0.1\\ 
   1790&4.4$\pm$0.3 &  23.1$\pm$0.1&  1.9$\pm$0.2 &  4.4$\pm$0.3   &  21.9$\pm$0.1 & 2.4$\pm$0.3 & 0.6$\pm$0.1&4.1$\pm$0.1 &   19.3$\pm$0.1 &  2.3$\pm$0.1\\ 
   2054& 5.0$\pm$0.3 &  23.2$\pm$0.1&  2.7$\pm$0.4 &  4.4$\pm$0.3   &  21.9$\pm$0.1 & 4.2$\pm$0.6 & 0.7$\pm$0.1& 5.0$\pm$0.1 &   19.5$\pm$0.1 &  3.7$\pm$0.2\\
   2147&  5.4$\pm$0.3 &  23.5$\pm$0.1&  3.6$\pm$0.5 &  5.3$\pm$0.3   &  22.2$\pm$0.1 & 5.3$\pm$0.8 & 0.7$\pm$0.1& 4.0$\pm$0.1 &   20.0$\pm$0.1 &  2.3$\pm$0.1\\
   2166&  3.6$\pm$0.2 &  22.6$\pm$0.1&  1.2$\pm$0.1 &  3.0$\pm$0.2   &  21.5$\pm$0.1 & 1.4$\pm$0.1 & 0.6$\pm$0.1& ---         &   ---            &  ---         \\
   2429&  2.1$\pm$0.4 &  24.3$\pm$0.2&  0.6$\pm$0.1 &  2.1$\pm$0.2   &  23.0$\pm$0.1 & 0.9$\pm$0.1& 0.6$\pm$0.1& 2.1$\pm$0.1&   21.0$\pm$0.1 &  0.9$\pm$0.1\\
   2809&  4.8$\pm$0.3 &  23.3$\pm$0.1&  1.2$\pm$0.1 &  3.2$\pm$0.2   & 22.7$\pm$0.1 & 1.4$\pm$0.2 & 0.6$\pm$0.1& --- & ---    & ---  &\\
\hline
\end{tabular}
}
\label{tab:tab_gal}
\end{table*}

As described in Section \ref{Structural parameters} we have derived the surface brightness 
profiles by fitting a S\'ersic law (equation \ref{eq:sersic}).

\begin{figure*}
\begin{center}
\includegraphics[width=4.3cm]{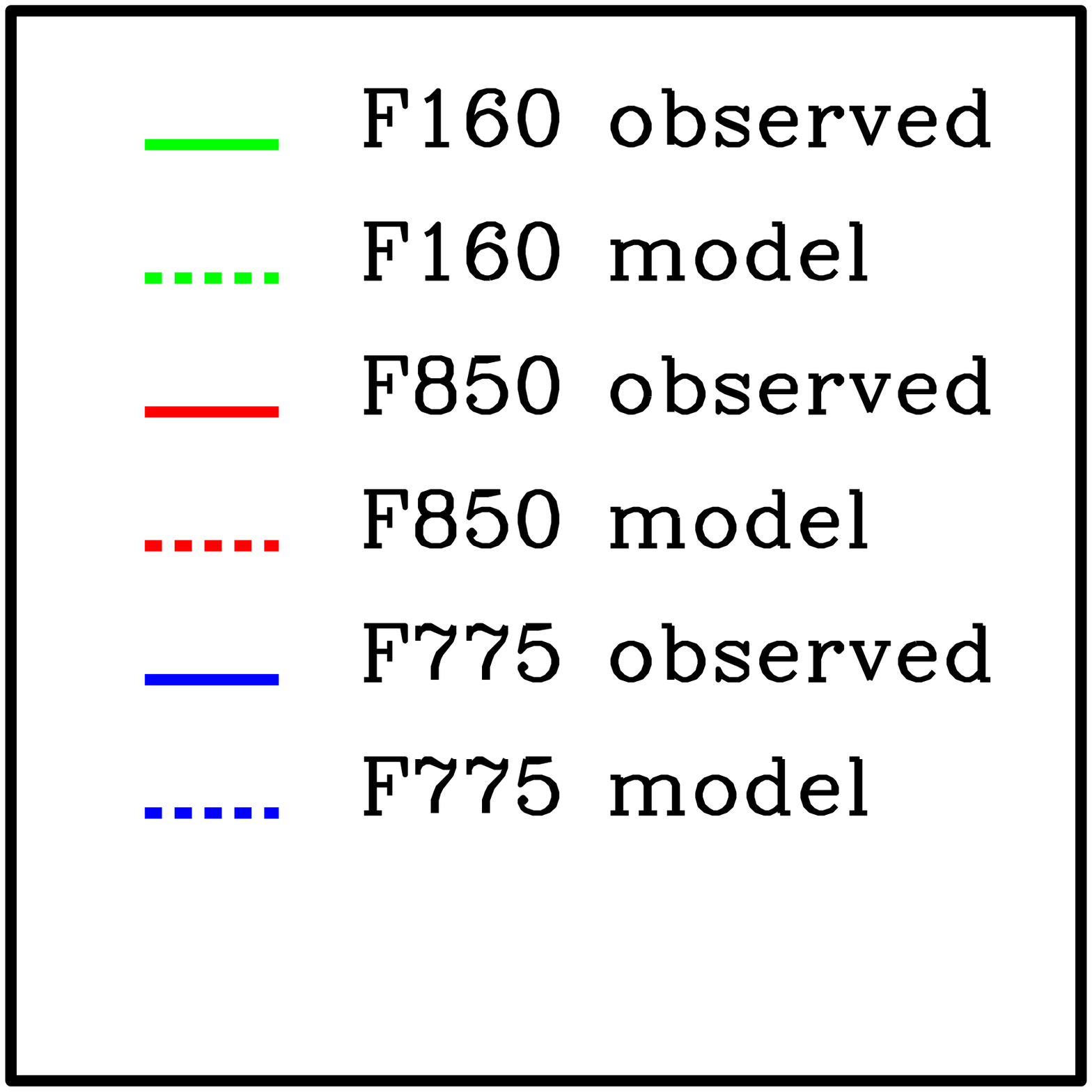}
\includegraphics[width=4.3cm]{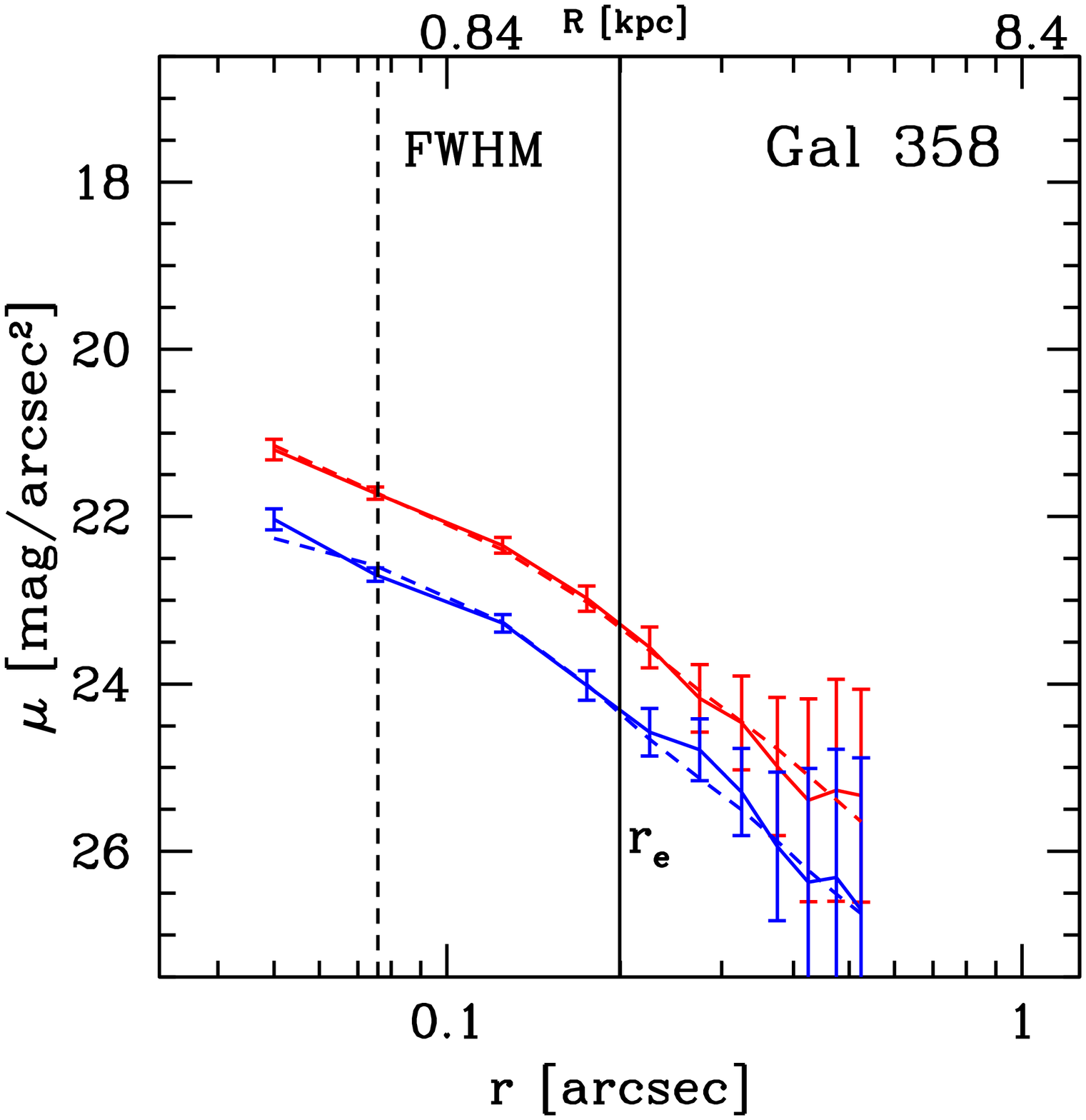}
\includegraphics[width=4.3cm]{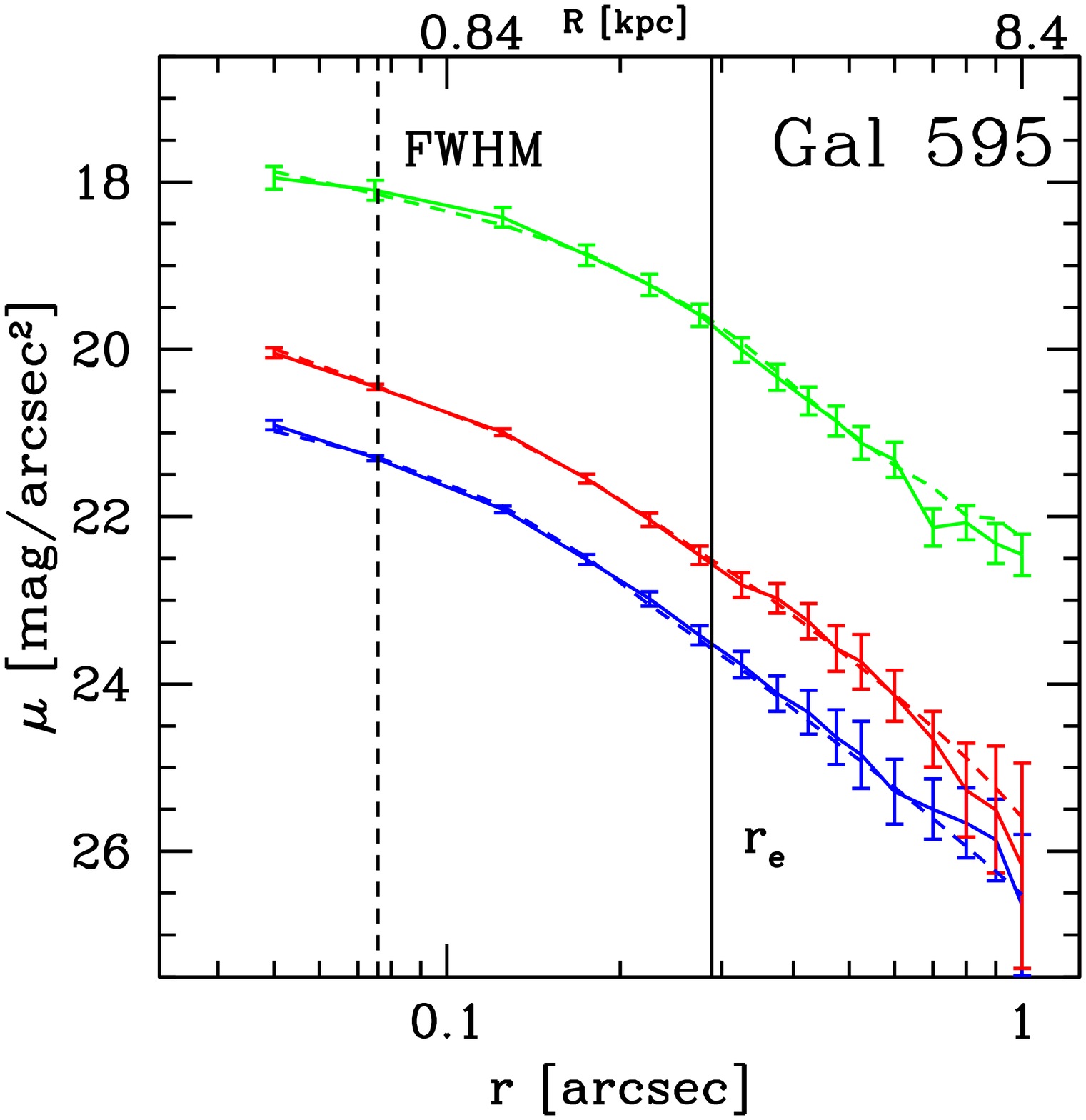}
\includegraphics[width=4.3cm]{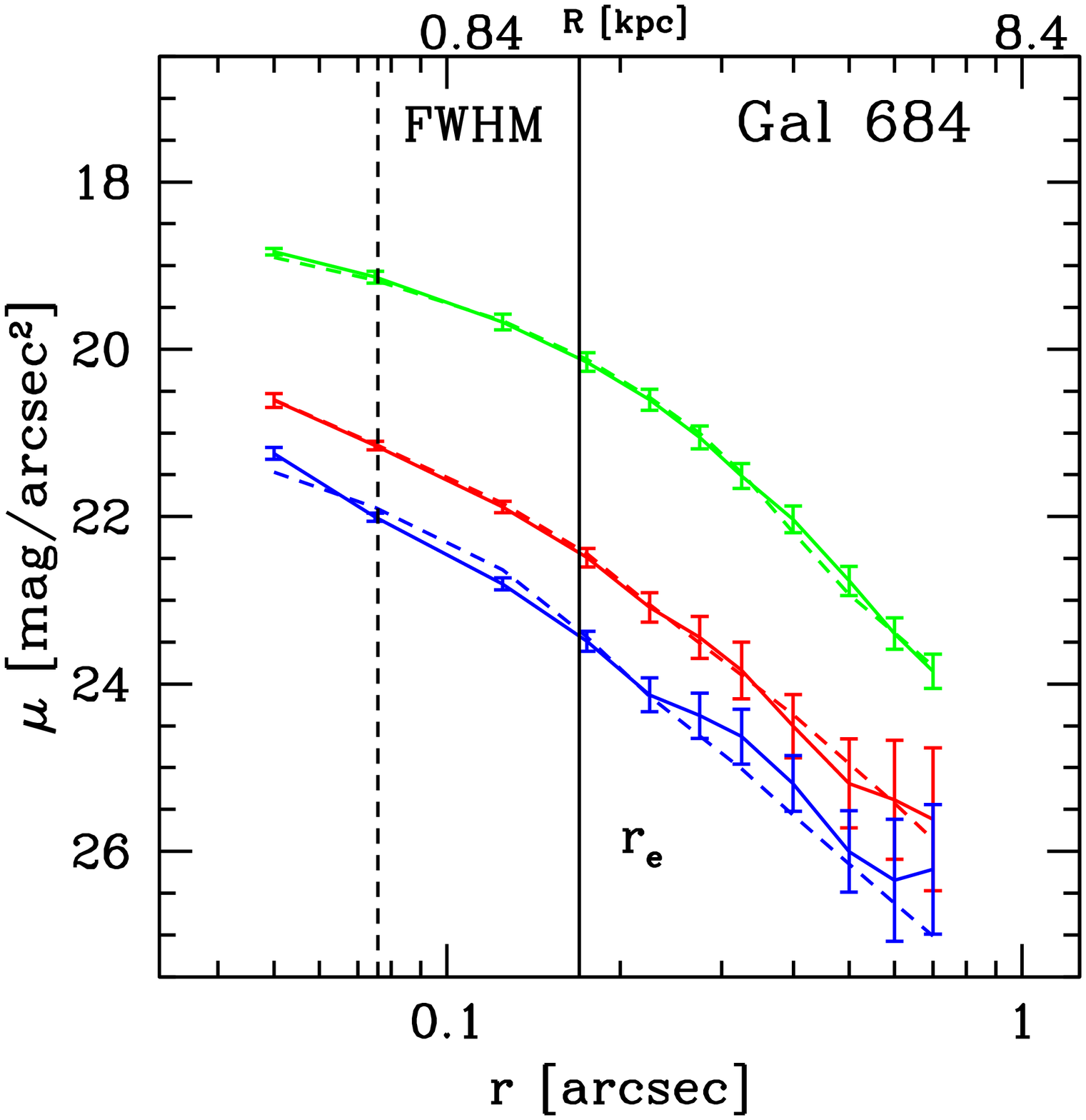}
\includegraphics[width=4.3cm]{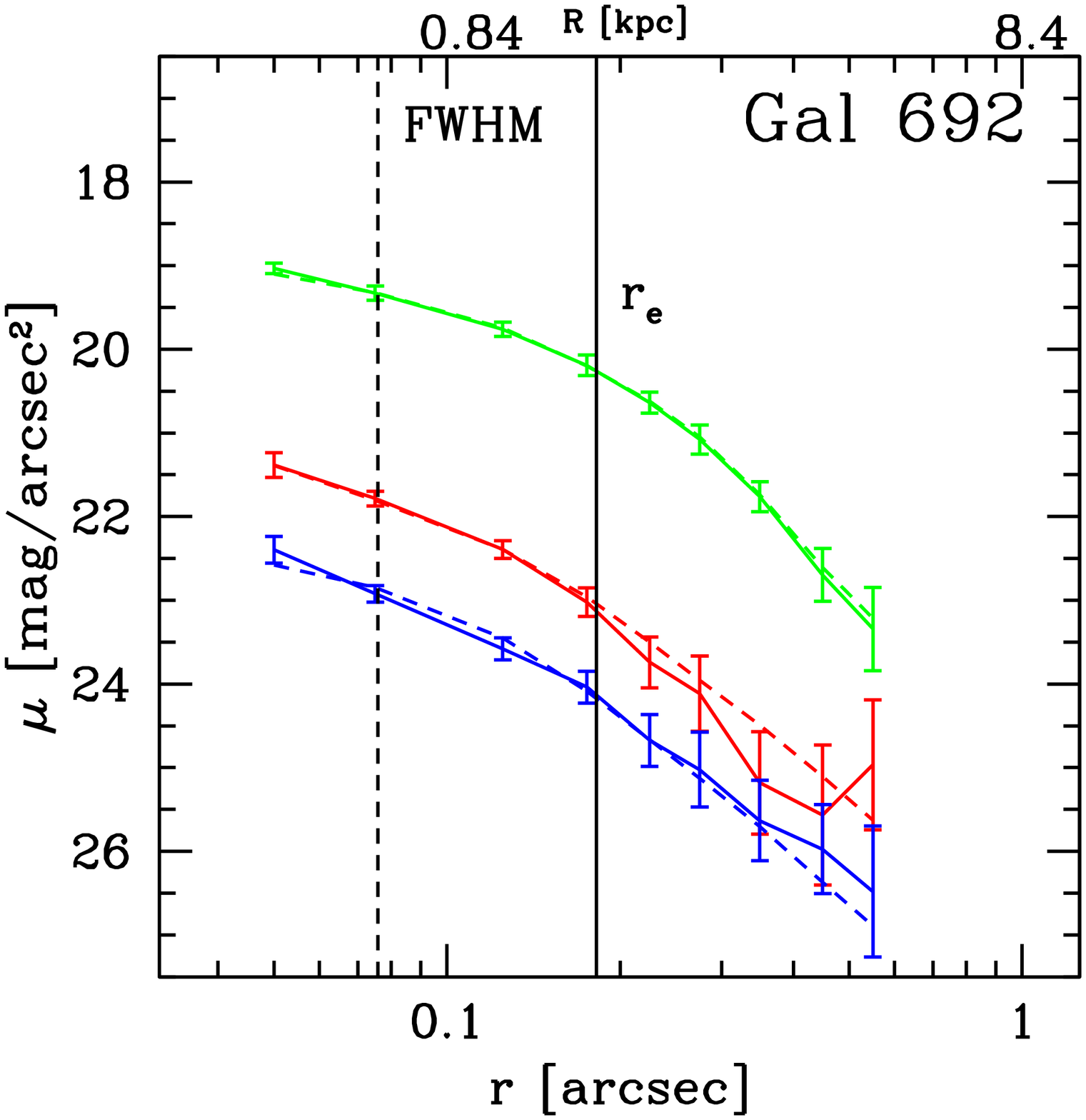}
\includegraphics[width=4.3cm]{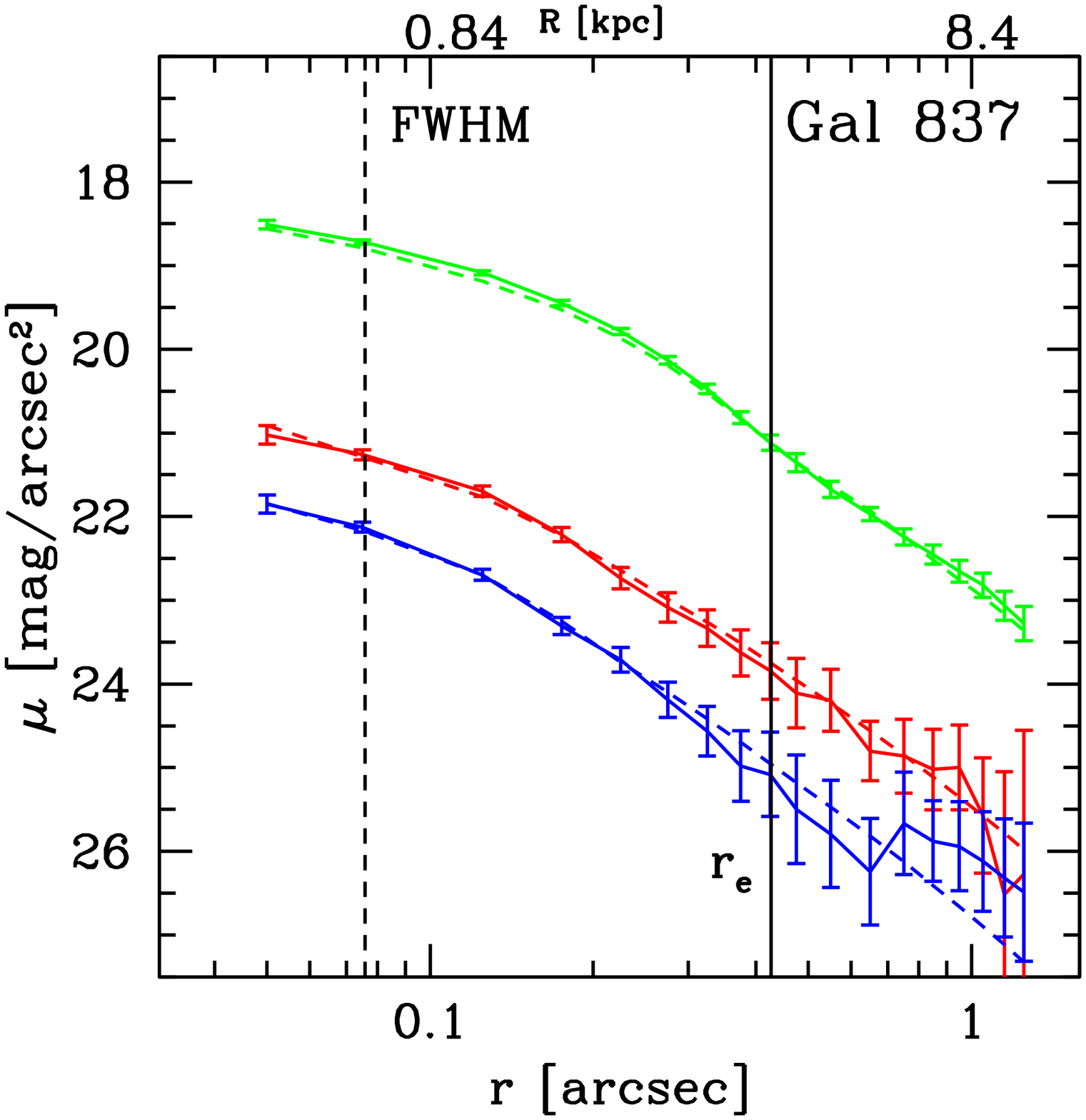}
\includegraphics[width=4.3cm]{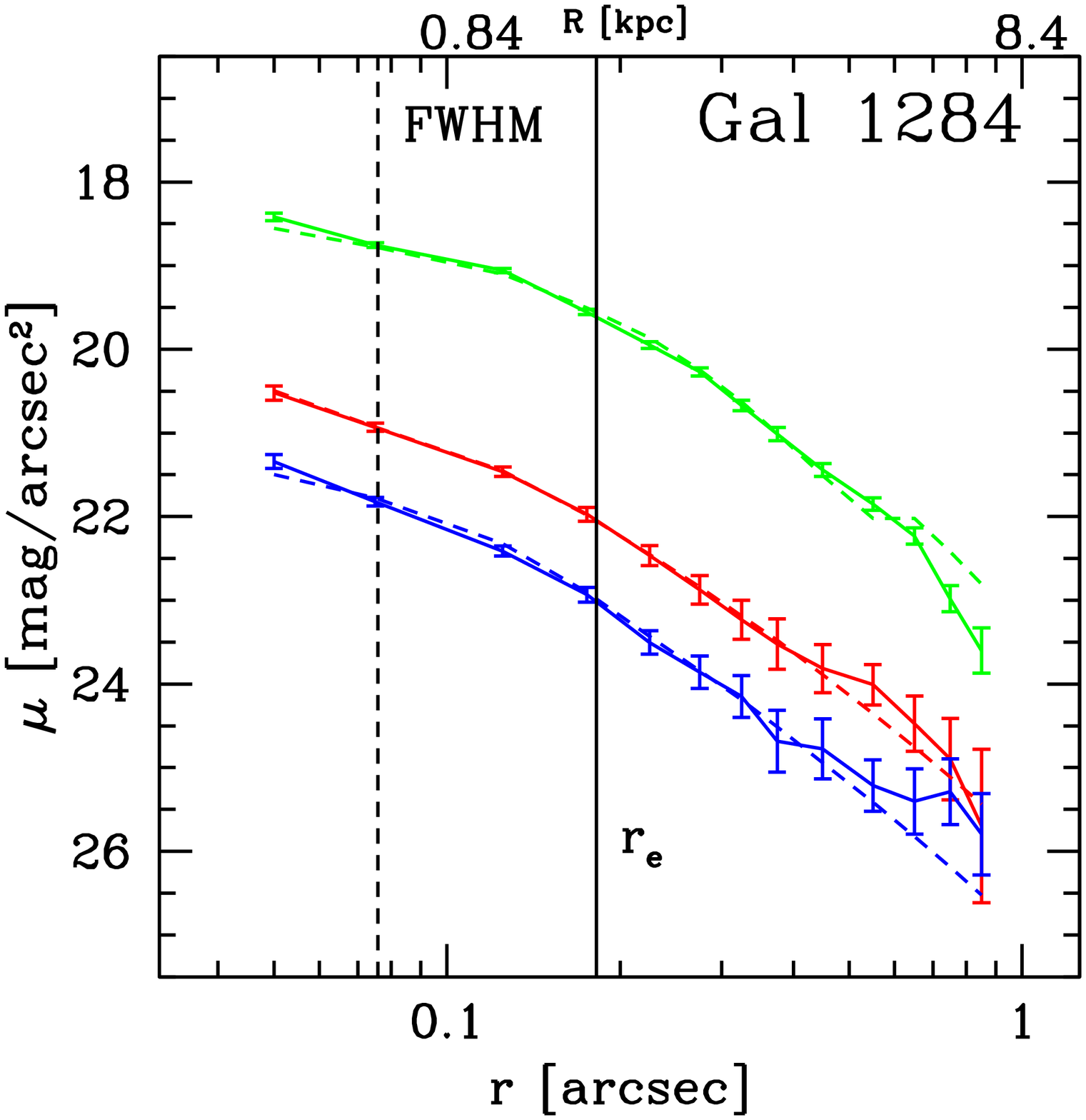}
\includegraphics[width=4.3cm]{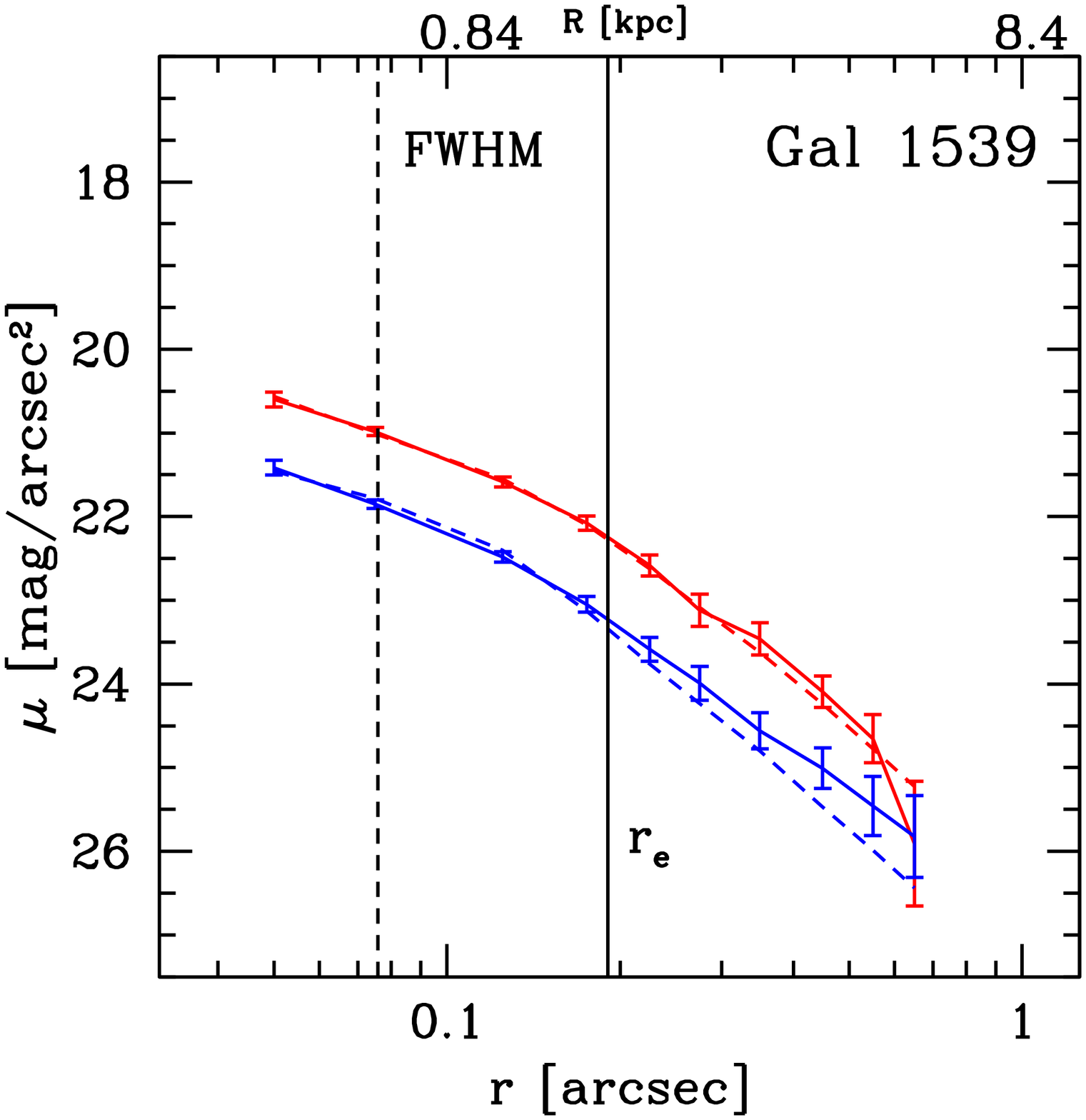}
\includegraphics[width=4.3cm]{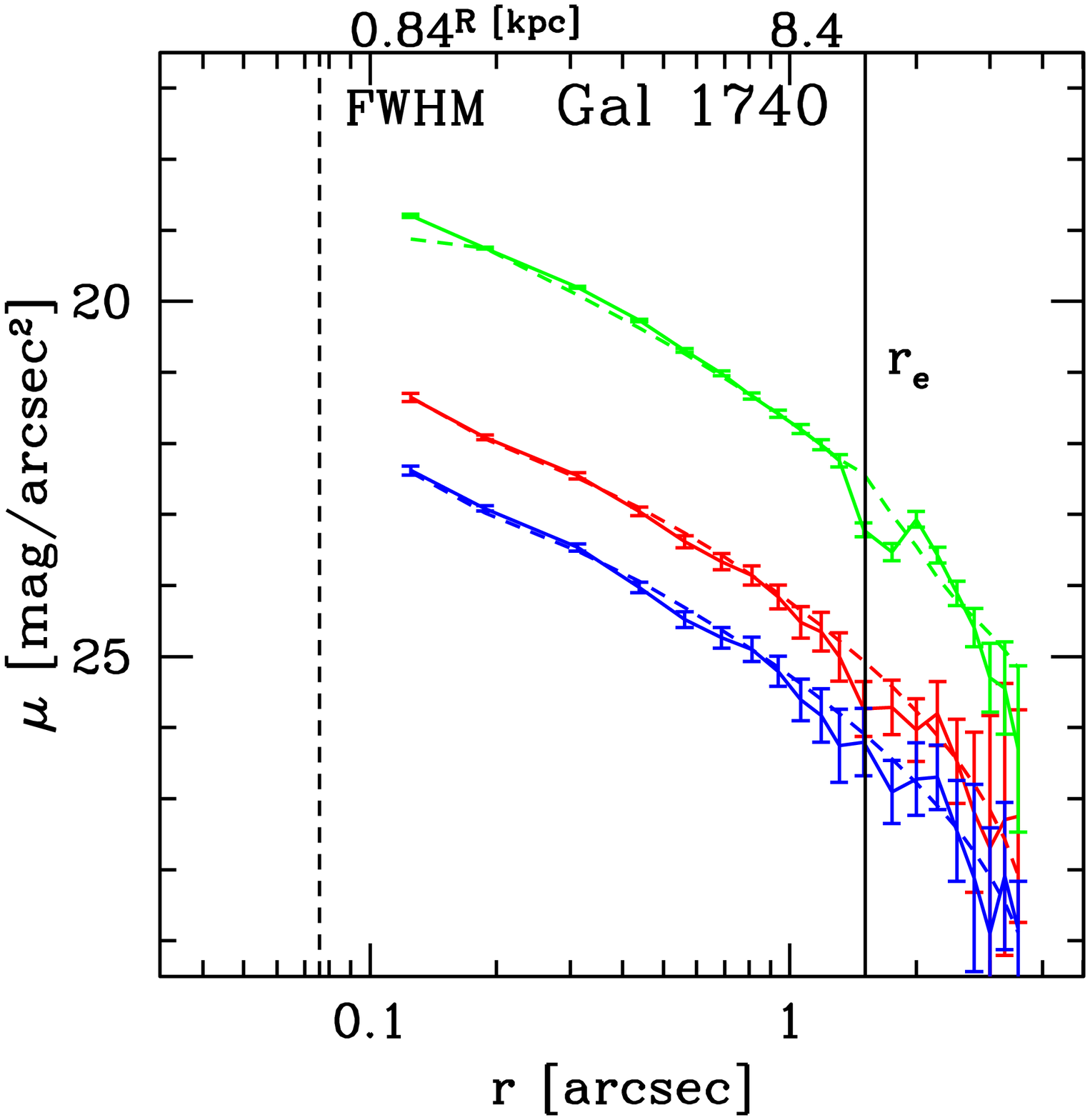}
\includegraphics[width=4.3cm]{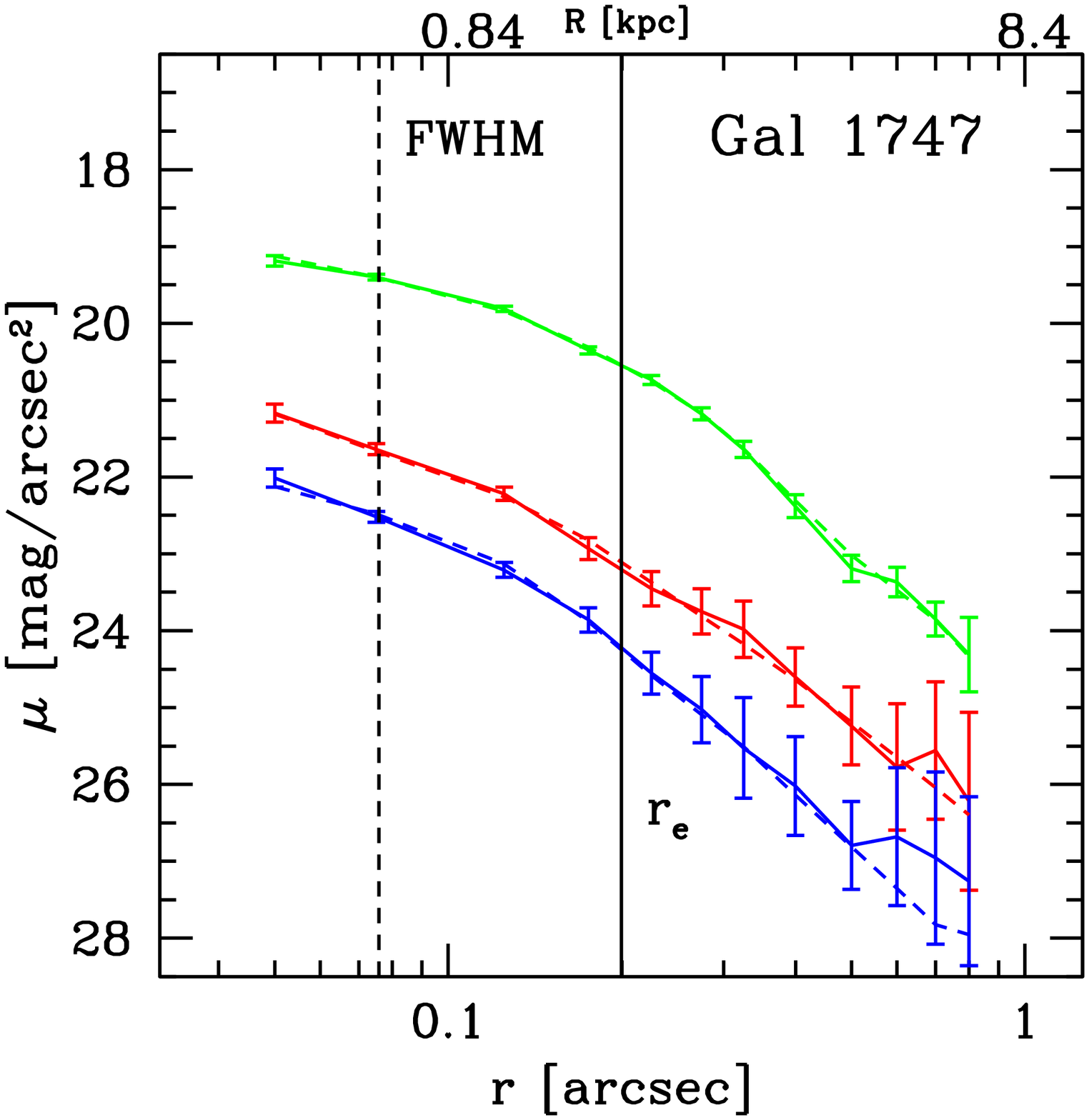}
\includegraphics[width=4.3cm]{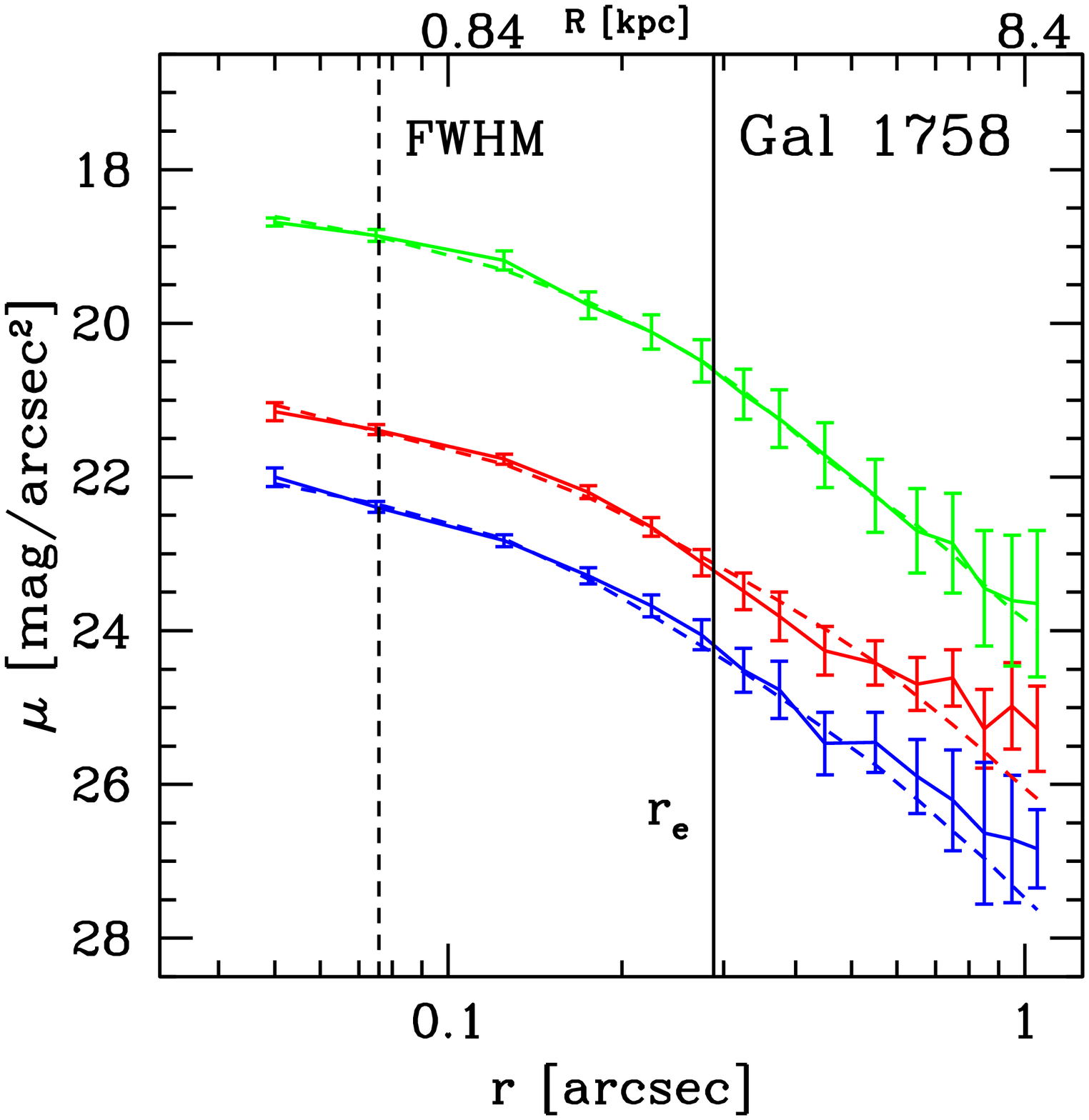}
\includegraphics[width=4.3cm]{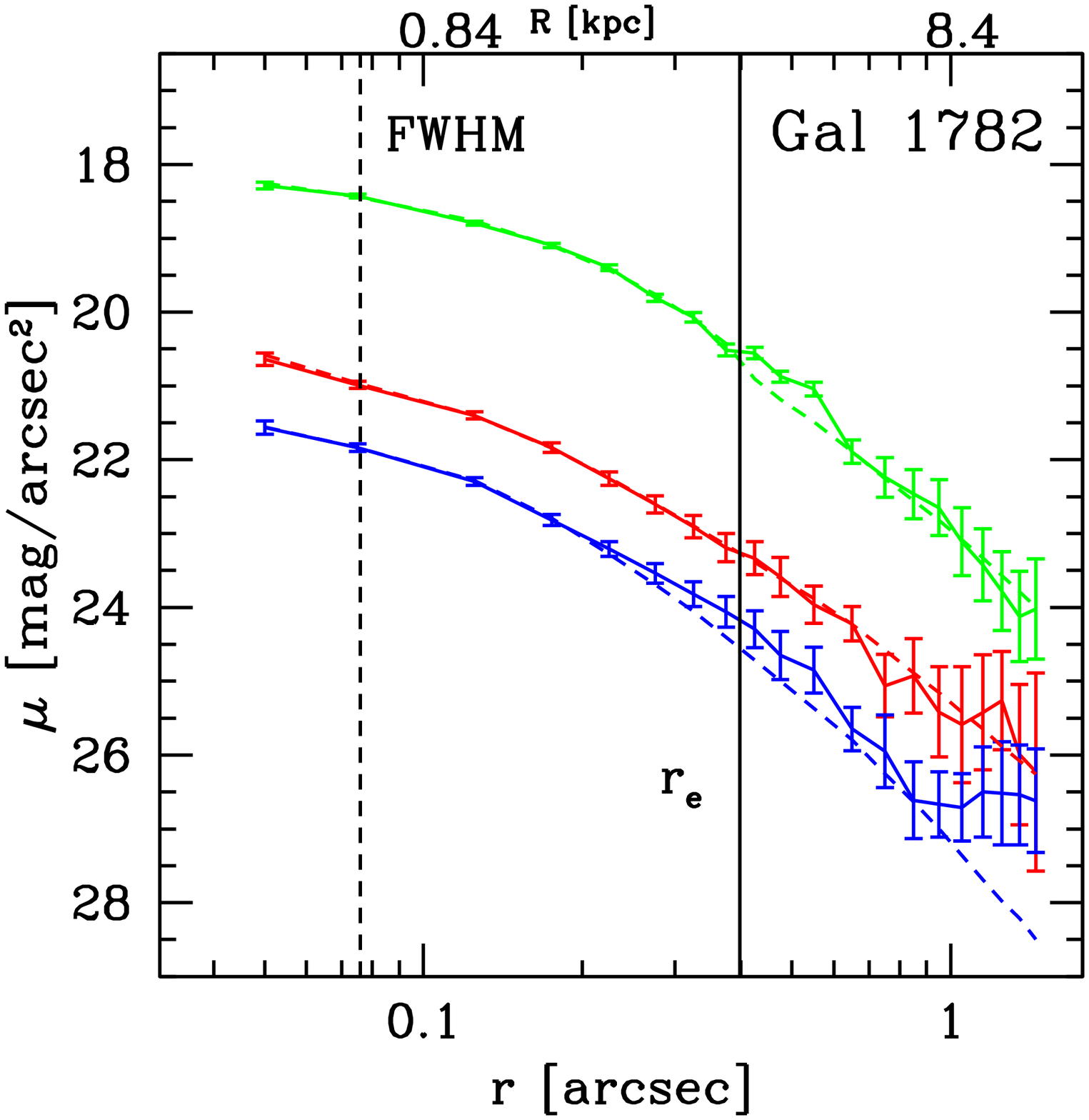}
\includegraphics[width=4.3cm]{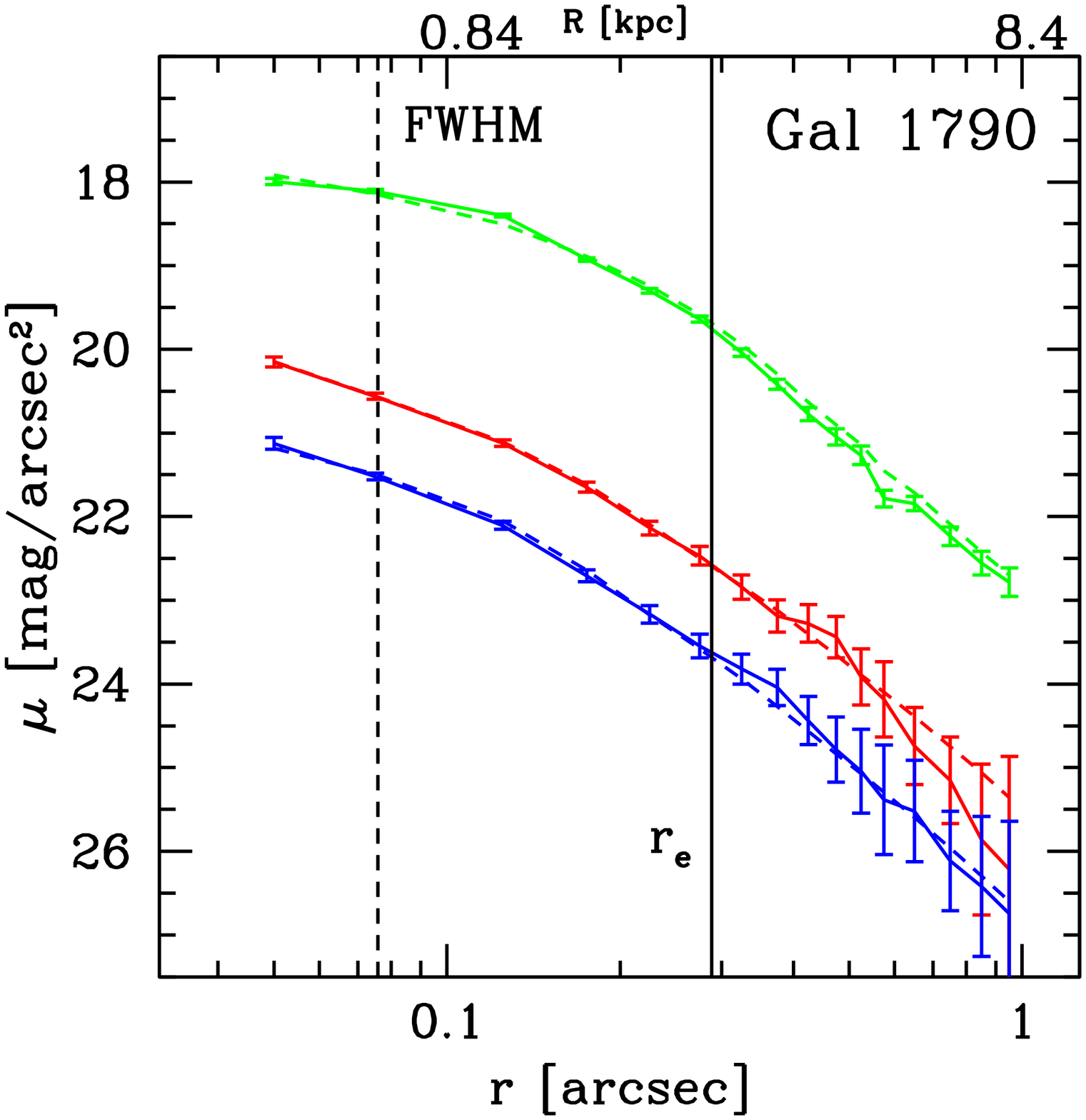}
\includegraphics[width=4.3cm]{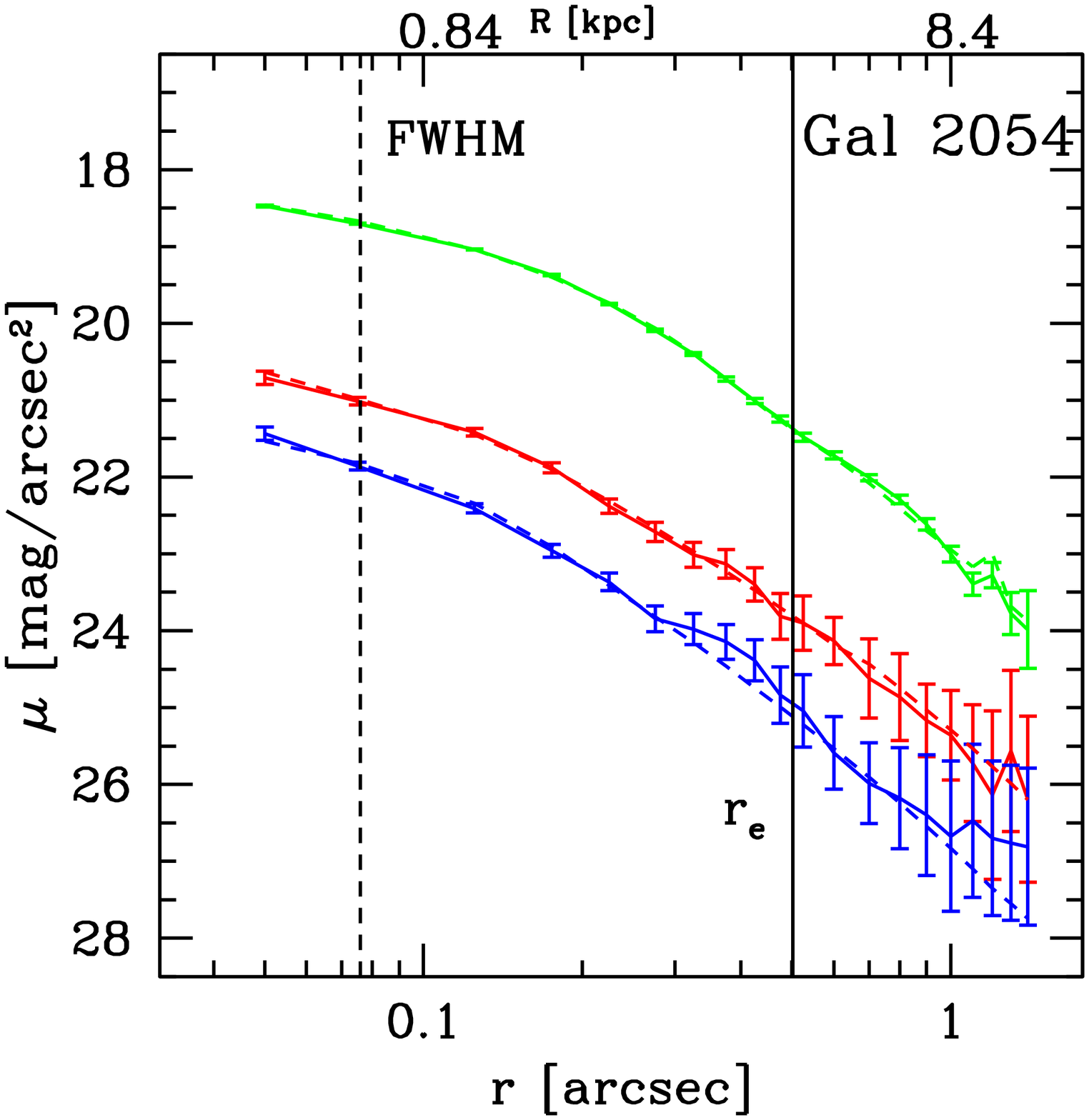}
\includegraphics[width=4.3cm]{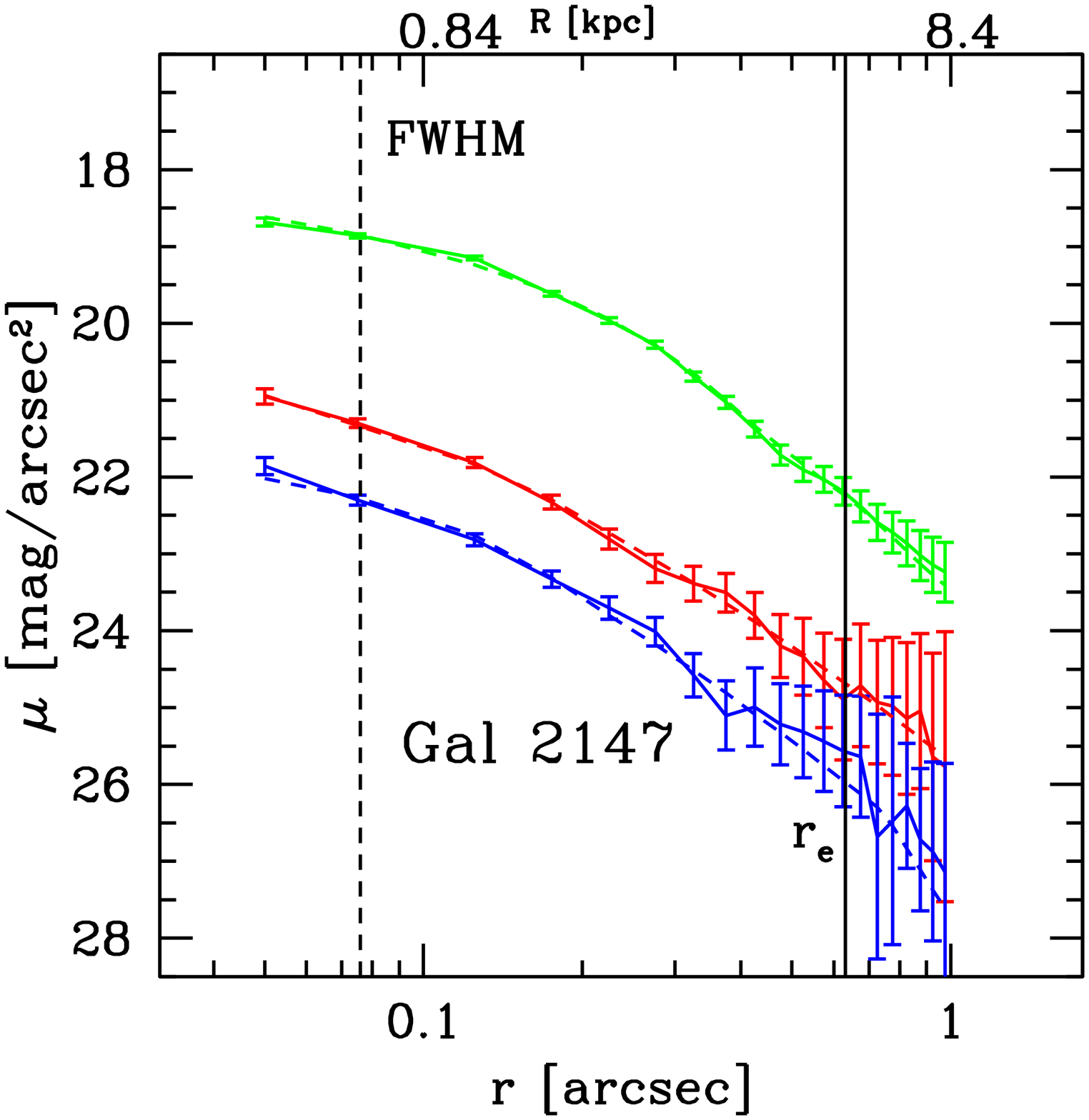}
\includegraphics[width=4.3cm]{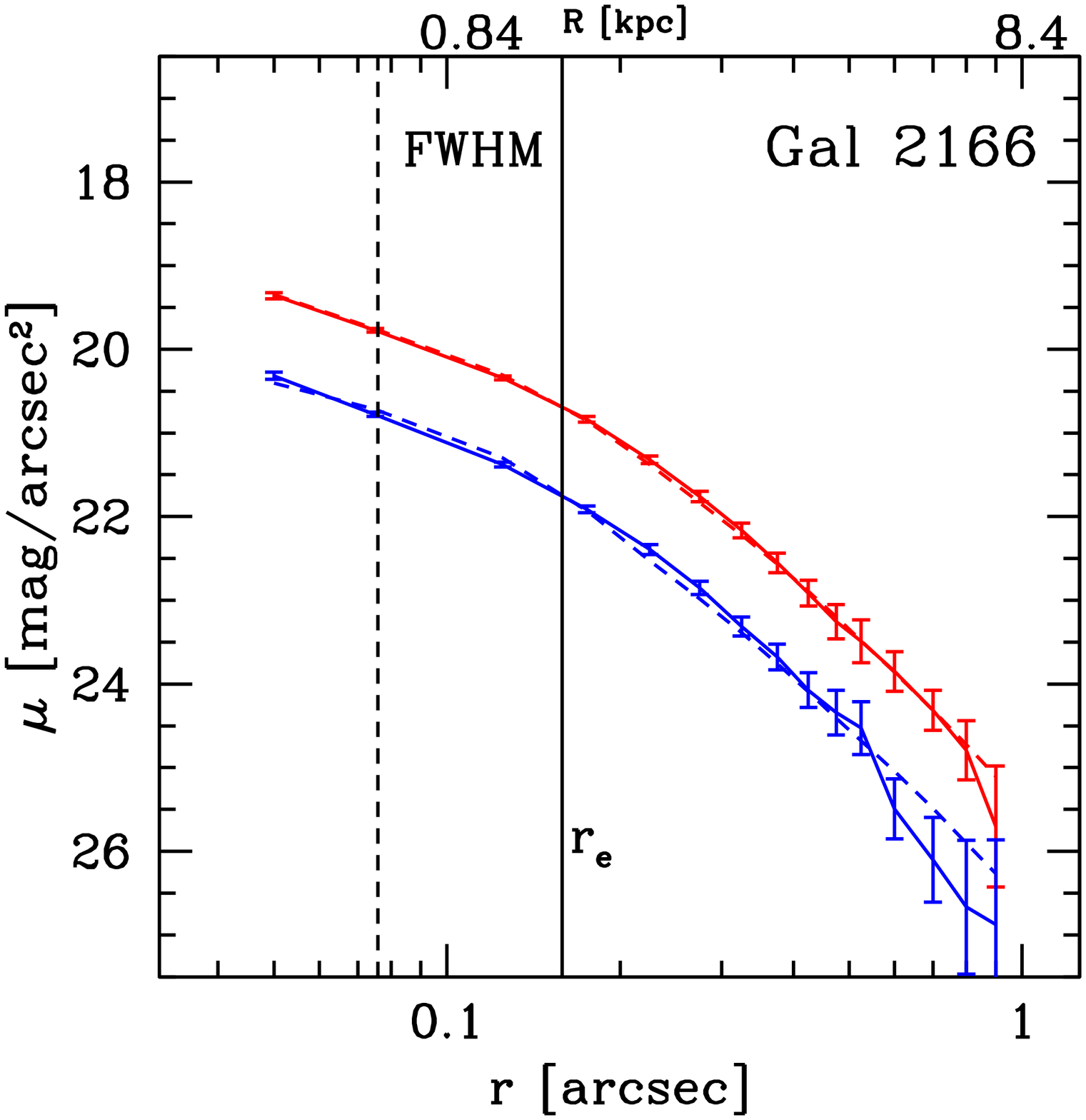}
\includegraphics[width=4.3cm]{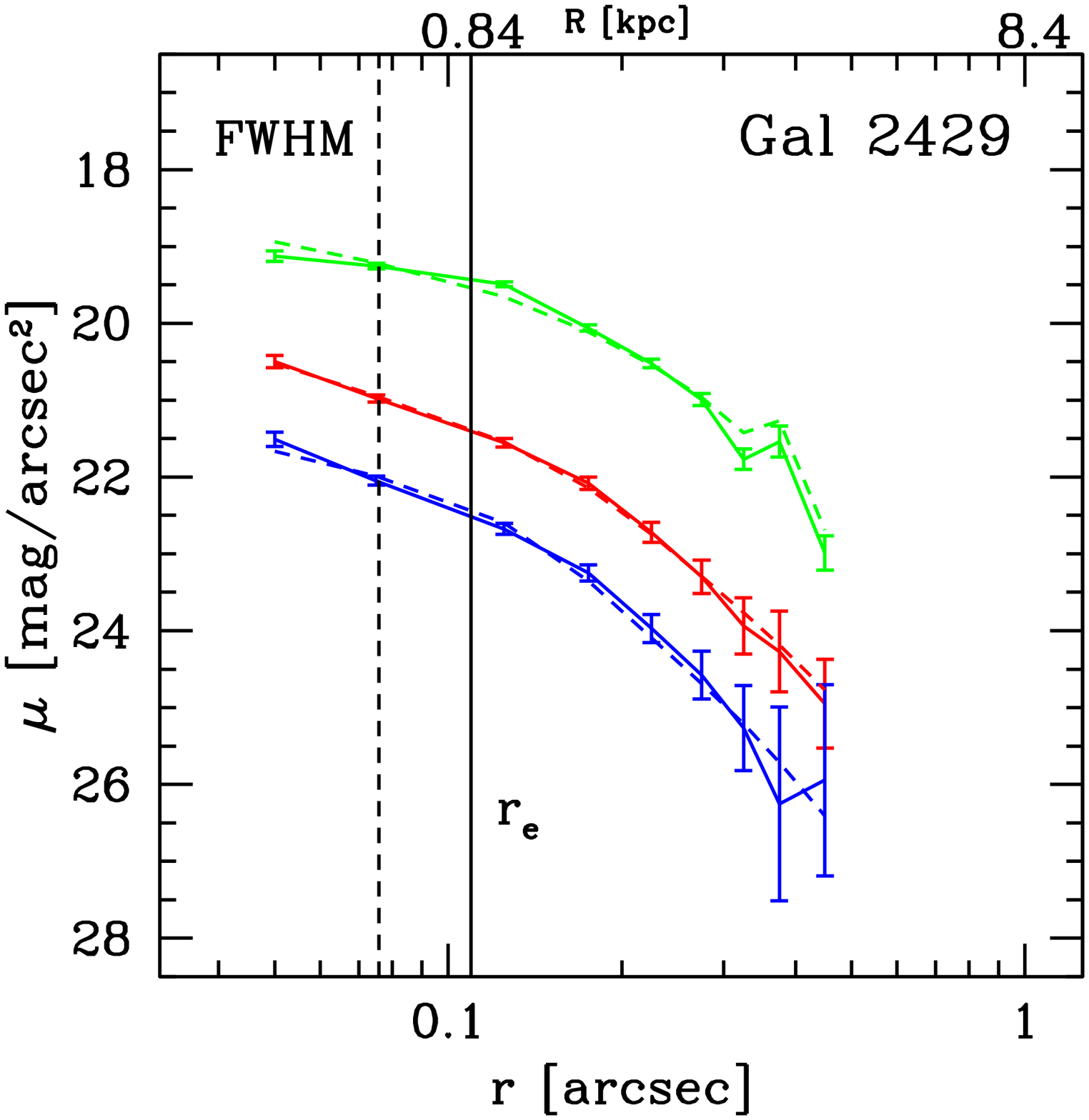}
\includegraphics[width=4.3cm]{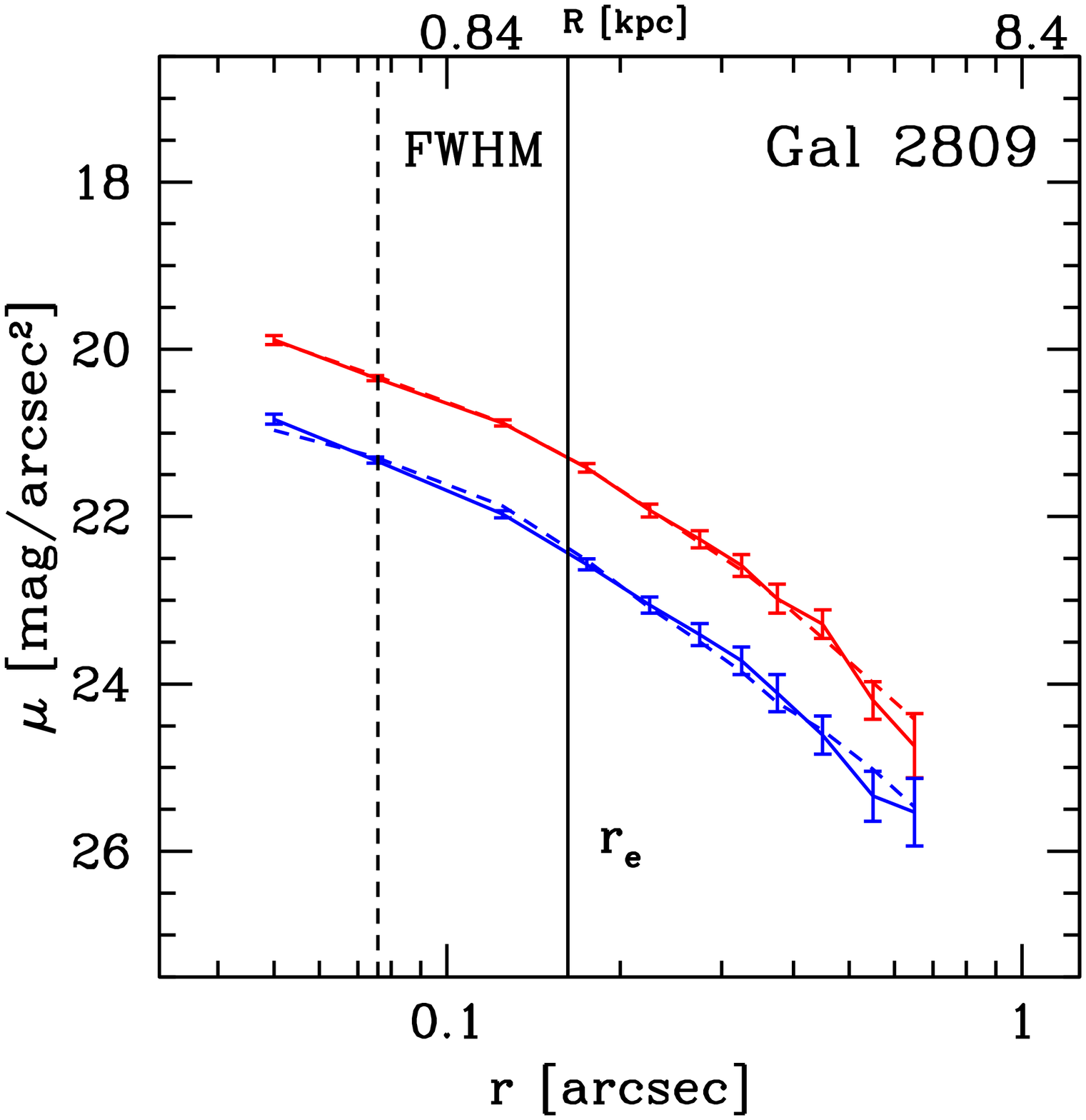}
\vskip -0.2truecm
\caption{Comparison between the observed brightness profiles (solid lines) and the models 
obtained with \textlcsc{Galfit} (dashed lines) in the three HST bands for the 17 ellipticals 
of the sample. The surface brightness profiles in the F775W, F850LP and F160W bands are 
plotted in blue, red and green, respectively. 
The surface brightness has been measured within circular annuli centred on each galaxy. 
The dashed black vertical line represents the radius of the FWHM of the PSF ($\sim 0.06$ arcsec), 
while the solid black vertical line corresponds to the effective radius of the galaxy as 
derived in F850LP band.
}
\label{fig:prof}
\end{center}
\end{figure*}

The fitting was performed over a box of 7.5 $\times$ 7.5 arcsec$^2$  centred on the centroid 
of each galaxy derived by SE\textlcsc{xtractor}. 
To fix the fitting box and the convolution box sizes 
(7.5 $\times$ 7.5 arcsec$^2$), we repeatedly fitted the surface brightness profiles by 
increasing the boxes size until the best-fitting parameters values converged. 
We masked sources distant from our targets using the segmentation image generated 
by SE\textlcsc{xtractor}, while we fitted simultaneously the sources close to the target galaxies 
in order to avoid the overlapping of the light profiles. For few galaxies 
(two in the F775W and in the F850LP images, two in the F160W band and one in all 
the three images), the fitting procedure was not able to converge. 
Hence, we re-performed the fit in that band several times for different fixed values 
of $n$ and for each of them we obtained a set of structural parameters as output. 
We chose as best solution the set of structural parameters for which the $\chi^2$ 
was the lowest and the residual map was lacking of residual structures. 
The structural parameters obtained through the fit are shown in Table \ref{tab:tab_gal}.
The goodness of the fits is shown in Fig. \ref{fig:prof}, where the observed surface 
brightness profiles of the galaxies are compared with the best-fitting models generated
 by \textlcsc{Galfit}. 
The surface brightness has been measured within circular annuli 
 centred on each galaxy and it is plotted in blue (F775W), red (F850LP) and green 
(F160W). 
The error bars are obtained by 
propagating the errors on fluxes measured within the concentric annuli. 
In all the cases, a range of at least five magnitudes in surface brightness has been 
covered in the profiles of the galaxies which extend up to at least $\sim$2$r_e$
for the largest galaxies. 
For the galaxies with a small $r_e$ ($\sim 0.1 - 0.2$ arcsec) the observed brightness 
profiles extend up to $\sim$ 3 - 4 $r_e$. 

\subsection{Simulations}
\label{simulations}

\begin{figure*}
\begin{center}
\includegraphics[width=7cm]{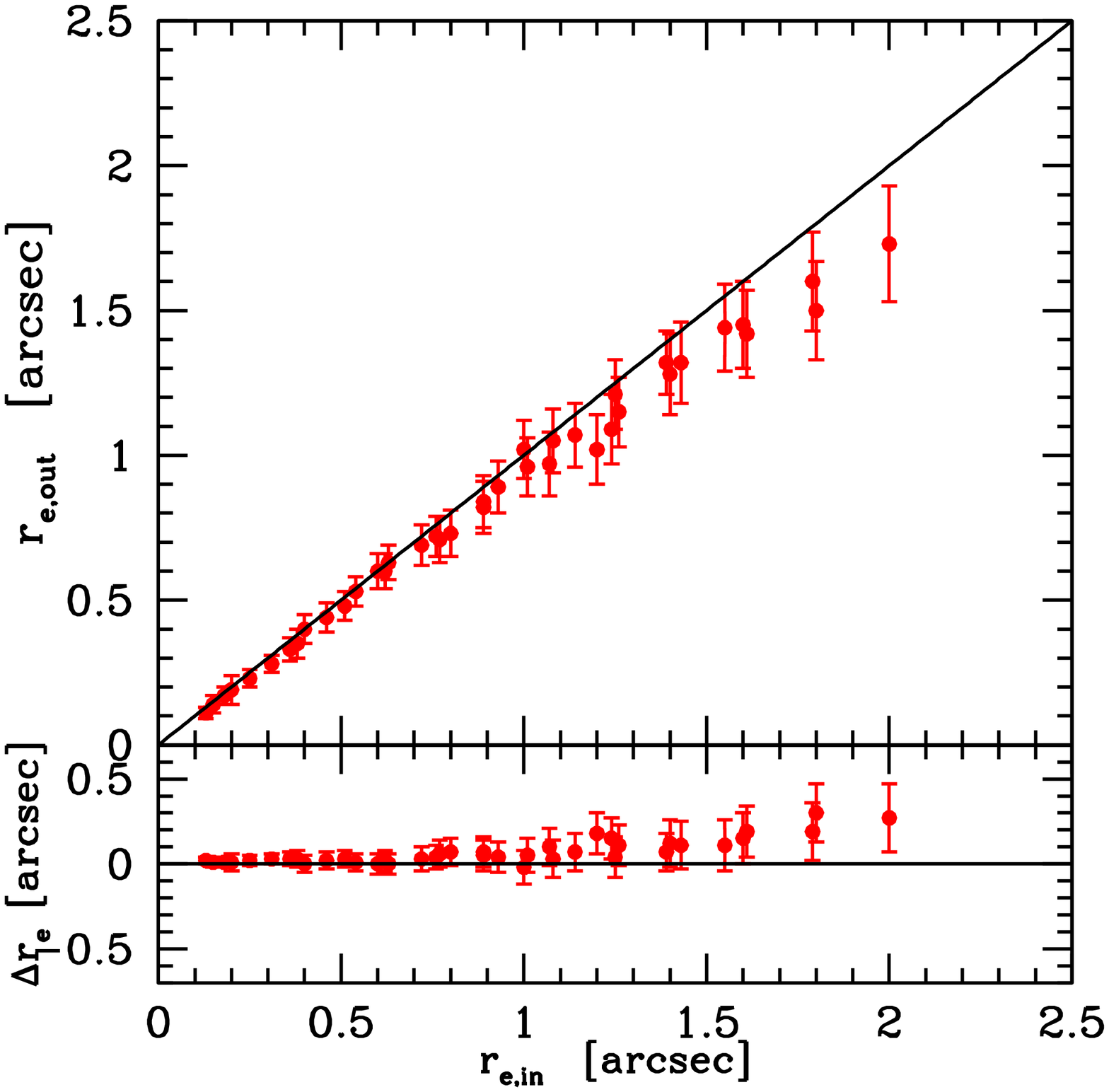}
\includegraphics[width=7cm]{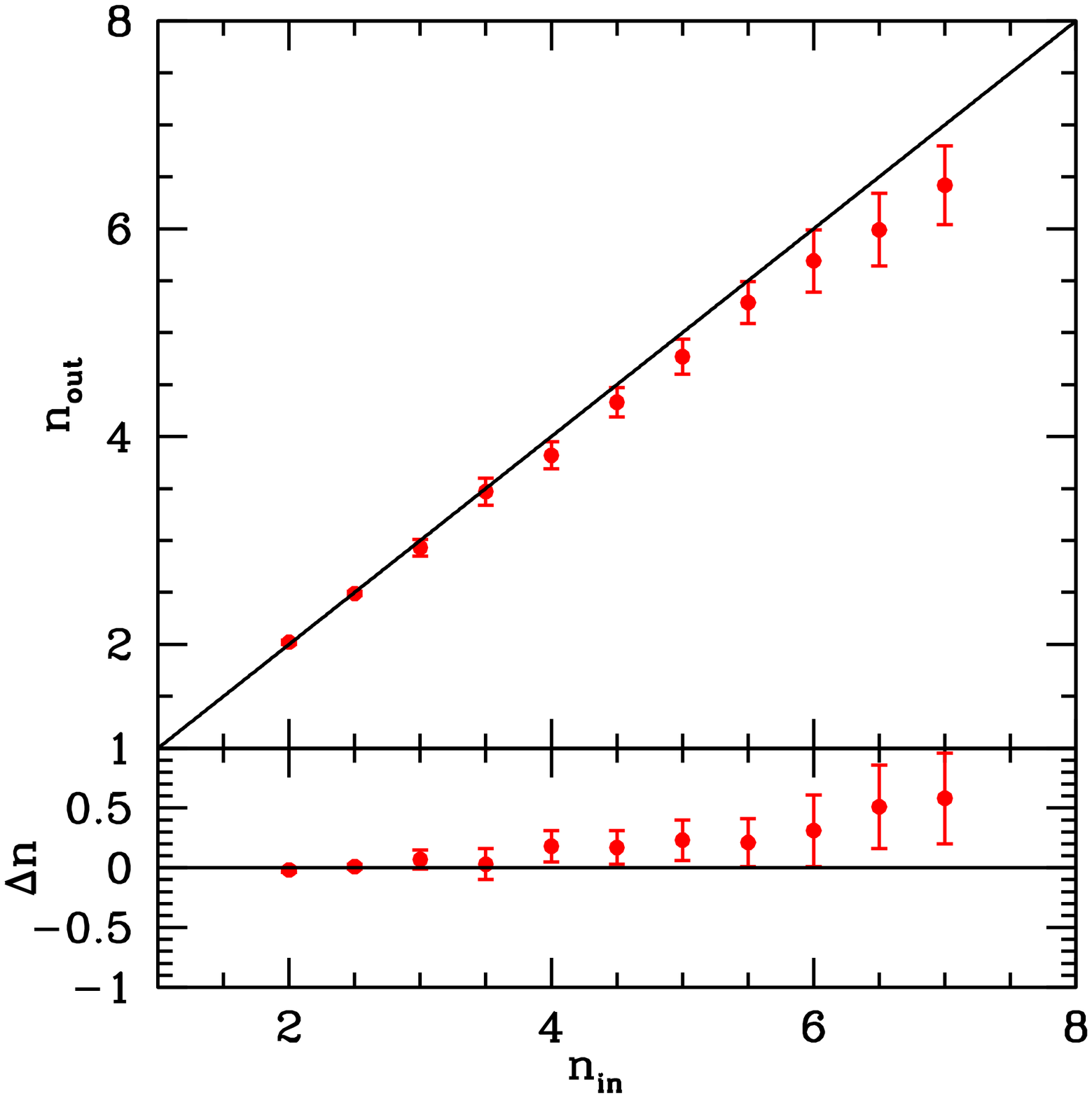}
\vskip -0.3truecm
\caption{Comparison between the effective radius ($r_{e,in}$) and the index of 
concentration ($n_{in}$) of the simulated galaxies and the effective radius ($r_{e,out}$) 
and the index of concentration ($n_{out}$) obtained through the fitting with the S\'{e}rsic 
profile (upper left-hand and right-hand panel, respectively). 
The differences $\Delta r_e = r_{e,in}-r_{e.out}$ and 
$\Delta n = n_{in}-n_{out}$ are plotted as a function of the input parameters 
(lower left-hand and right-hand panel, respectively).
}
\label{fig:sim}
\end{center}
\end{figure*}

In order to test for the absence of possible systematics introduced by the fitting 
algorithm and to derive reliable statistical errors for the structural parameters, 
we performed the same fitting procedure used for real galaxies to a set of simulated 
galaxies. 
In particular, to test for the absence of systematics, we generated with \textlcsc{Galfit} 
a set of $\sim 8000$ galaxies described by a S\'ersic profile with effective 
radius $r_e$, index of concentration $n$ and magnitude  $m$ assigned in the ranges 
$0.1 < r_e < 2$ arcsec, $2 < n < 7$ and $20 < m < 24.5$ mag. 
The simulated galaxies have been 
convolved with the ACS-F850LP PSF and embedded in the real background, constructed by 
cutting different portions devoid of sources from the real image. For each structural 
parameters we verified how the output values vary as a function of the input values, 
keeping fixed the other two structural parameters. In Fig. \ref{fig:sim} we show the 
output values $r_{e,out}$ and $n_{out}$ versus the input values $r_{e,in}$ and $n_{in}$ 
(left panel and right panel, respectively). 
For $n < 6$ and $r_e < 1$ arcsec \textlcsc{Galfit} recovers exactly the input values. 
For $n > 6$ and $r_e > 1$ arcsec the output $r_e$ and $n$ are, on average,  underestimated 
of ~8 per cent and ~15 per cent, respectively. 
These underestimates are negligible compared to the statistical errors and they are 
possibly due to the fact that, for galaxies having larger surface brightness profile 
tails, we miss a fraction of the profile since it is dominated by the background. 
This would drop the brightness profile and favour lower values of $n$ in the fitting.
Since the galaxies of our sample have $r_e < 1$ arcsec and $n < 6$, we did not apply any 
correction to the values of the structural parameters obtained by the fitting. 
  
We also verified for each structural parameter whether a dependence of the output 
values on the input values of the other structural parameters was present 
(e.g., output $r_e$ versus input $n$ and $m$) and no systematics were found.

To derive the statistical uncertainties of each parameter, that is how the background 
RMS affects the estimates of the structural parameters, we have considered a subsample 
of the simulated galaxies and we embedded each of them in 10 different real backgrounds. 
Then, for each galaxy, we have run \textlcsc{Galfit} to derive the best-fitting parameters 
for the 10 different backgrounds. 
The standard deviation of the mean has been adopted as indicator of the statistical 
fluctuations. 

\section{Colour Gradients}
\label{colour gradients}

\begin{figure*}
\begin{center}
\includegraphics[width=3.3cm]{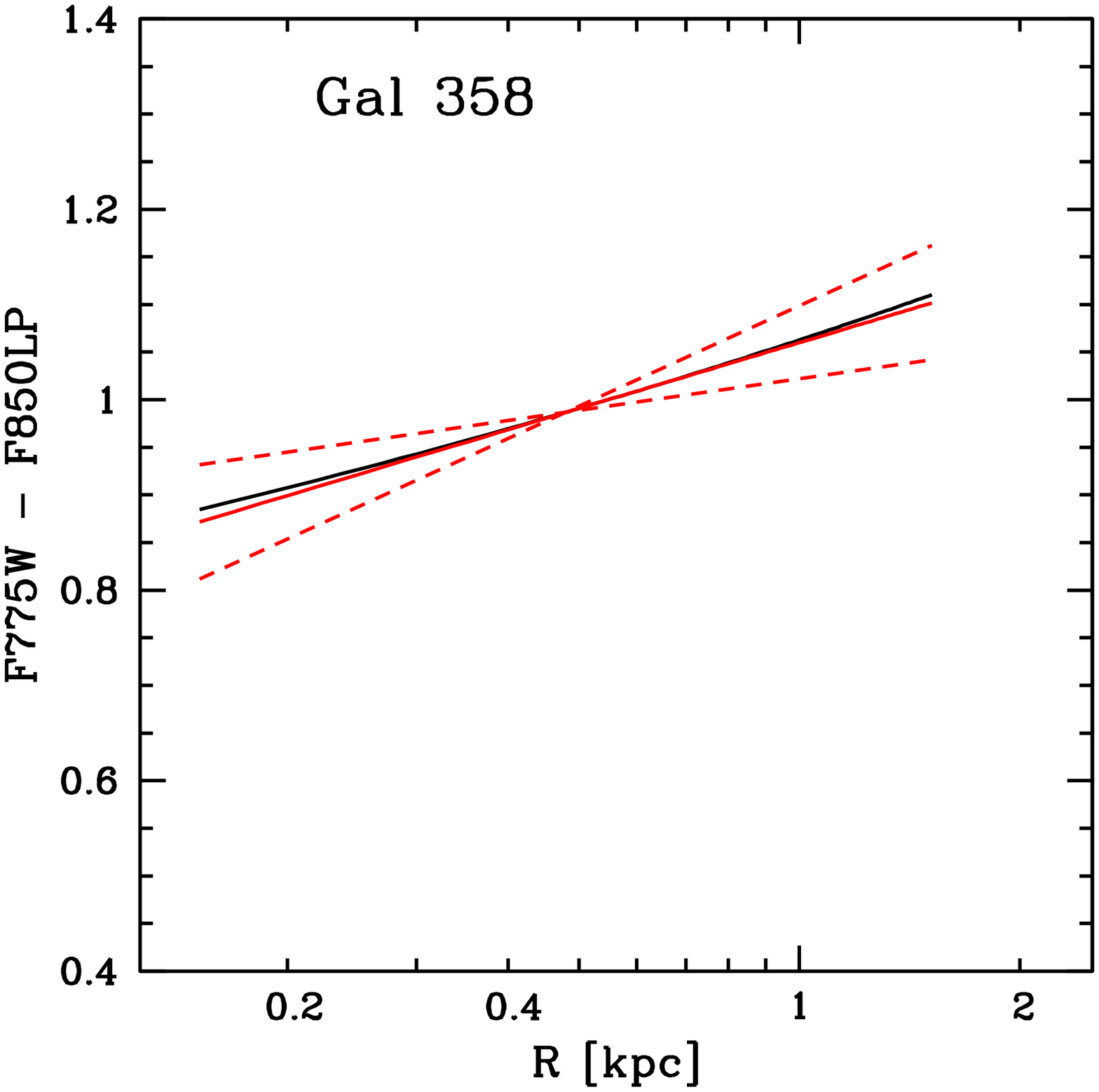}
\includegraphics[width=3.3cm]{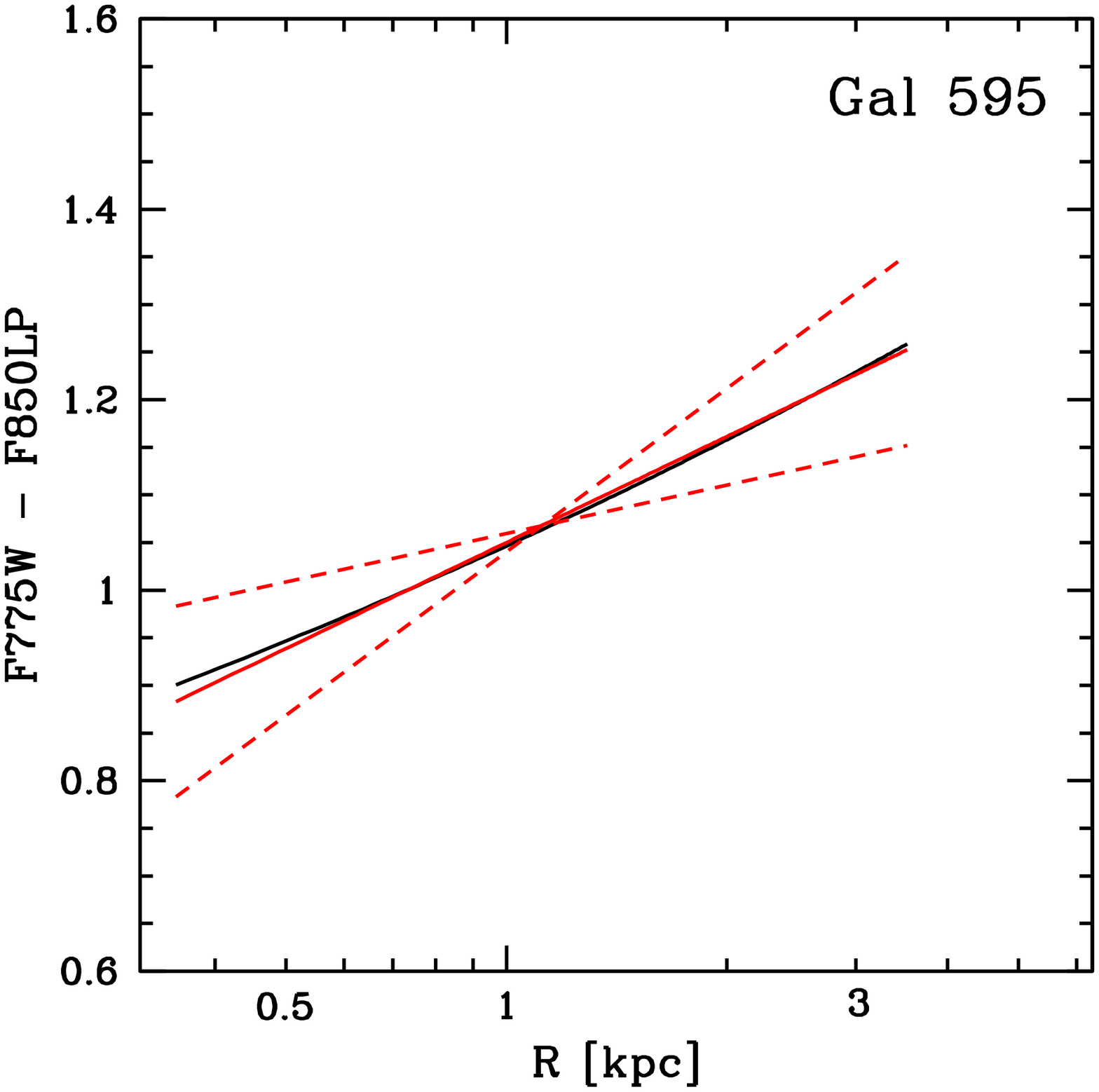}
\includegraphics[width=3.3cm]{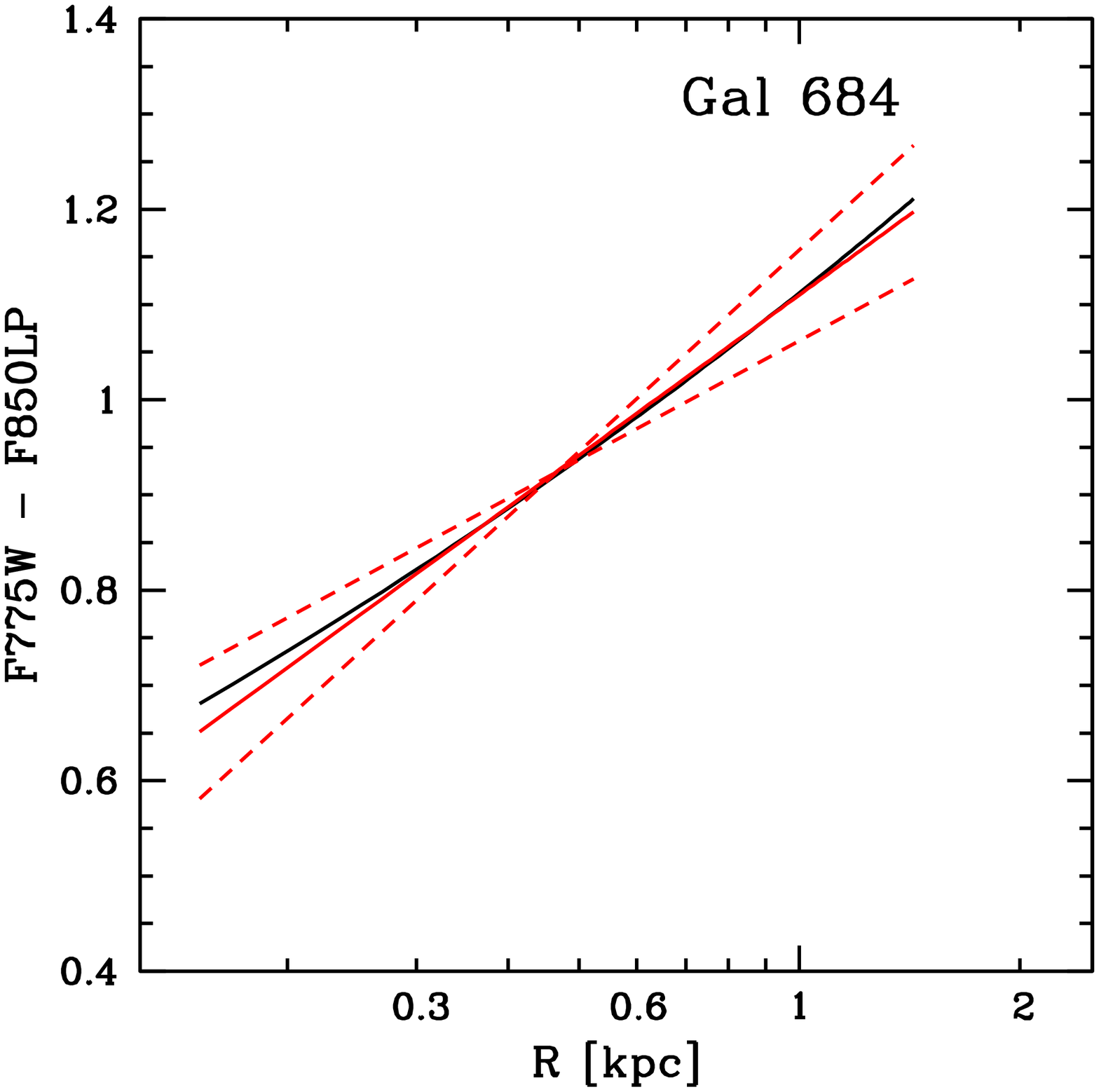}
\includegraphics[width=3.3cm]{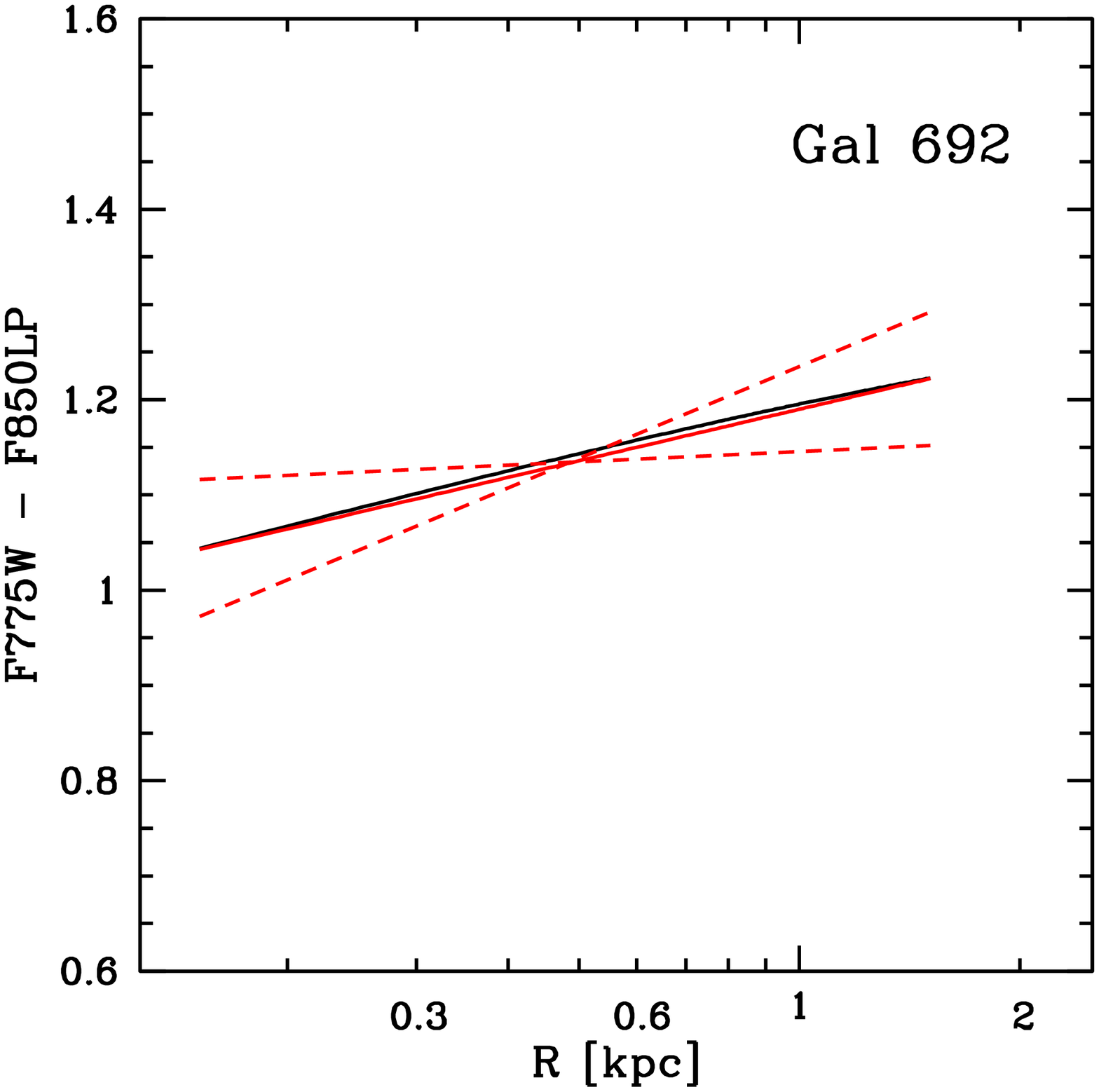}
\includegraphics[width=3.3cm]{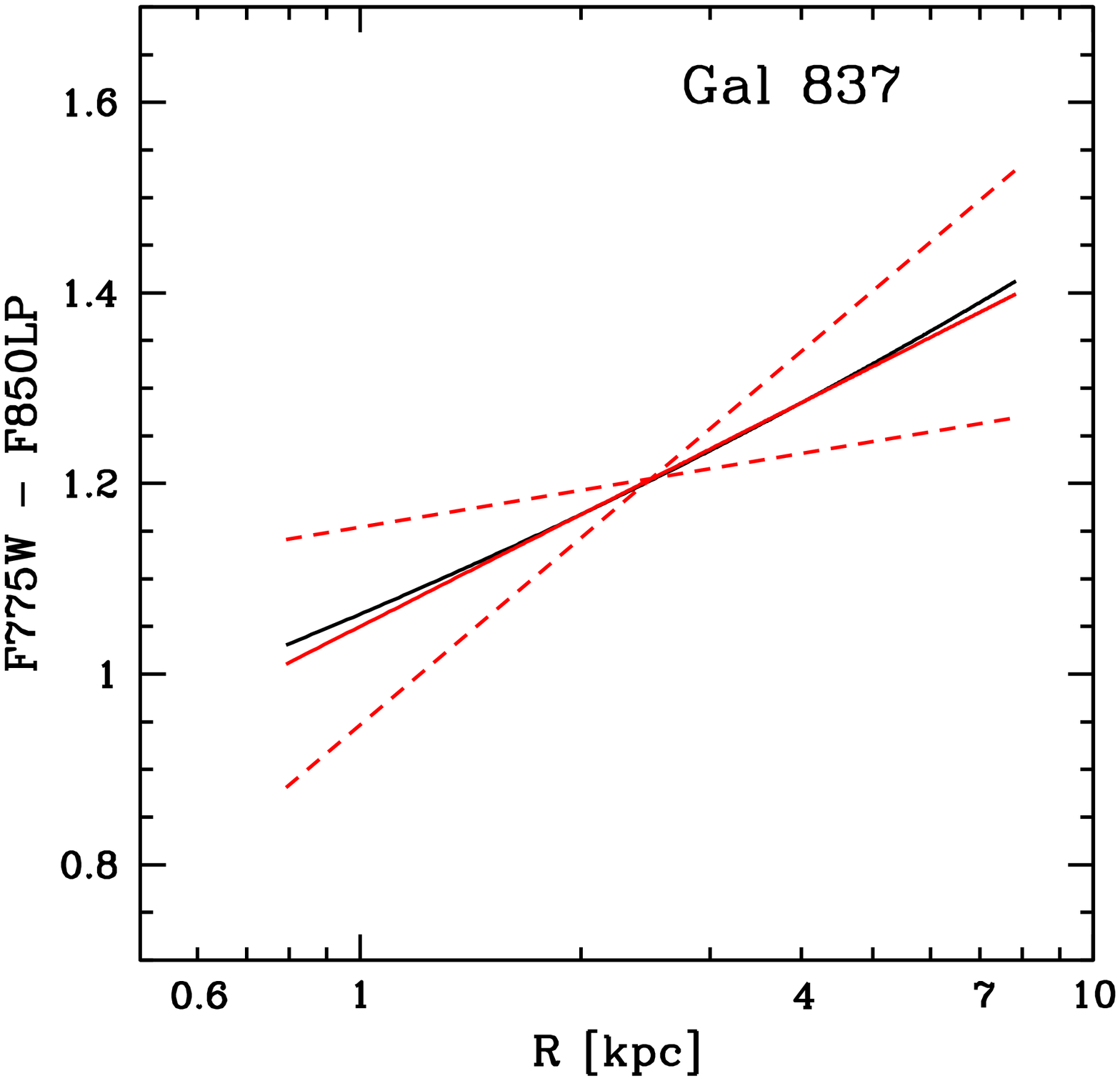}
\includegraphics[width=3.3cm]{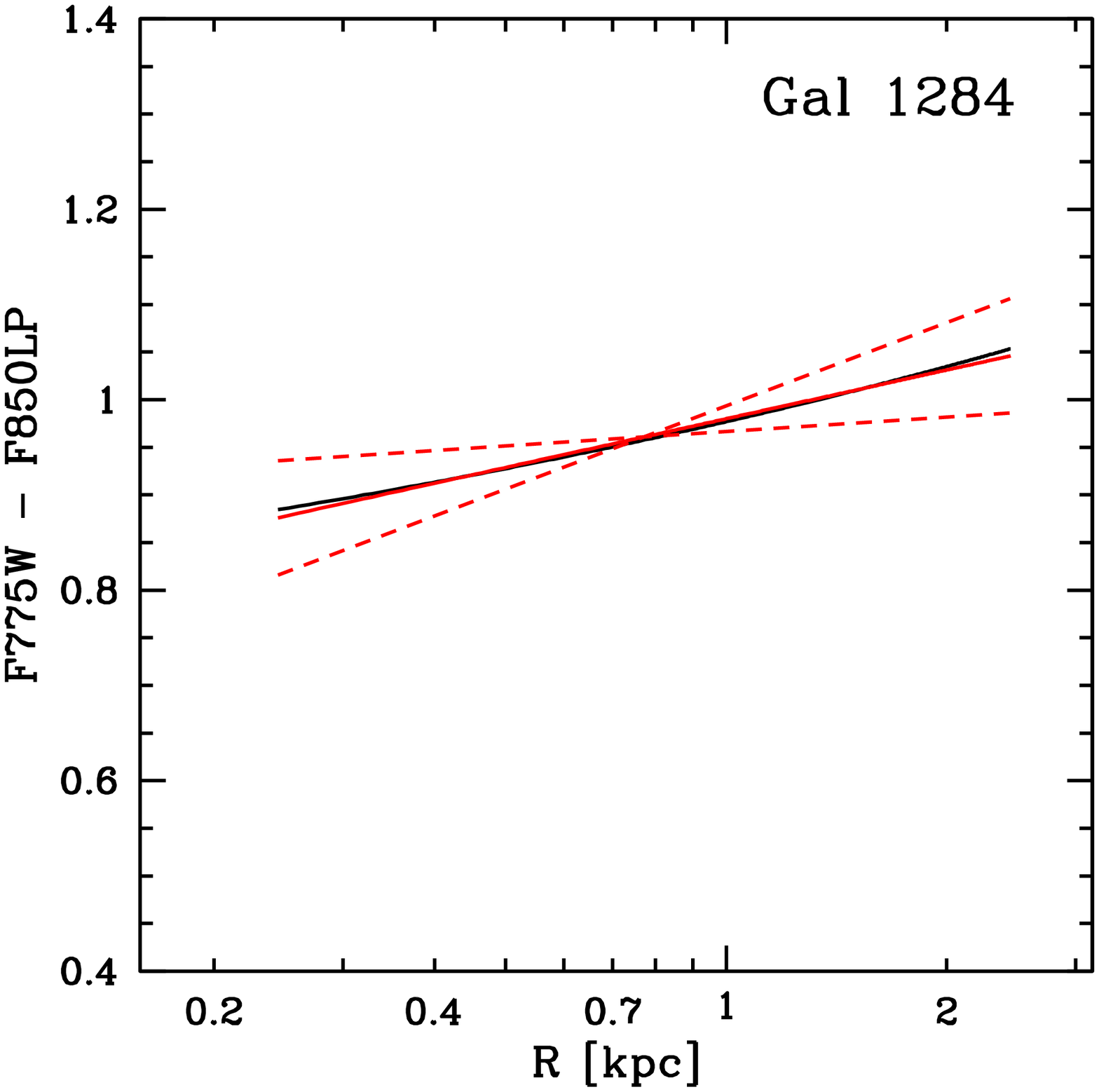}
\includegraphics[width=3.3cm]{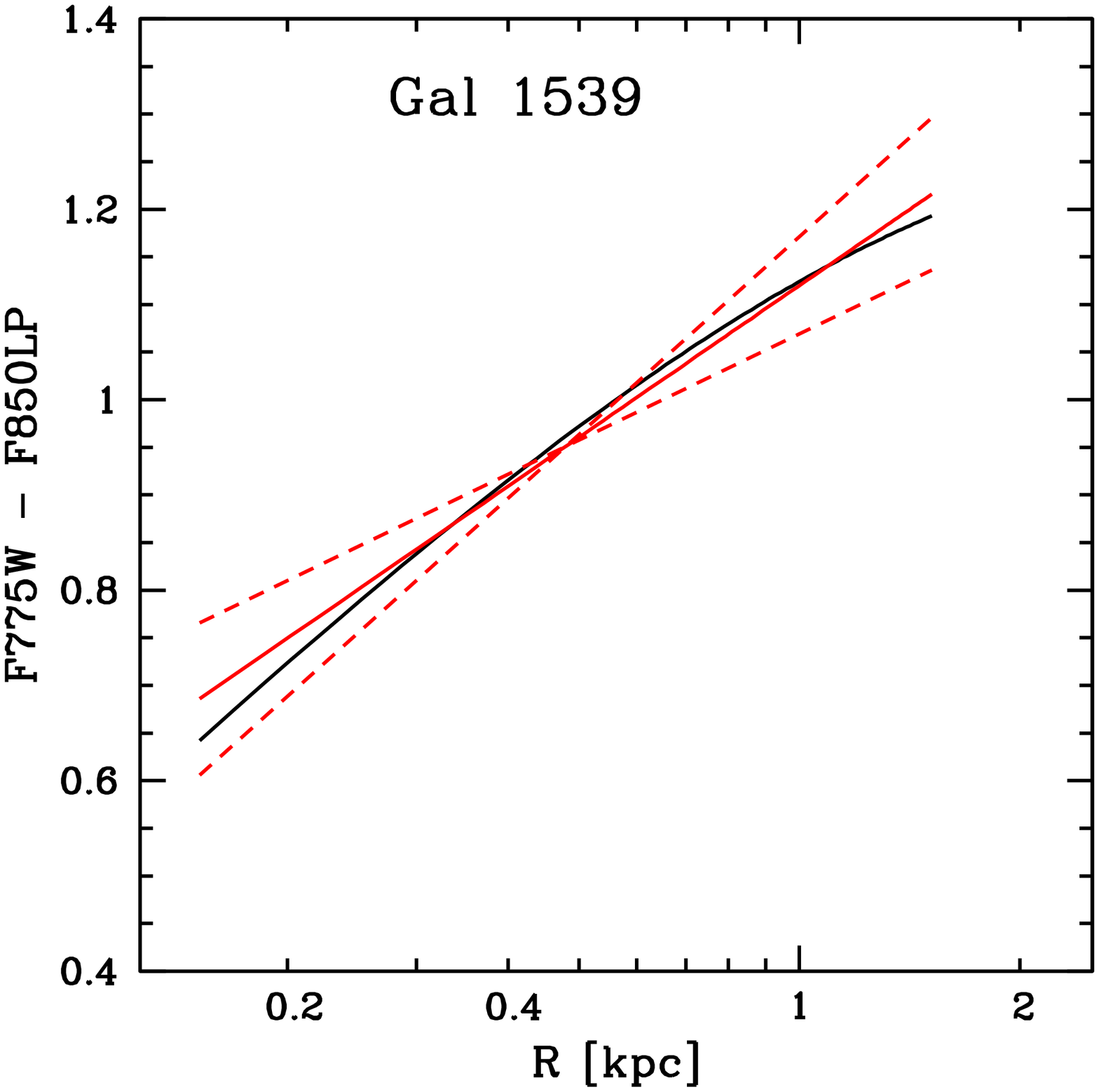}
\includegraphics[width=3.3cm]{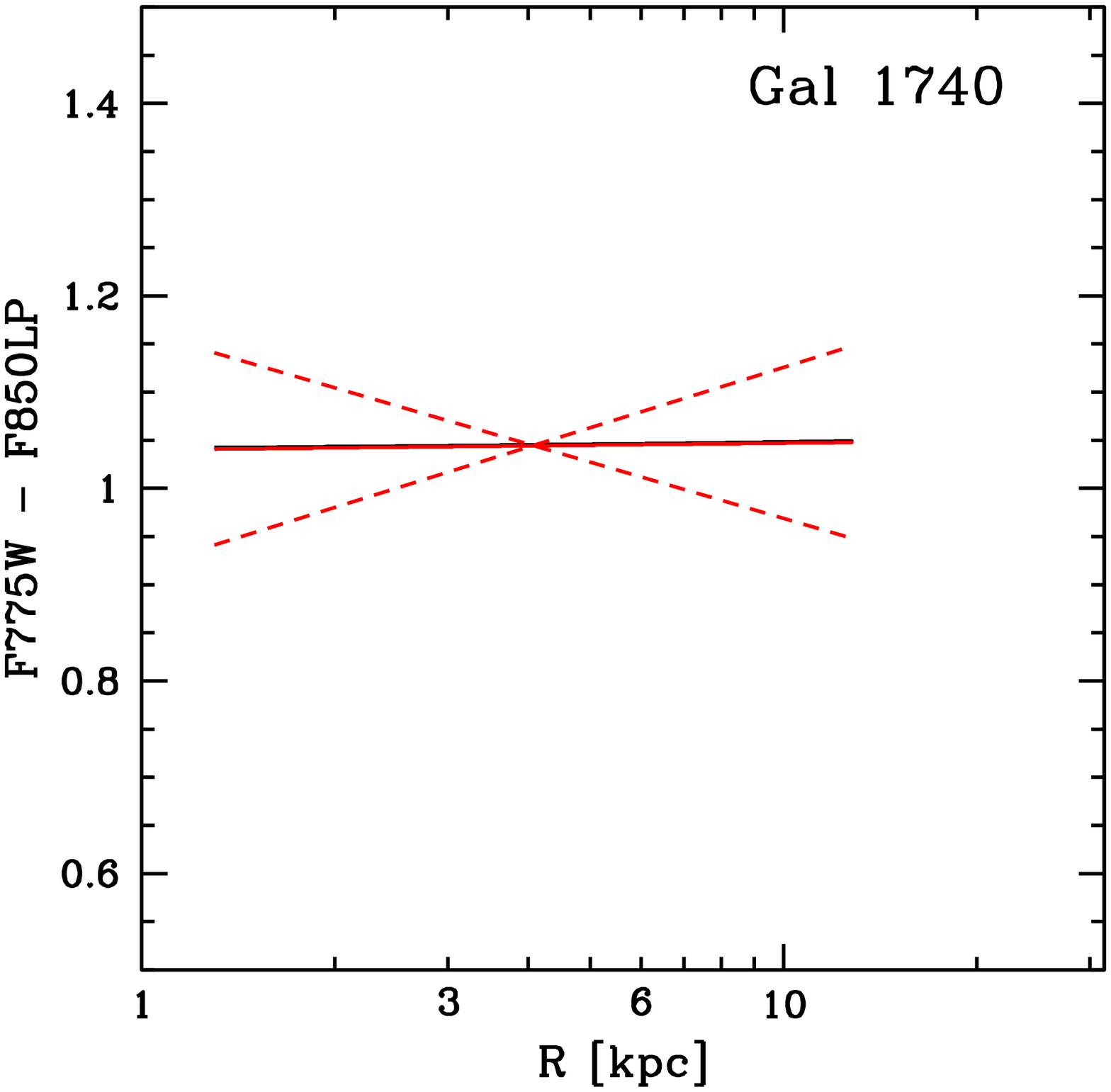}
\includegraphics[width=3.3cm]{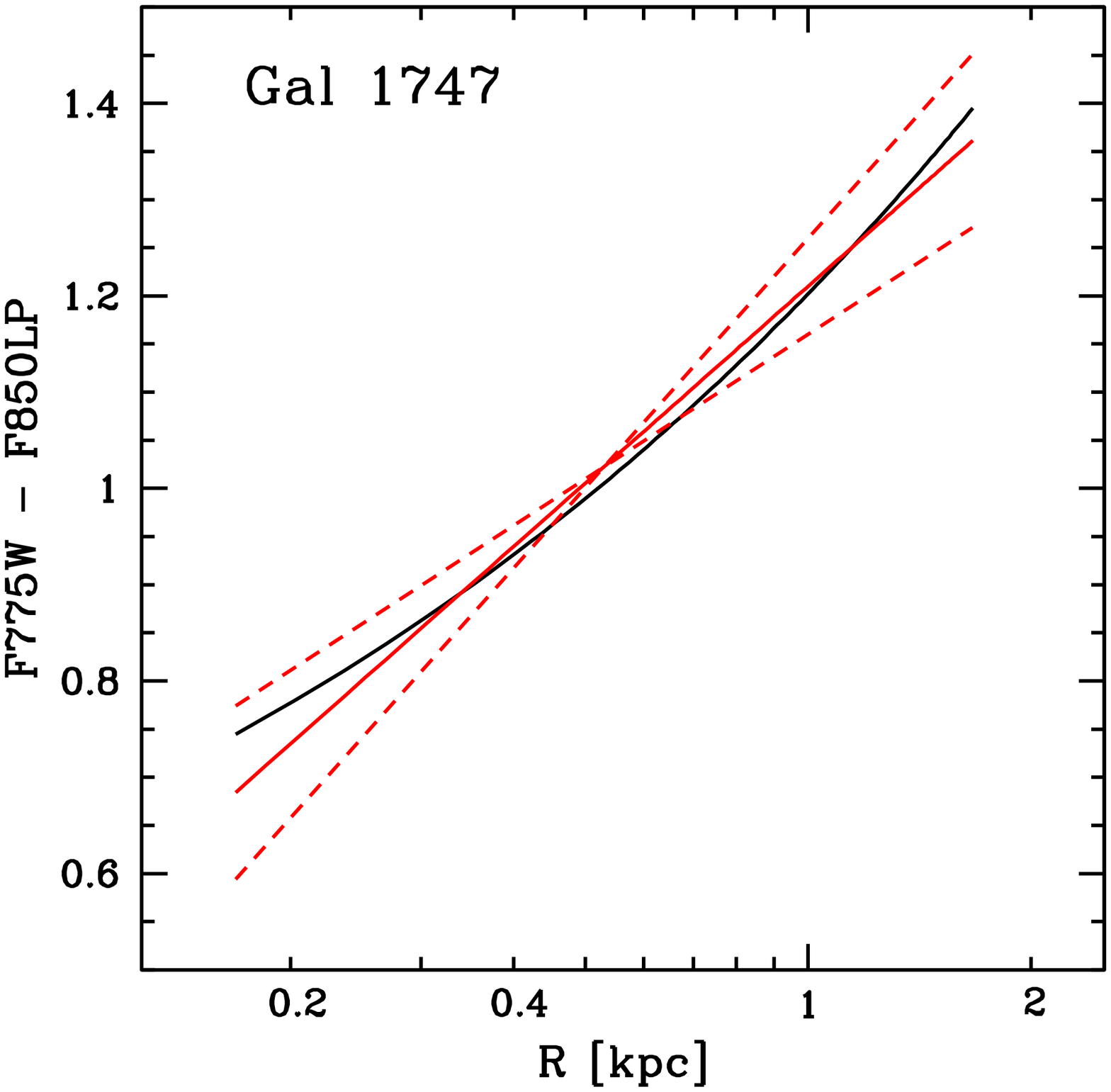}
\includegraphics[width=3.3cm]{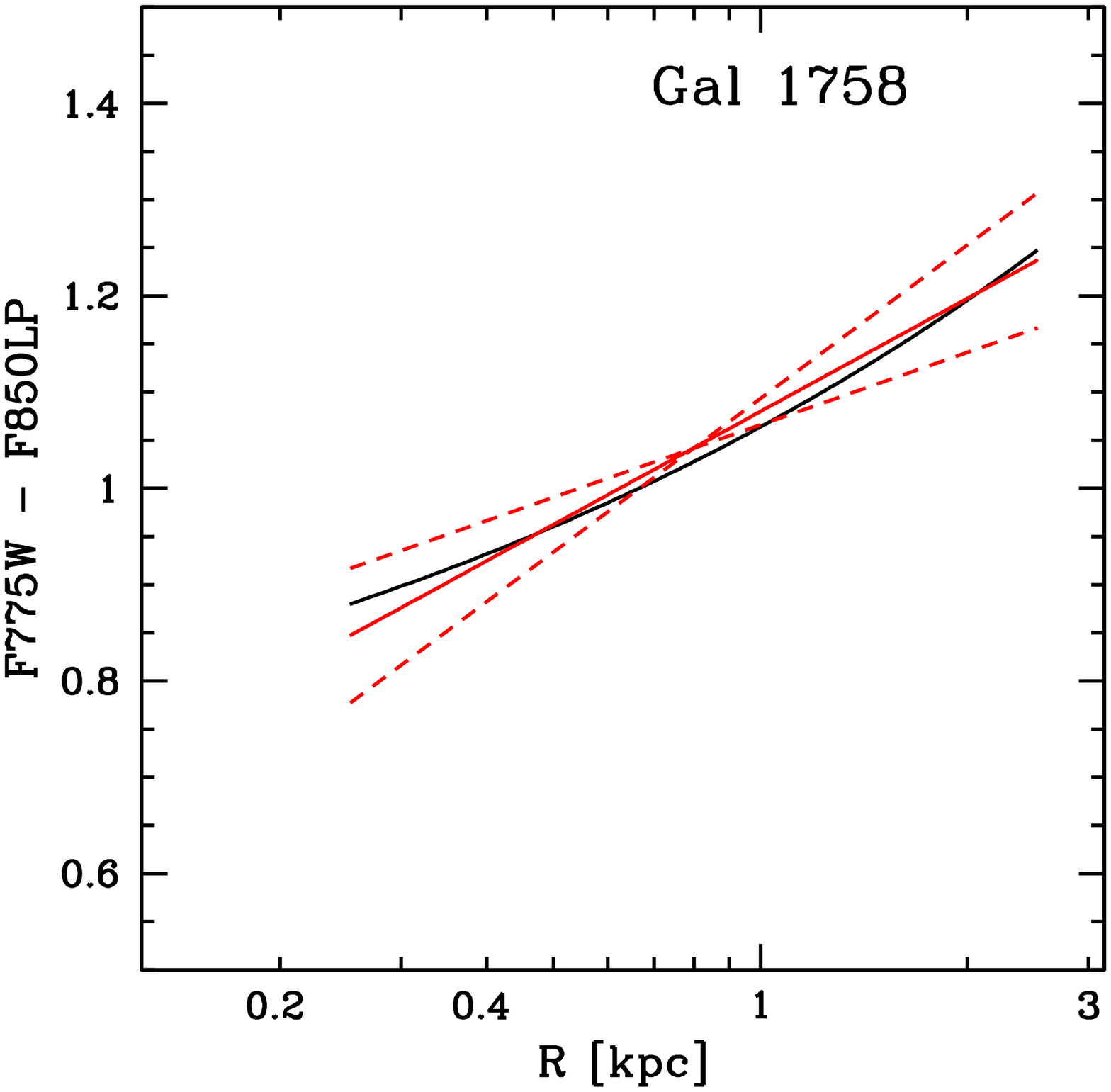}
\includegraphics[width=3.3cm]{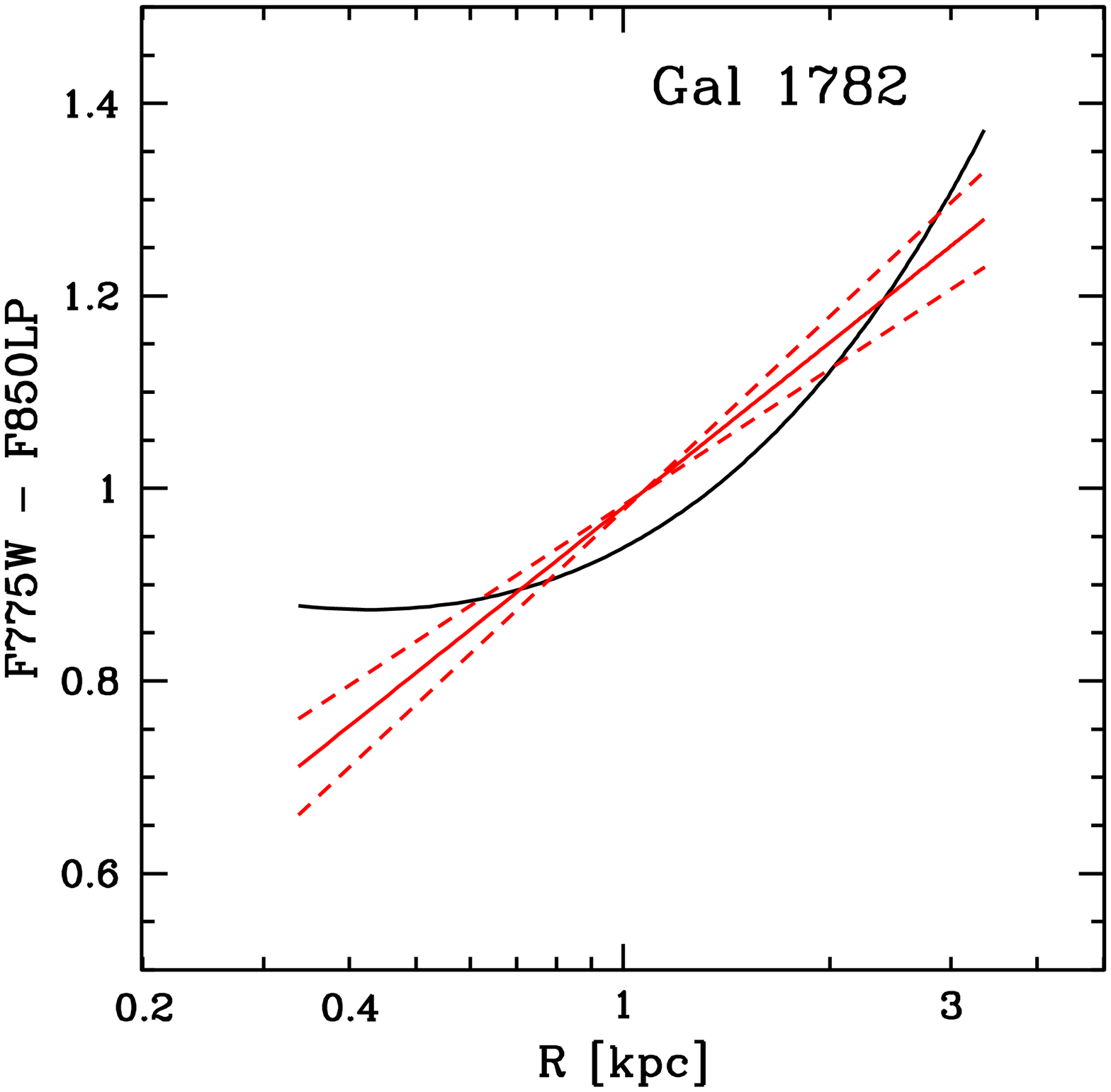}
\includegraphics[width=3.3cm]{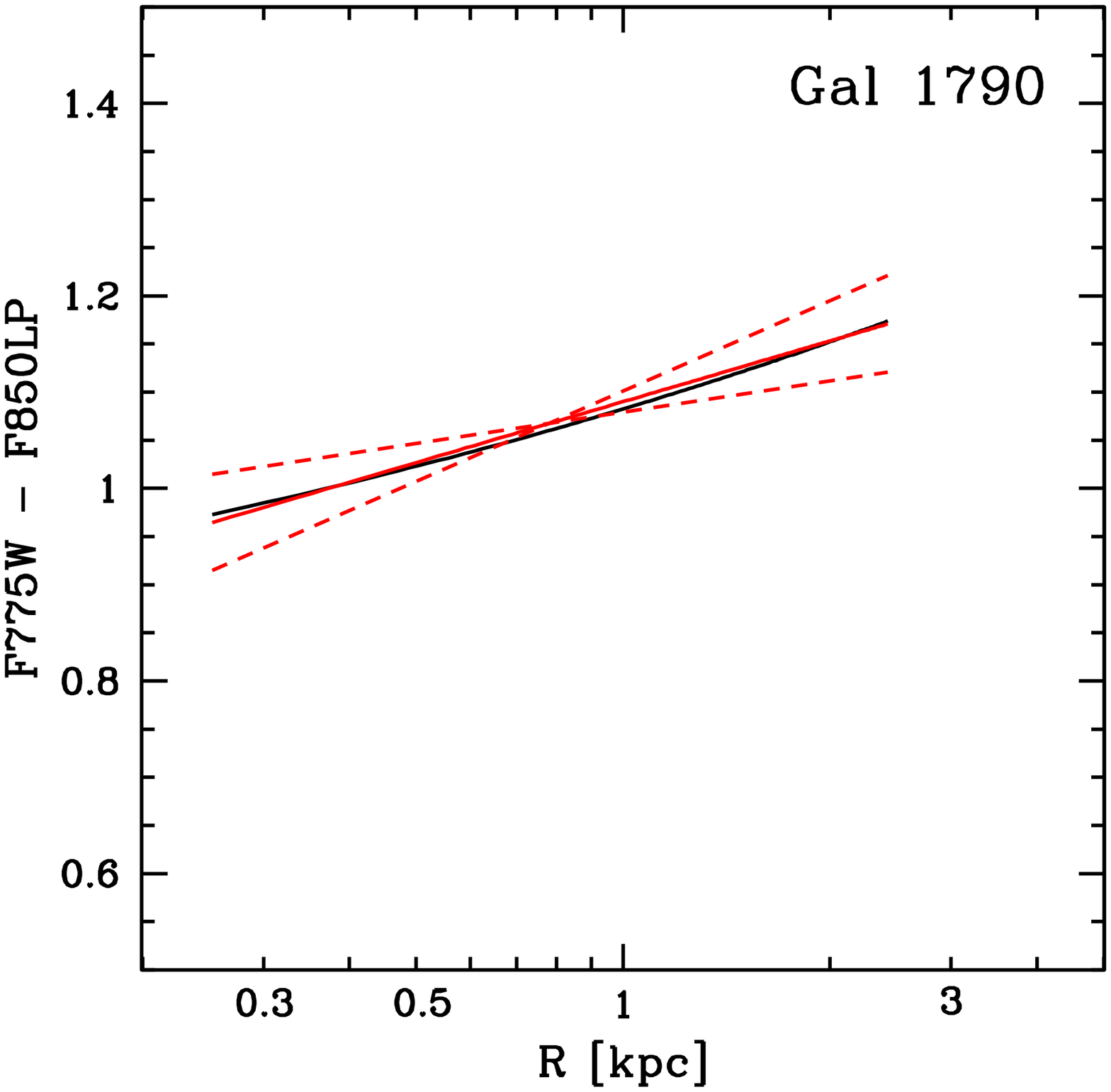}
\includegraphics[width=3.3cm]{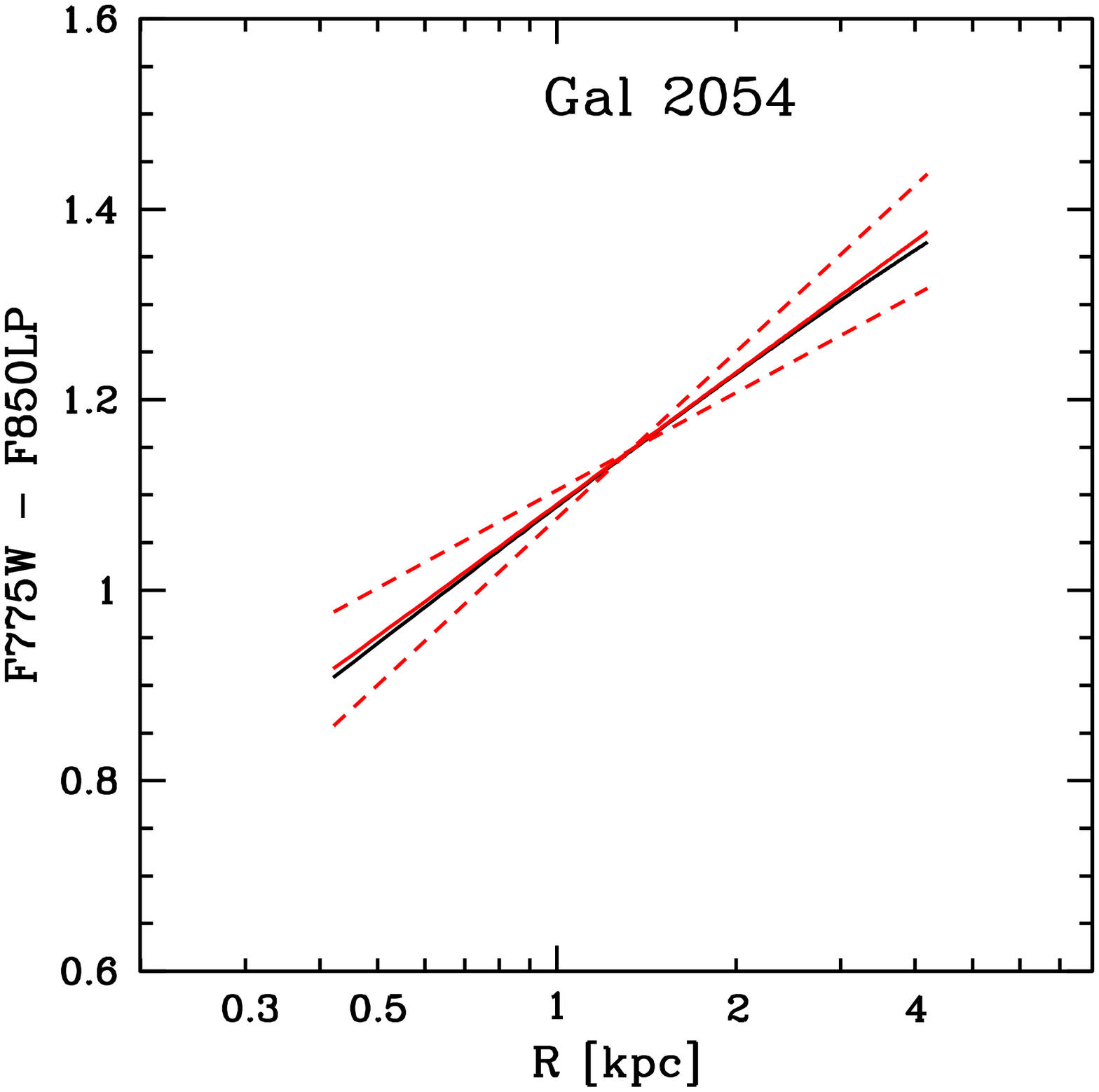}
\includegraphics[width=3.3cm]{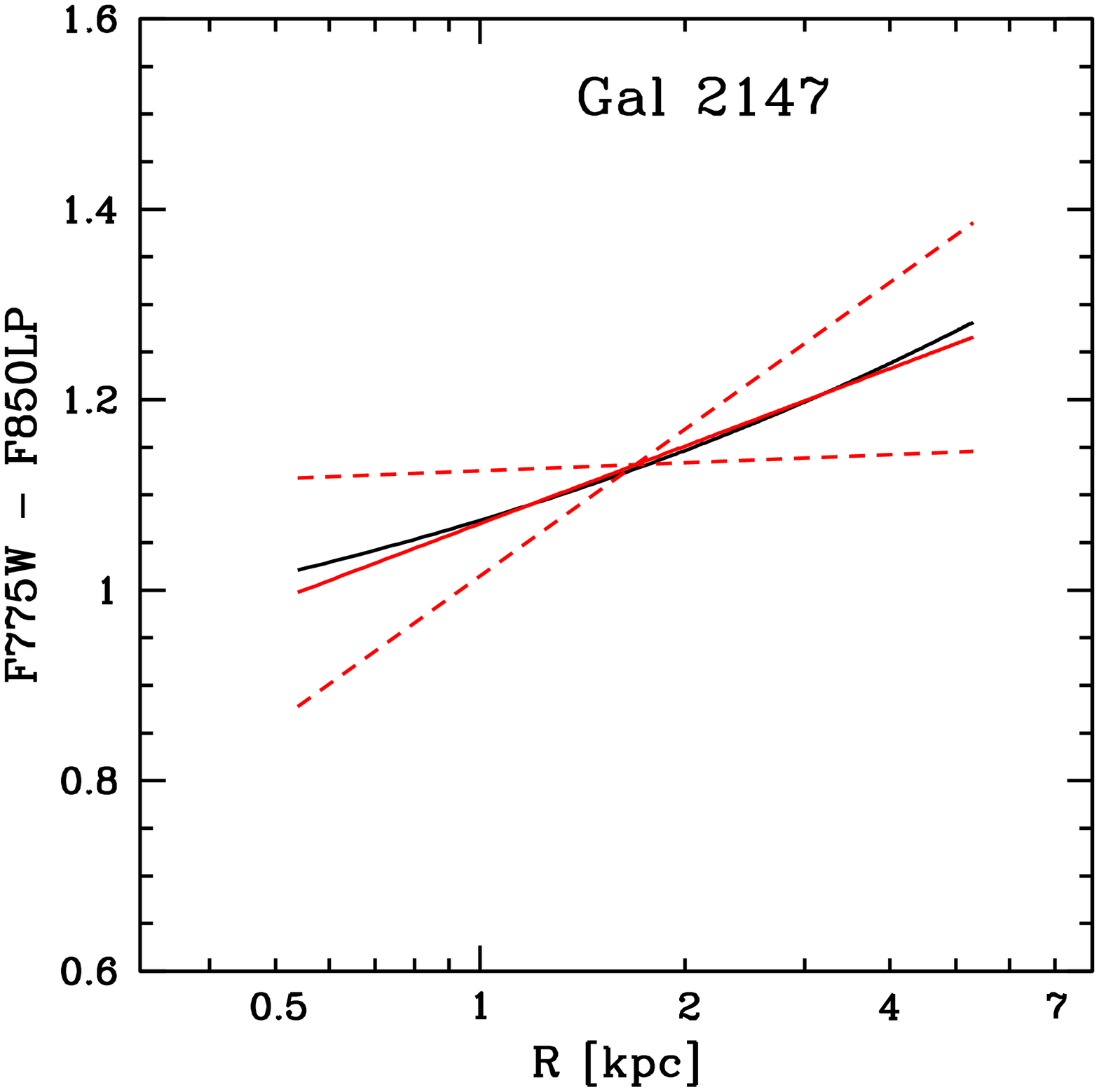}
\includegraphics[width=3.3cm]{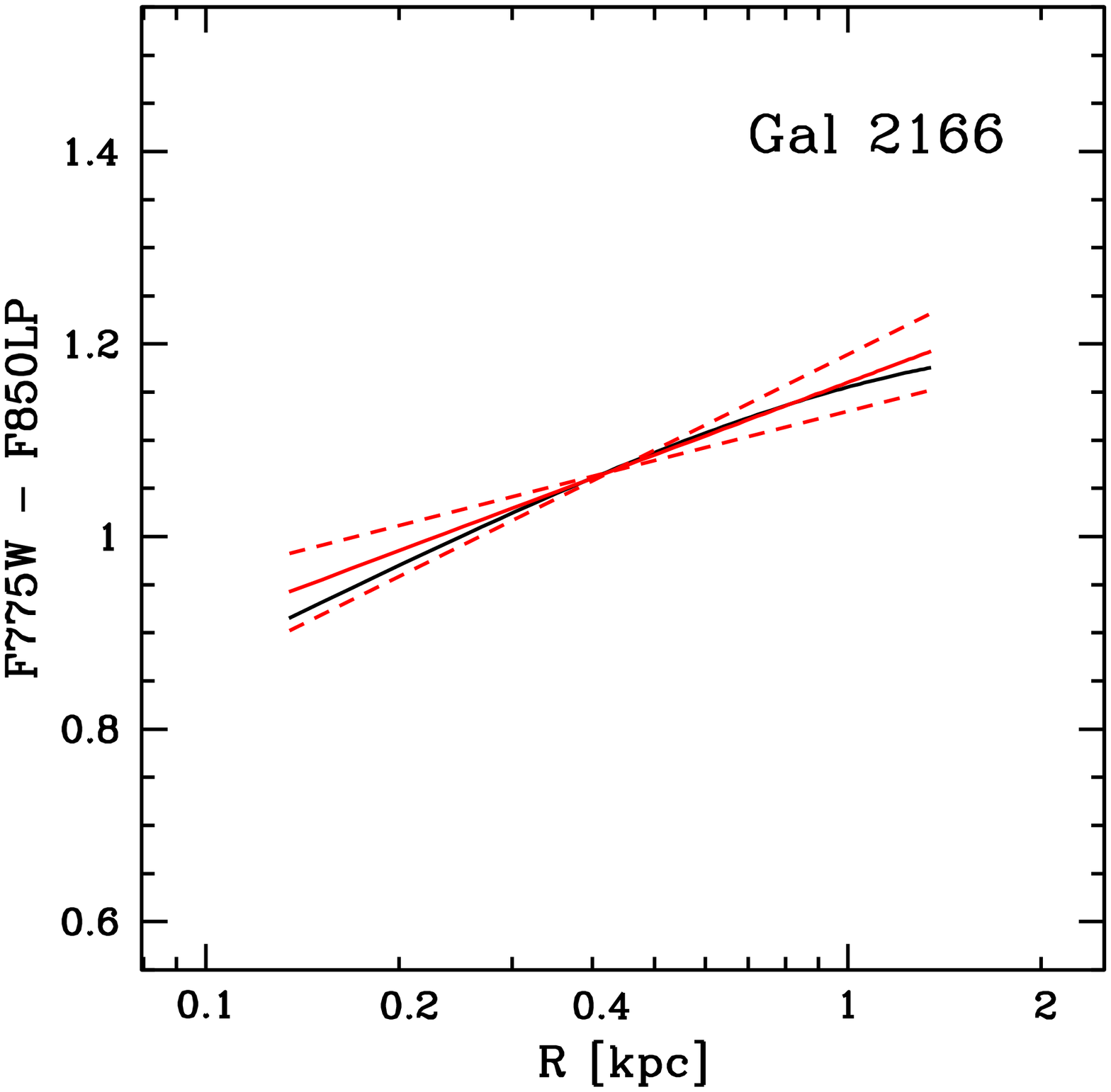}
\includegraphics[width=3.3cm]{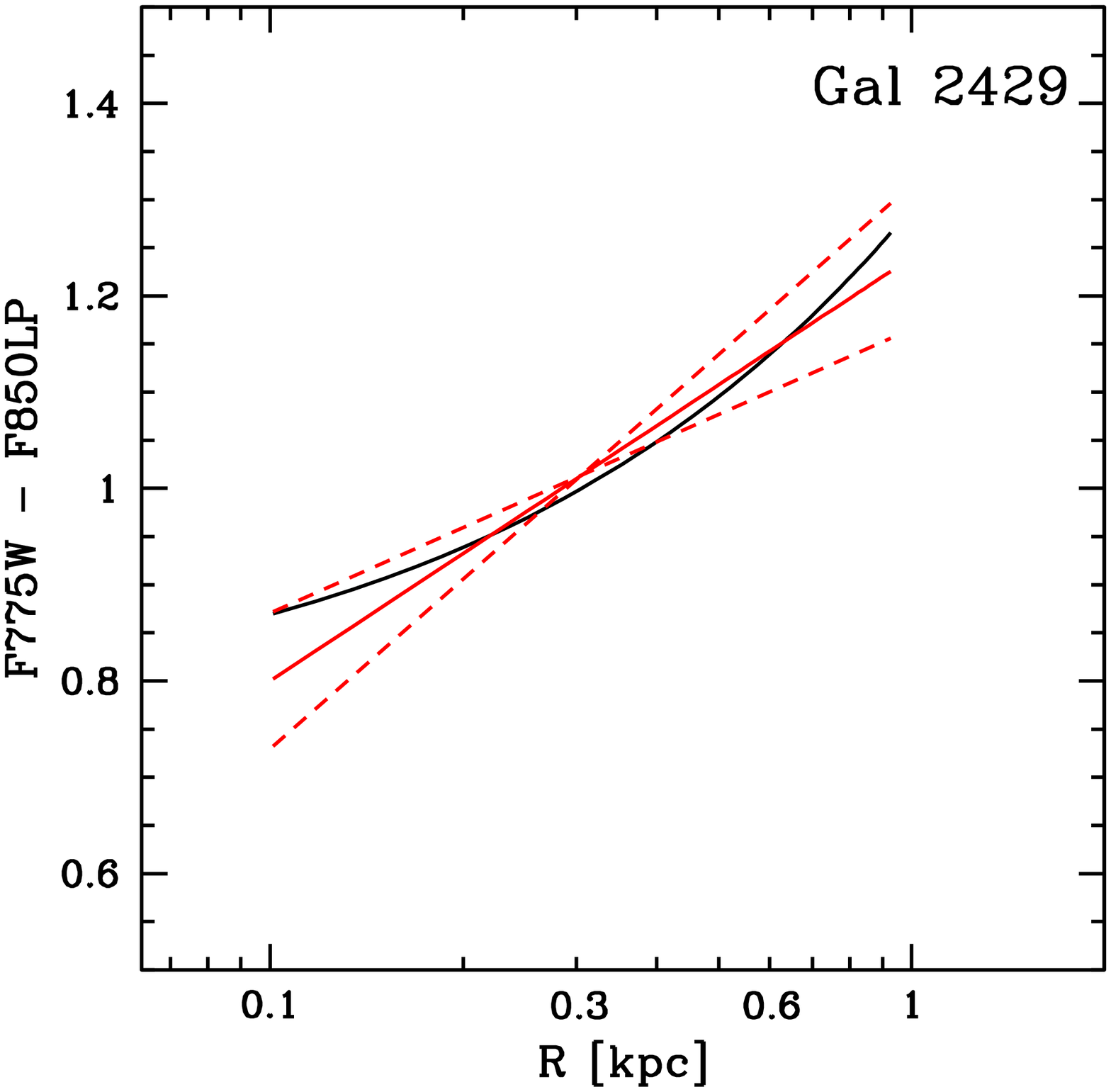}
\includegraphics[width=3.3cm]{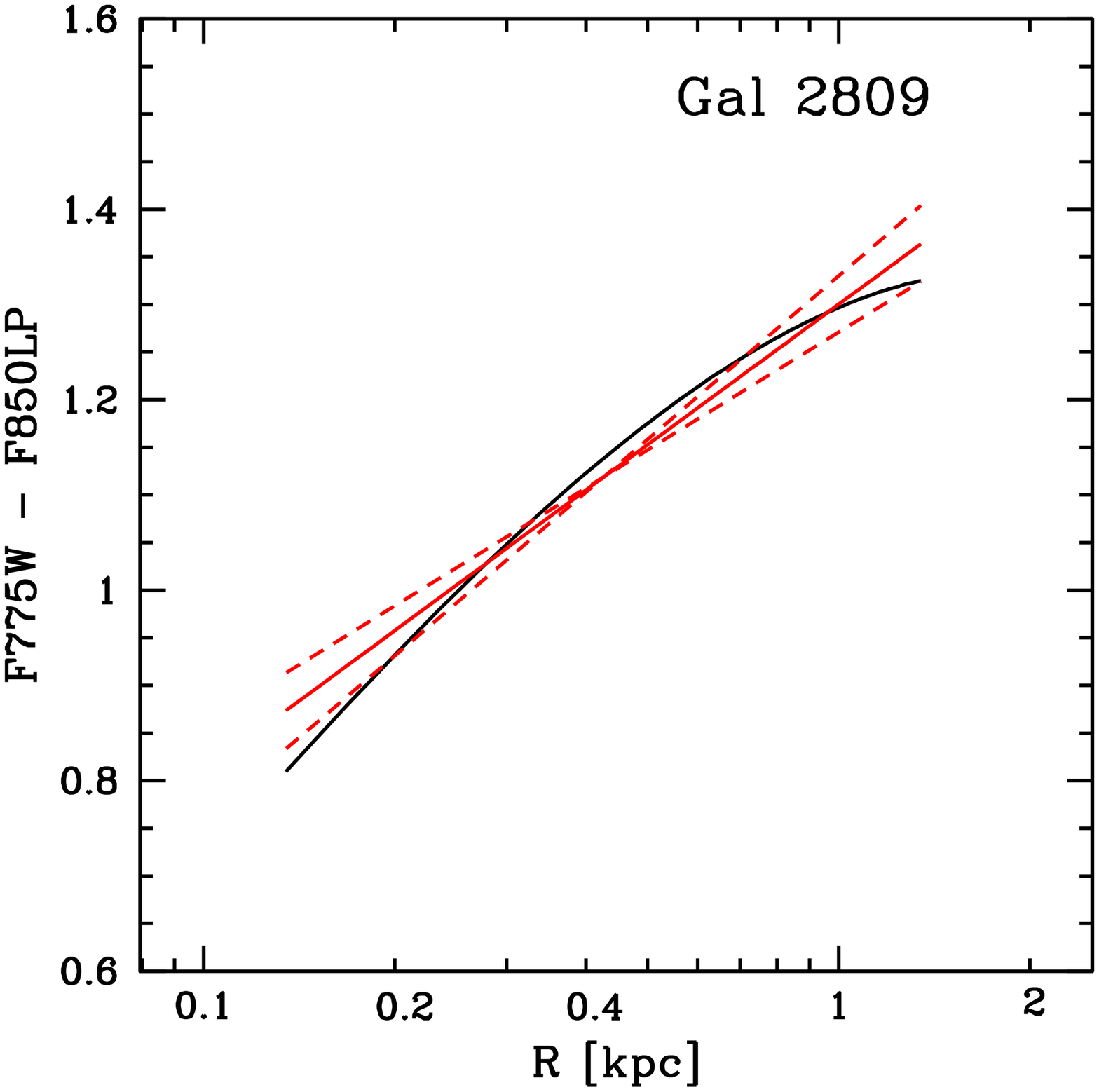}
\vskip -0.2truecm
\caption{The F775W - F850LP colour gradients for the galaxies of our sample. 
Black lines represent the deconvolved colour profiles between 0.1R$_e$ and 
1R$_e$ and the red solid lines are the best-fitted lines to the models. 
The red dashed lines set 1$\sigma$ errors. 
For all the galaxies in the \textit{y}-axis $\Delta$colour = 1 mag. 
The transformation to obtain the UV-U colours is: UV-U = (F775W-F850LP) - 0.8 mag.
}
\label{fig:col1}
\end{center}
\end{figure*}

\begin{figure*}
\begin{center}
\includegraphics[width=3.3cm]{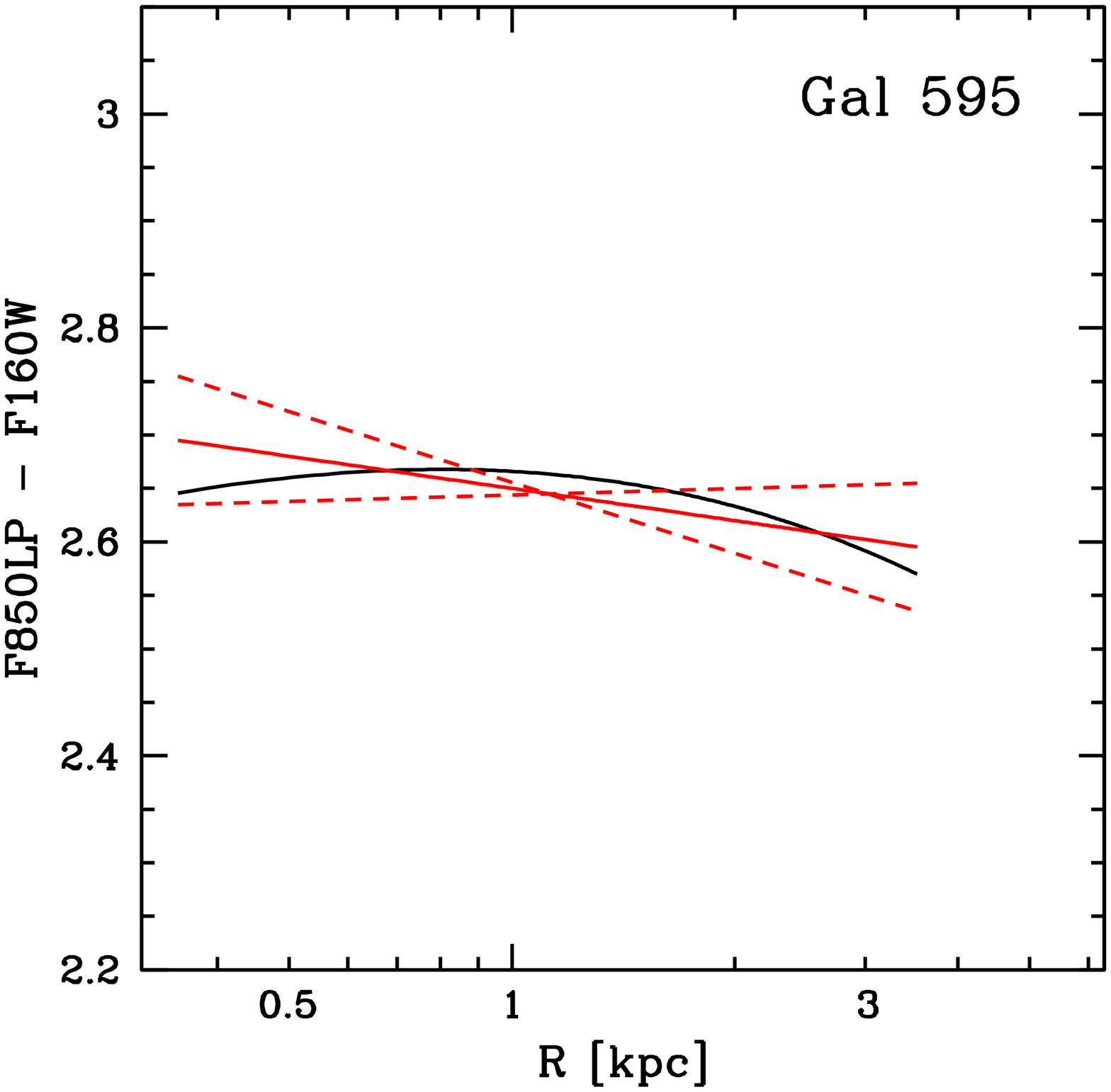}
\includegraphics[width=3.3cm]{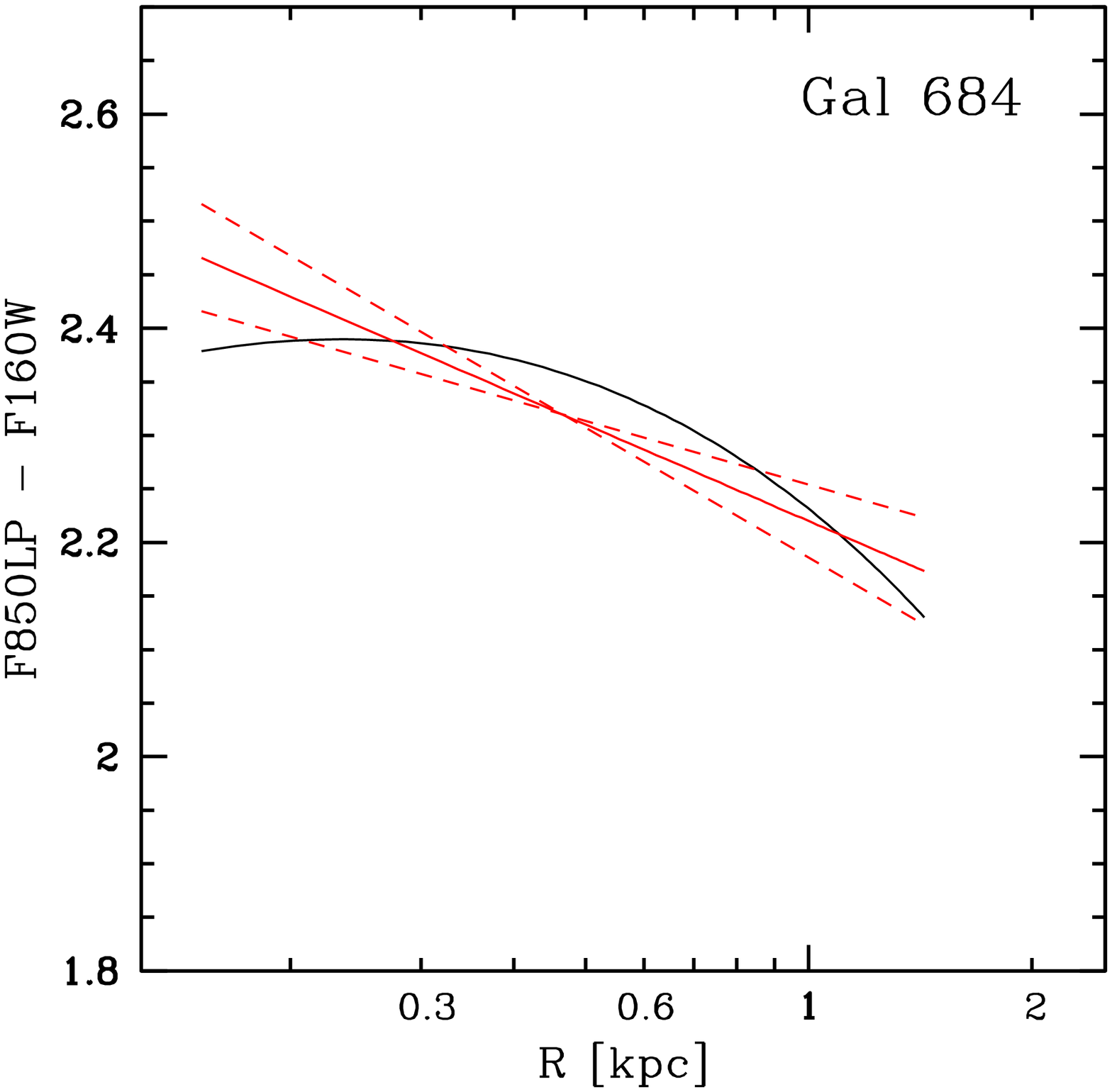}
\includegraphics[width=3.3cm]{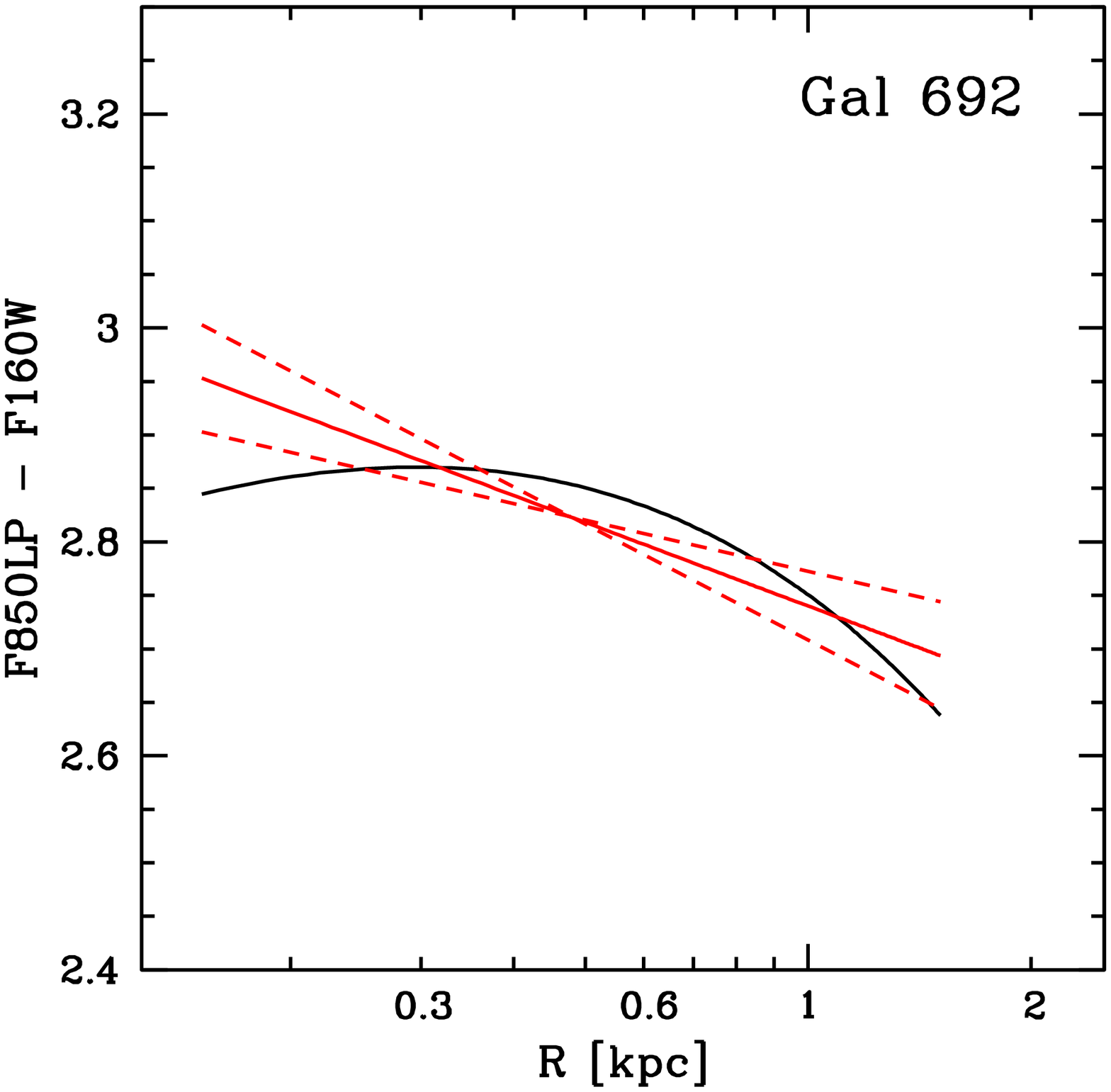}
\includegraphics[width=3.3cm]{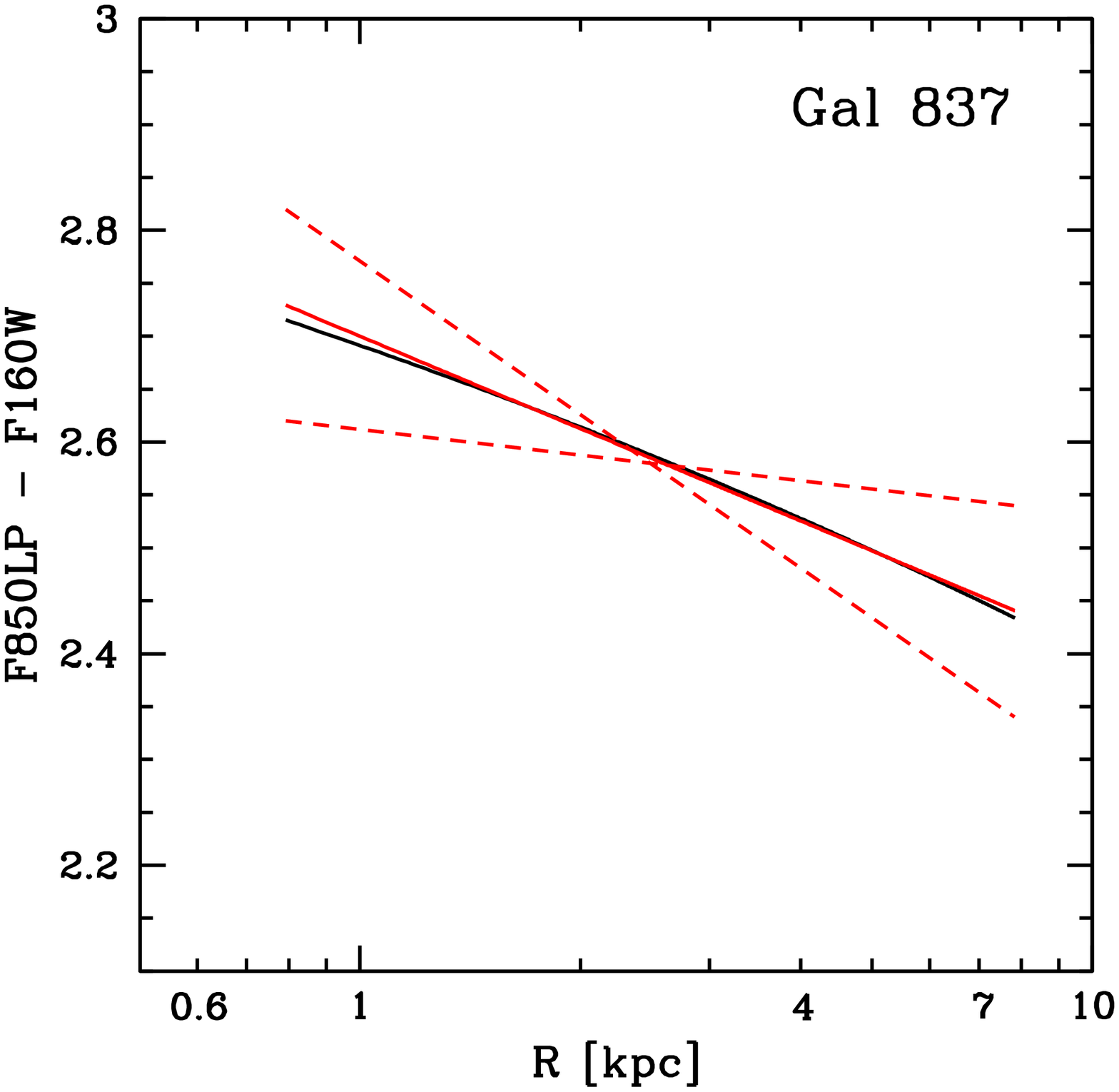}
\includegraphics[width=3.3cm]{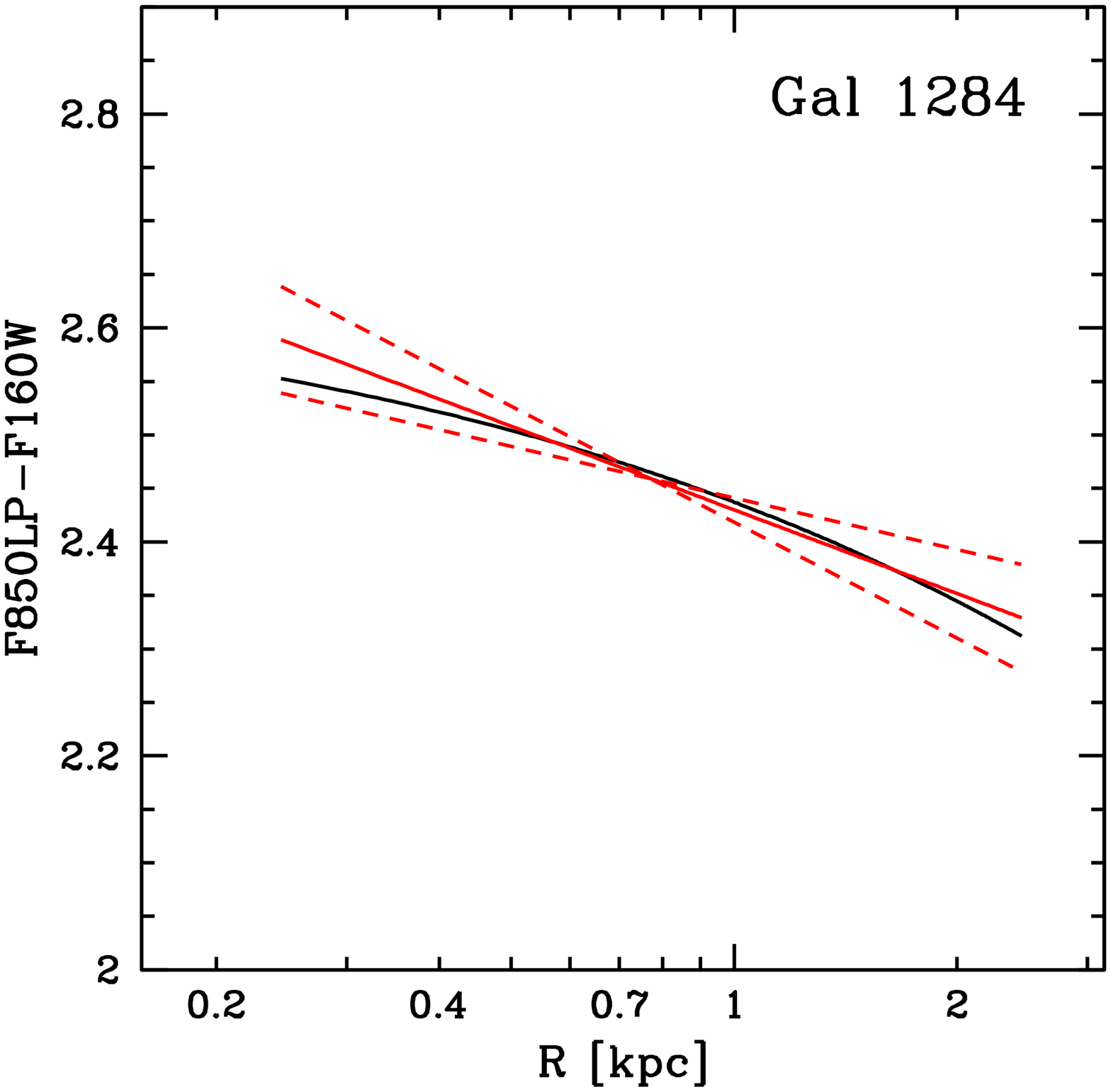}
\includegraphics[width=3.3cm]{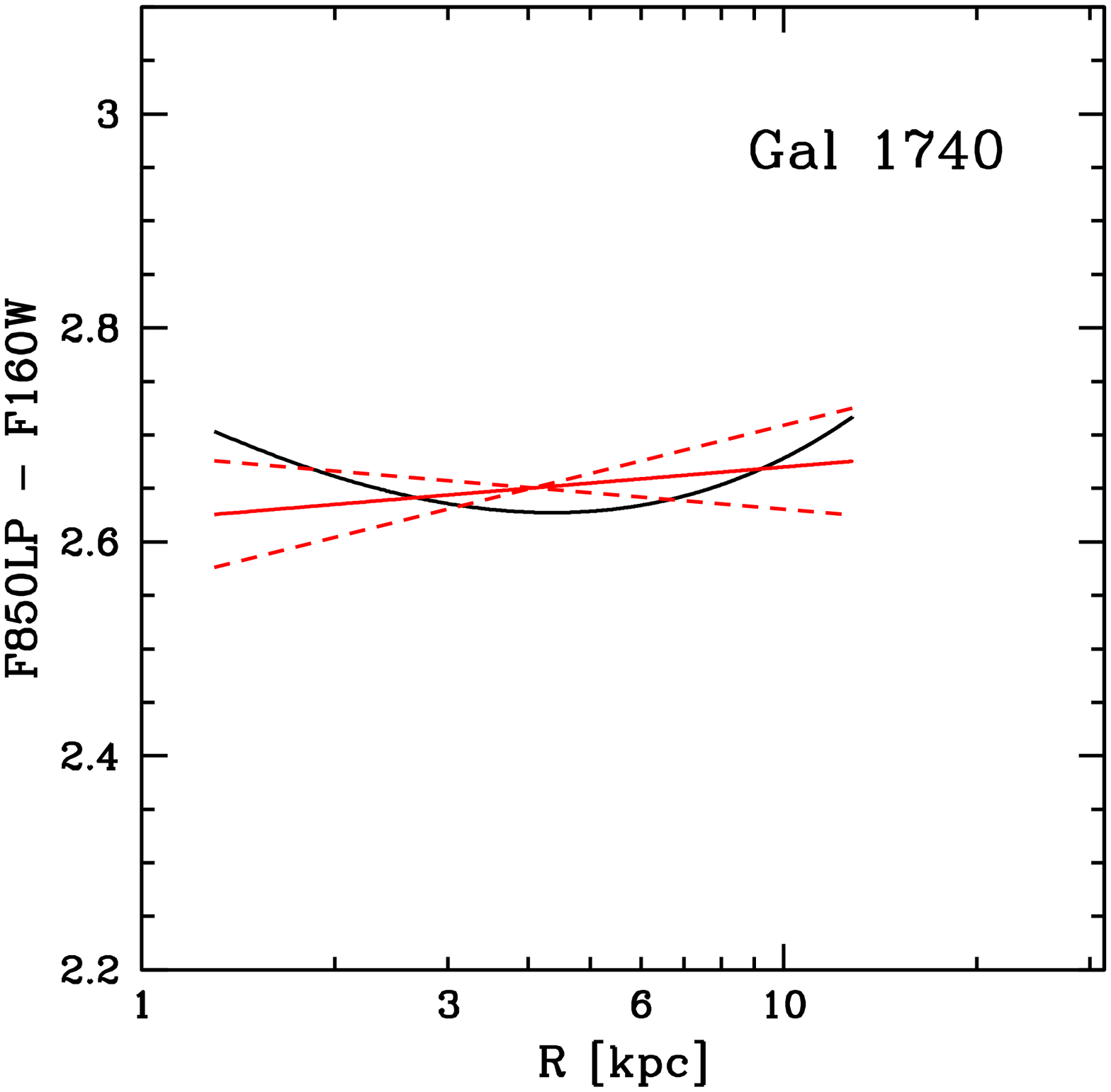}
\includegraphics[width=3.3cm]{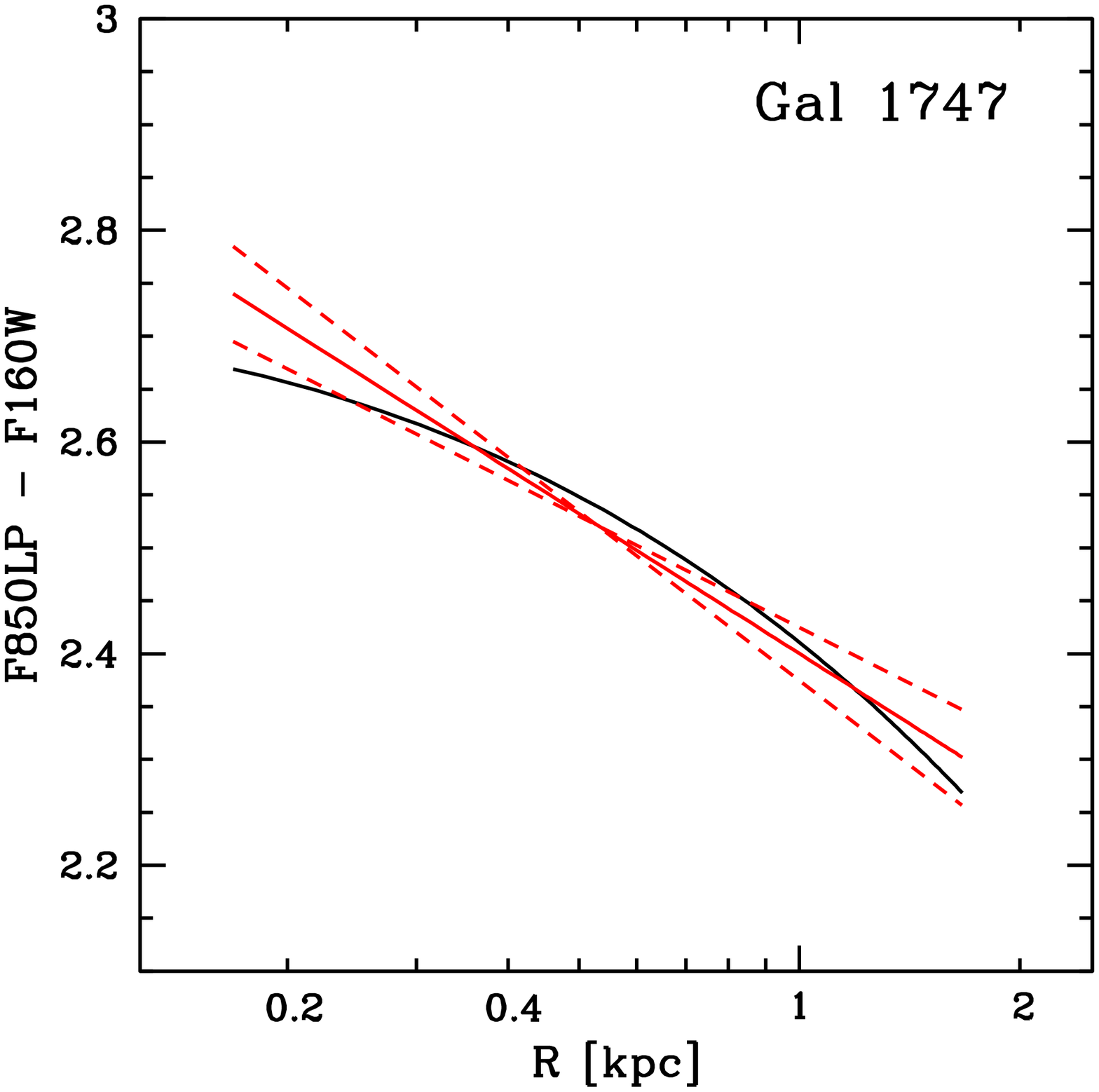}
\includegraphics[width=3.3cm]{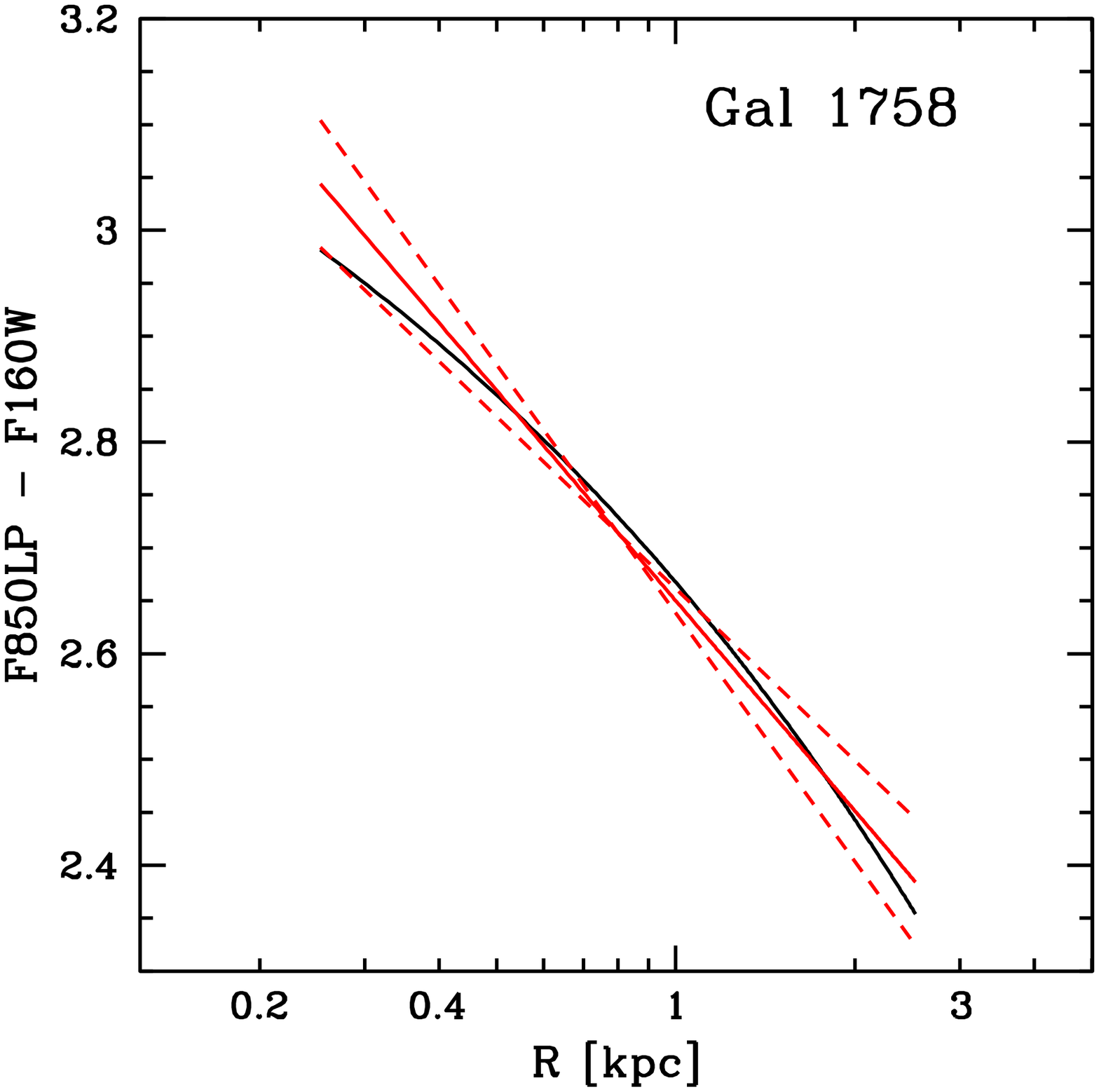}
\includegraphics[width=3.3cm]{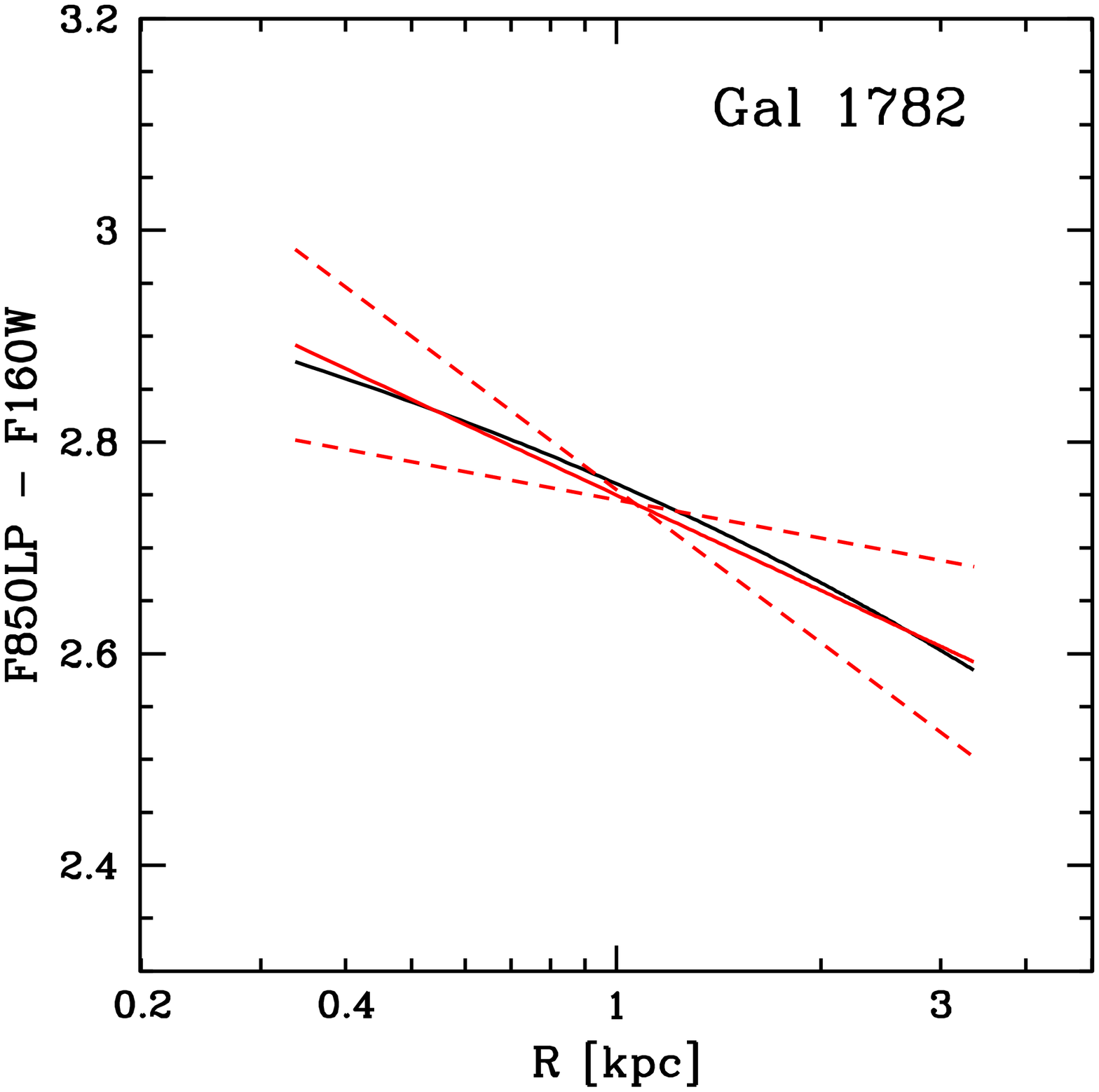}
\includegraphics[width=3.3cm]{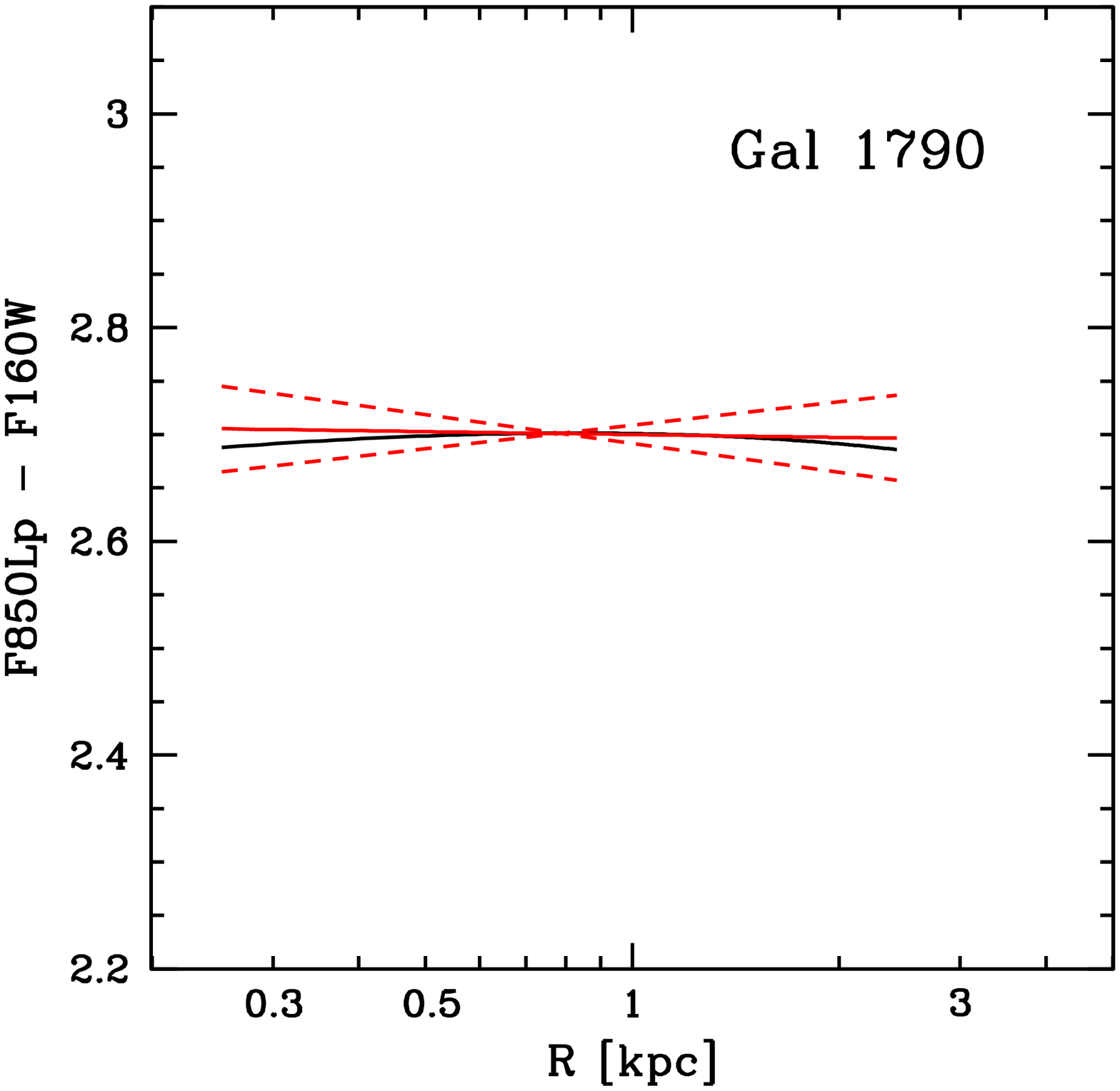}
\includegraphics[width=3.3cm]{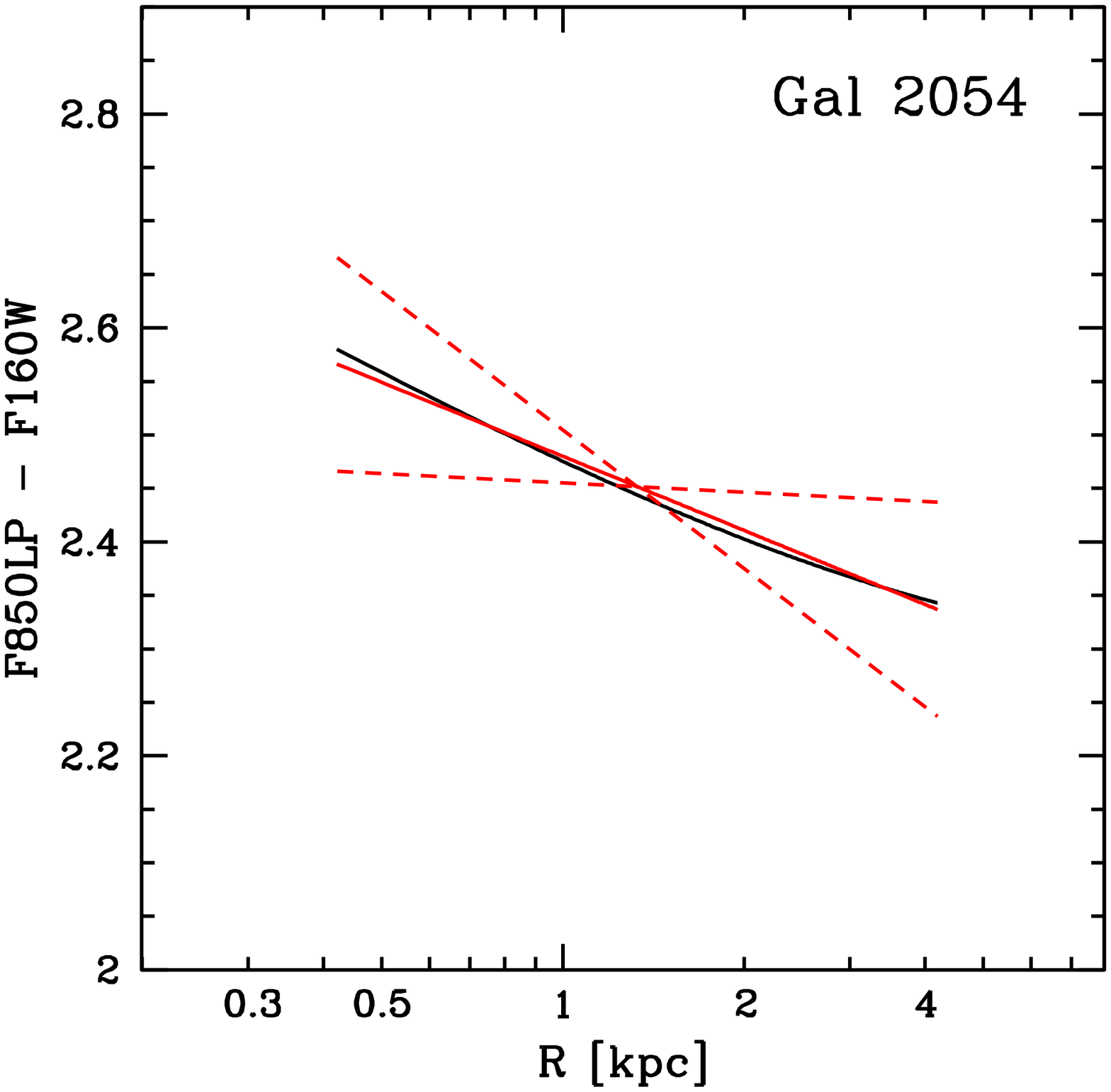}
\includegraphics[width=3.3cm]{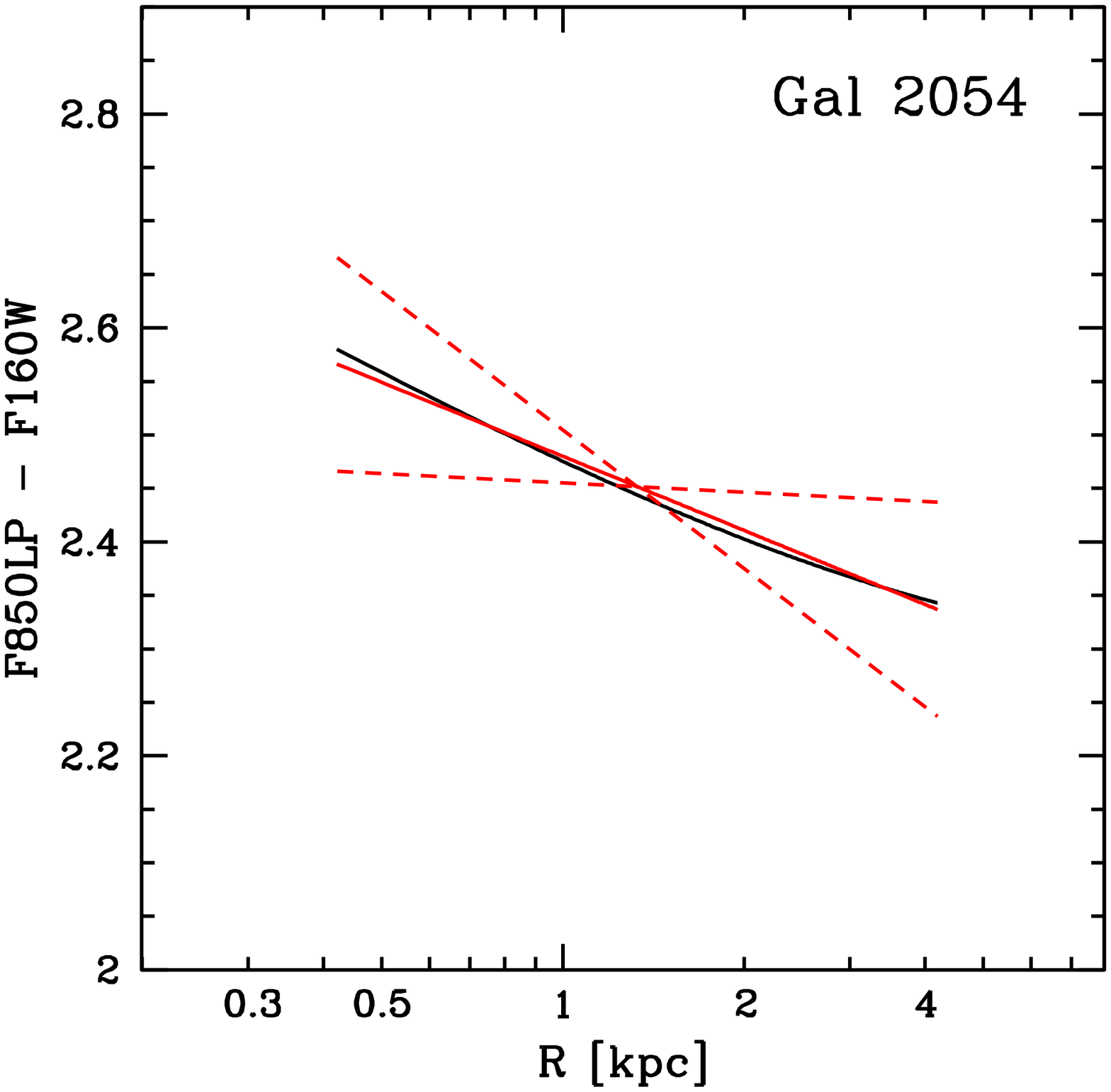}
\includegraphics[width=3.3cm]{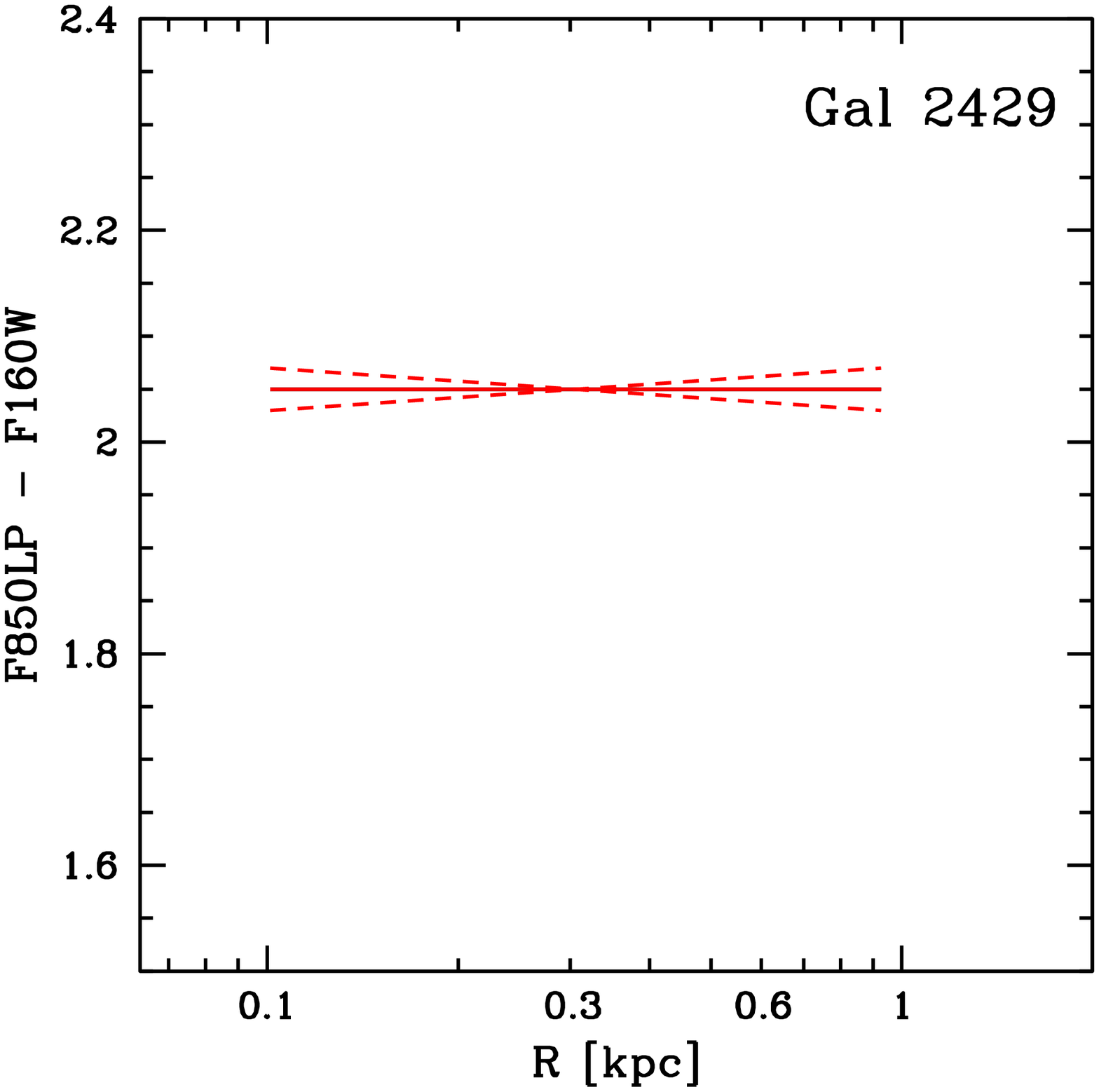}
\vskip -0.2truecm
\caption{The F850LP - F160W colour gradients for the galaxies of our sample. 
Black lines represent the deconvolved colour profiles between 0.1R$_e$ and 1R$_e$ 
and the red solid lines are the best-fitted lines to the models. 
The red dashed lines set 1$\sigma$ errors. 
For all the galaxies in the \textit{y}-axis $\Delta$colour = 0.9 mag. 
The transformation to obtain the U-R colours is: U-R = (F850LP-F160W) - 1.1 mag.
}
\label{fig:col2}
\end{center}
\end{figure*}

In this section, we show the colour gradients derived for the whole sample. 
In particular, in Fig. \ref{fig:col1} we show the F775W - F850LP colour 
gradients and in Fig. \ref{fig:col2} we show the F850LP - F160W colour 
gradients. 
In all the figures, the variation $\Delta$colour on the \textit{y}-axis is 1 mag for the
F775W-F850LP colour and it is 0.9 mag for the the F850LP-F160W colour.

\section{Comparison with Chan et al. 2016}
\label{Chan}

In this Appendix we compare our sample with the one by \cite{chan16}. 
The criteria adopted to select the two samples are different. 
In our analysis, we adopted the ACS-F850LP image for the object detection 
with SExtractor and as reference in the morphological and structural analysis,
given its high resolution ($\simeq 0.11$ arcsec) and small pixel scale 
(0.05 arcsec pixel$^{-1}$) when compared to the WFC3-F160W image (FWHM$\simeq 0.22$ 
arcsec; 0.123 arcsec pixel$^{-1}$).
Furthermore, we selected galaxies (see Section 2.2) at magnitudes F850LP$<24$,
where the sample is 100 per cent complete, according to their F775W-F850LP colour and
on the basis of their elliptical morphology.
Chan et al., instead, used the WFC3-F160W image for object detection with 
SE\textlcsc{xtractor} and as reference, and selected 36 passive galaxies according to 
their red sequence colour from the colour-magnitude diagram ((F850LP-F160W) 
versus F160W). 
They also applied a magnitude cut of $H_{160}$ < 22.5 mag, which corresponds to a 
completeness of $\sim$ 95 per cent. 

\begin{figure*}
\includegraphics[width=8cm]{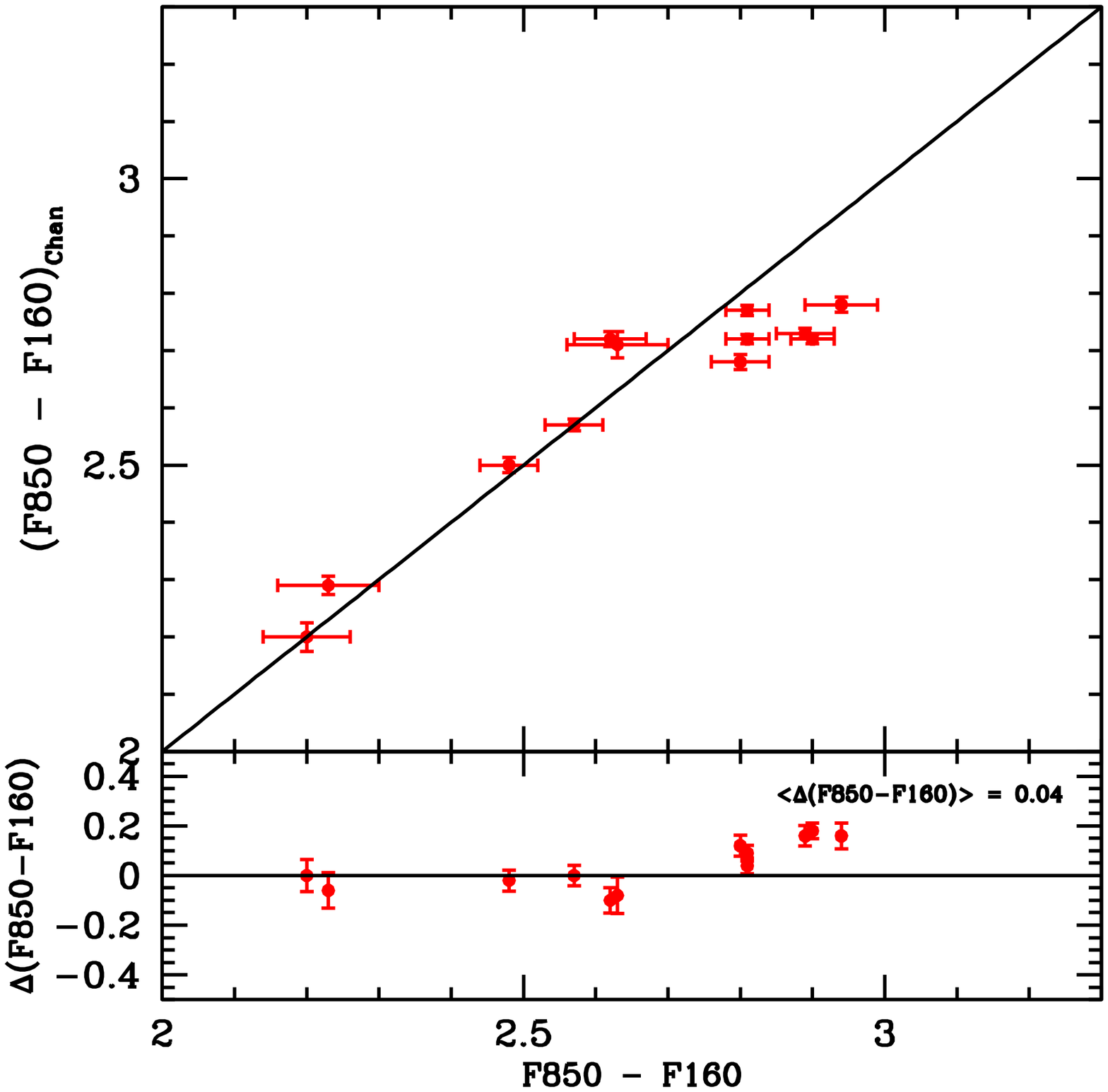}
 \caption{Comparison between the F850LP-F160W colours derived 
  by Chan et al. (2016) and our F850LP-F160W colours for the 12 galaxies 
  in common (upper panel). 
The difference $\Delta (F850-F160)$=(F850-F160) - (F850-F160)$_{Chan}$ is 
plotted as a function of our colours in the lower panel. 
}
\label{fig:comp_col}
\end{figure*}

We verified that 12 out to the 17 ellipticals of our sample are in common with 
Chan et al. 
Four of our galaxies (358, 1539, 2166, 2809) are not included in their
sample since they fall outside the smaller field covered by the WFC3 
(5 arcmin$^2$ instead of 11 arcmin$^2$ of the ACS).
Moreover, also galaxy 1747 is not included in their sample.
We verified that its F850LP-F160W colour would be consistent with their colour 
selection. 
However, its magnitude (F160W = 22.6 mag, in our sample)
seems to be slightly fainter than the cut they applied. 
This could be the reason why it is not included in their sample.
In Fig. \ref{fig:comp_col} we show the comparison between the F850LP-F160W 
 colour of the 12 galaxies in common, after having transformed their colours from AB to Vega magnitudes. 
We find that our F850LP-F160W colours (mean value 2.66$\pm$0.25 mag) are 
consistent with those derived by Chan et al. (mean value 2.62$\pm$0.19 mag).

\begin{figure*}
\begin{center}
\includegraphics[width=8 cm]{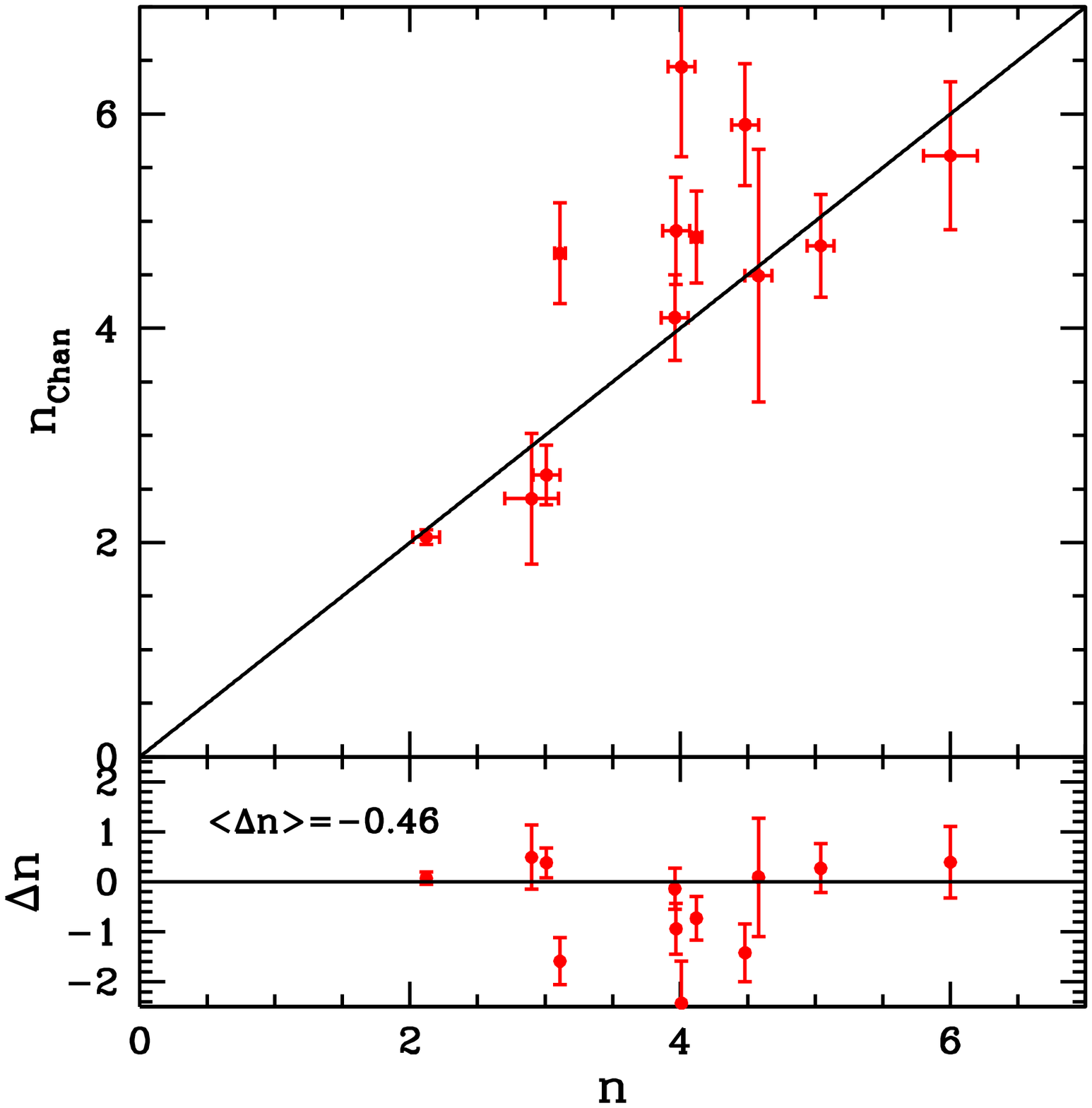}
\includegraphics[width=8 cm]{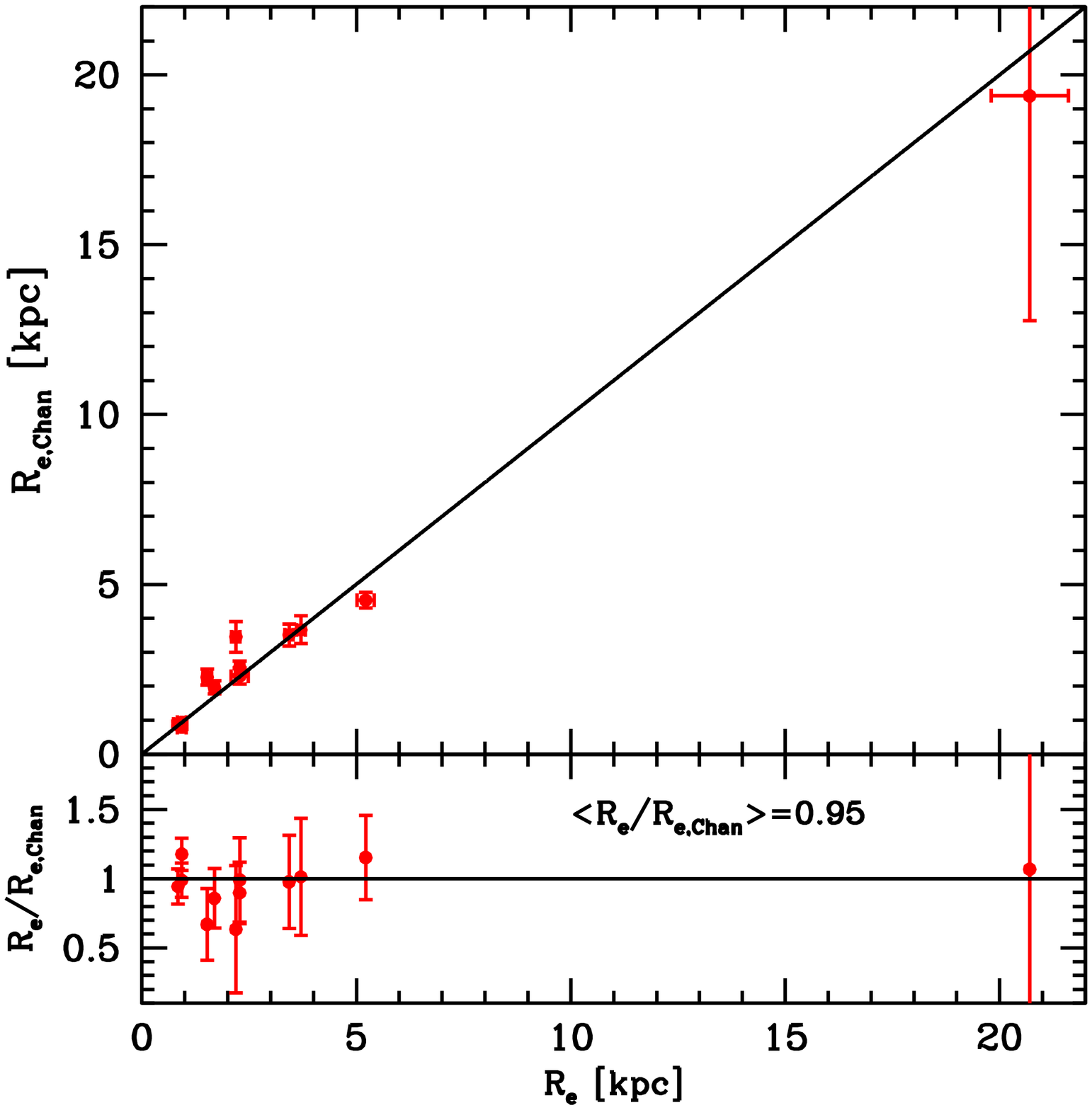}
\caption{\textit{Left panel}: comparison between the S\'ersic indices $n$ derived by Chan et al. (2016) and our S\'ersic indices $n$ for the 12 galaxies in common (upper panel). The difference $\Delta$n=$n$ - $n_{Chan}$ is plotted as a function of our $n$ in the lower panel.
\textit{Right panel}: comparison between R$_{e,Chan}$ derived by Chan et al. (2016) in the F160W band and our R$_e$ for the 12 galaxies in common (upper panel). The ratio R$_e$/R$_{e,Chan}$ is plotted as a function of our R$_e$ in the lower panel.}
\label{fig:comp_para}
\end{center}
\end{figure*}

The structural parameters of the galaxies have been derived, in both the 
analyses, using the software \textlcsc{Galfit}. 
Since the authors do not report the structural parameters derived in the 
F775W and F850LP bands, we only compare the structural parameters derived in 
the F160W band. 
The results are shown in Fig. \ref{fig:comp_para}, where in the left panel 
we compare the S\'ersic indices $n$ and in the right panel we compare the 
effective radii R$_e$. 
We found a good agreement between our and their estimate of the effective 
radii (<R$_e$/R$_{e,Chan}>\simeq$1) and also between the indices $n$. 
For three galaxies (595, 684 and 837) we obtained from the fit a 
smaller value of the index $n$ (< $n$ - $n_{Chan}$> = -1.78). 
However, we verified through the simulations that an eventual underestimate 
of $n$ does not affect the estimate of the effective radii and, hence, of the 
colour gradients.

\begin{figure*}
\includegraphics[width=8cm]{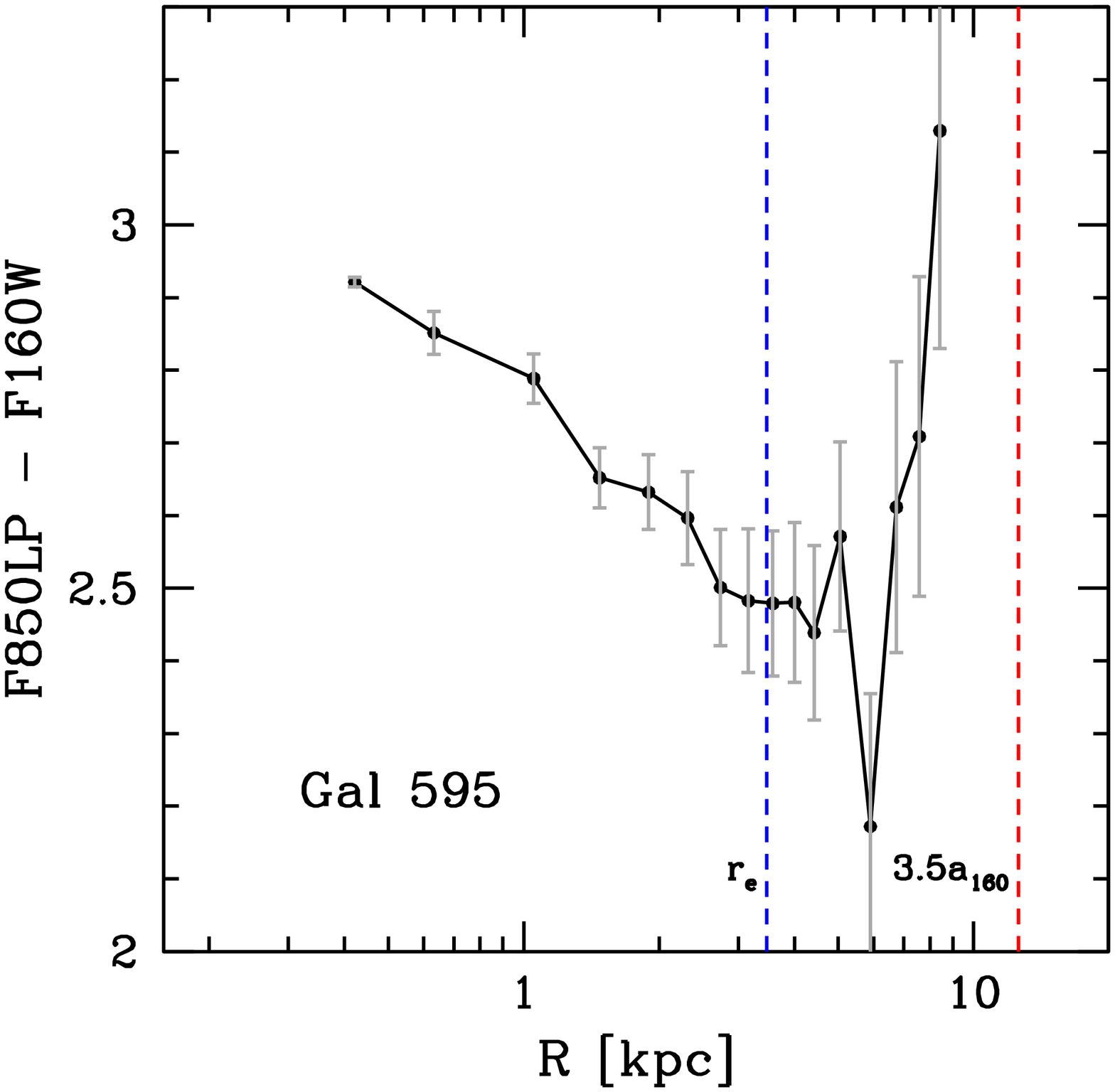}
\includegraphics[width=8cm]{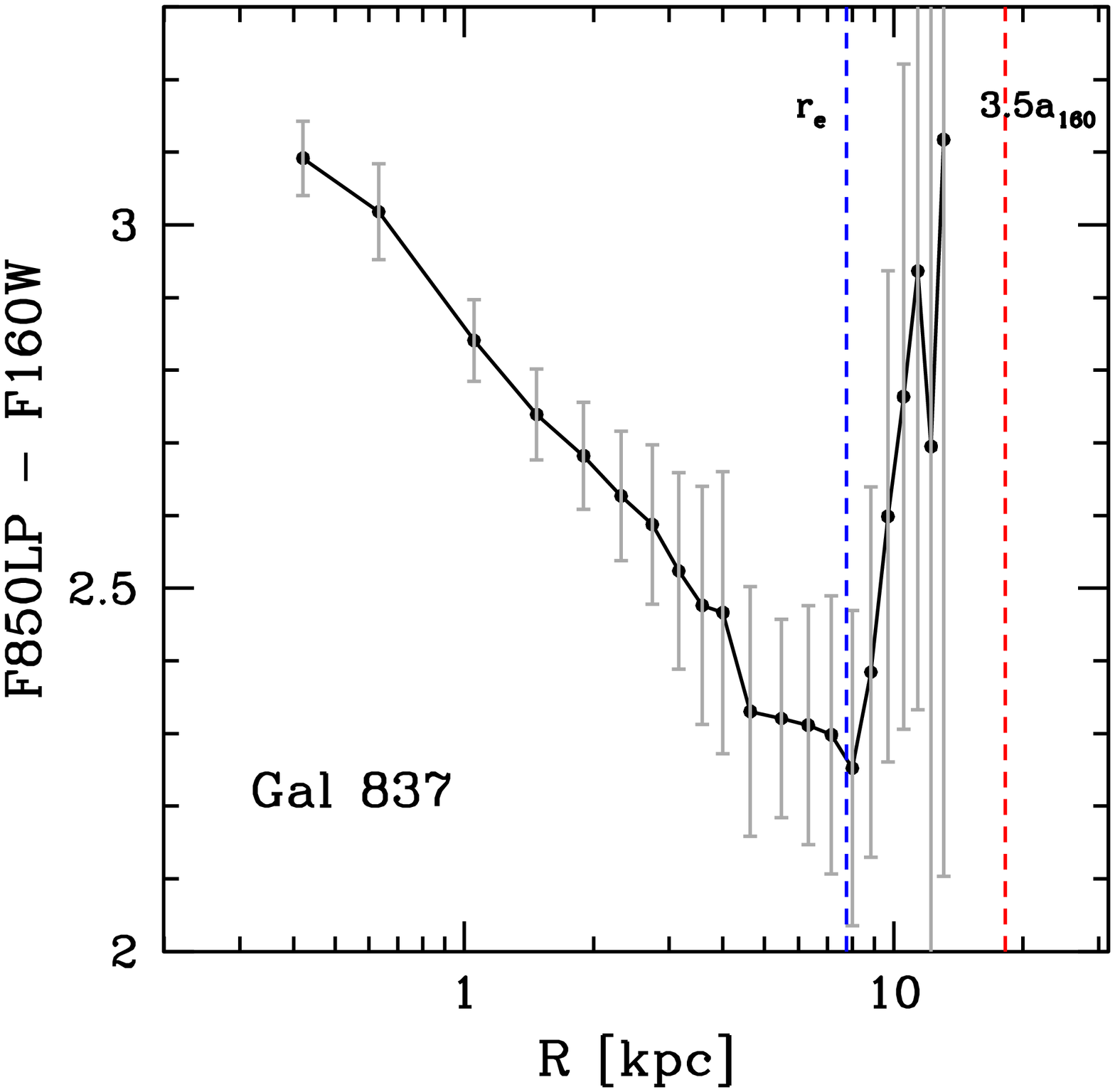}
\caption{Observed F850LP - F160W colour profiles derived by measuring the fluxes within circular apertures centred on each galaxy. Blue dashed line represents the effective radius of the galaxies as we derived in the F850LP band, while red dashed line represents 3.5a (major axis) as derived in the F160W band by Chan et al.}
\label{fig:comp_gal}
\end{figure*}

Finally, we compared the colour gradients. 
Chan et al. derived only the F850LP-F160W gradients
from the observed colour profiles up to 3.5$a_e$ ($a_e$=major axis), 
whereas we derived colour gradients through three different methods.
Actually, we can not properly compare their observed colour gradients with ours since 
they used elliptical apertures taking as reference the F160W image, whereas we 
measured the fluxes within circular aperture starting from F850LP image.
The different apertures used can affect the estimates of the colour gradients, 
causing a scatter between the measured colour gradients.
Most importantly, as discussed in Section \ref{Color gradients}, reliable 
estimates of colour gradients can be derived within the region where 
the S/N is high in all the bands considered.
Fig. \ref{fig:comp_gal} shows the observed F850LP - F160W colour profiles
up to 3.5$a_e$ for galaxies 595 and 837, having effective radius of 3.5
and 7.8 kpc. 
Blue dashed line represents the effective radius of the galaxies as derived in 
the F850LP band, while red dashed line represents 3.5$a_e$ as derived in the F160W 
band by Chan et al.
We can see that for R $\gae$ 1.5R$_e$ the colour profiles of the two galaxies 
are dominated by the sky noise of the F850LP band, which introduces a spurious 
contribution in the estimate of the colour gradients.
This is the reason why we derived the colour gradients at most up to 2R$_e$. 
For R $\gae$ 2R$_e$, the estimates of the colour gradients are not reliable 
given the depth reached by the ACS image.

\begin{figure*}
\includegraphics[width=8 cm]{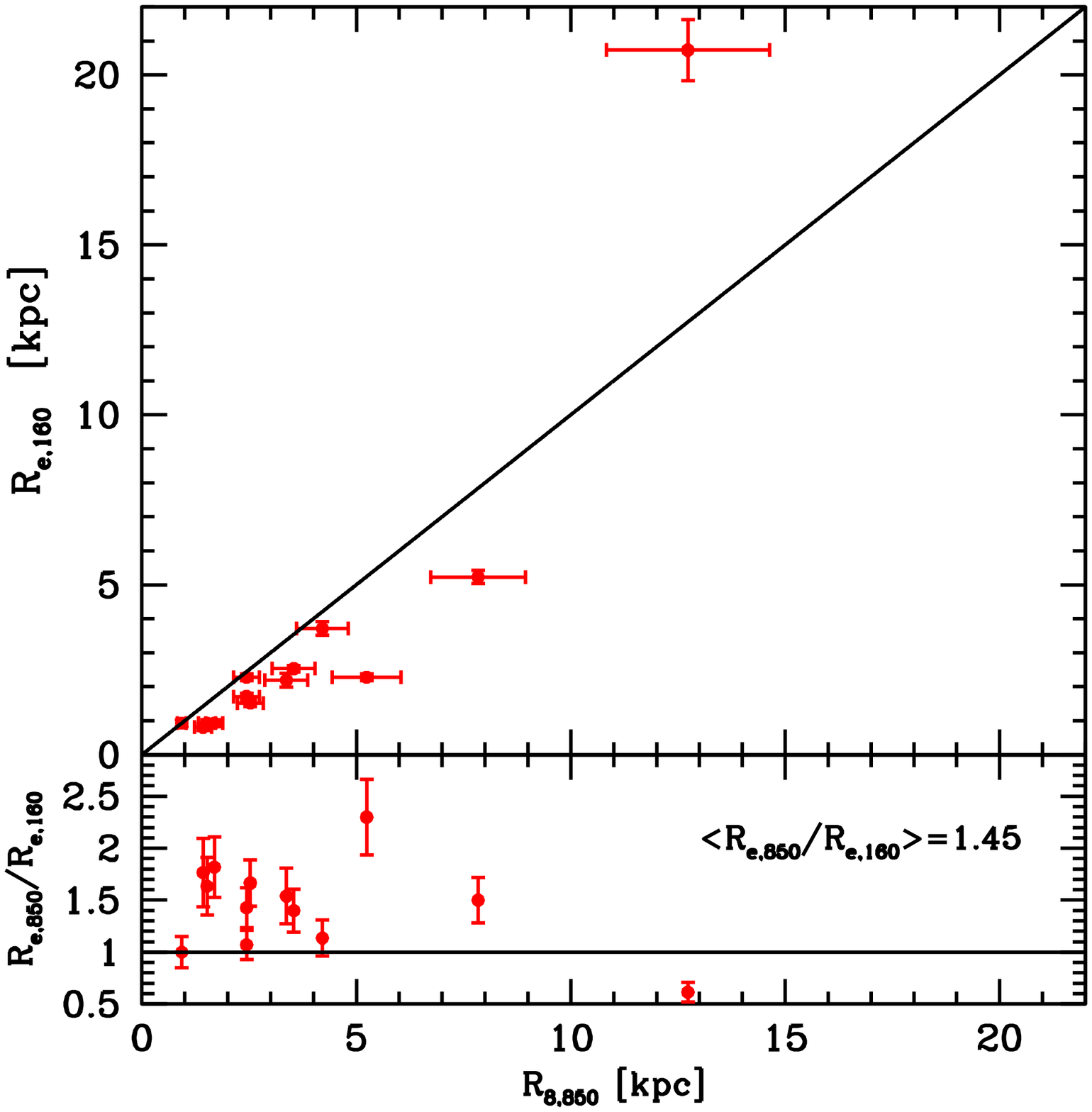}
\includegraphics[width=8 cm]{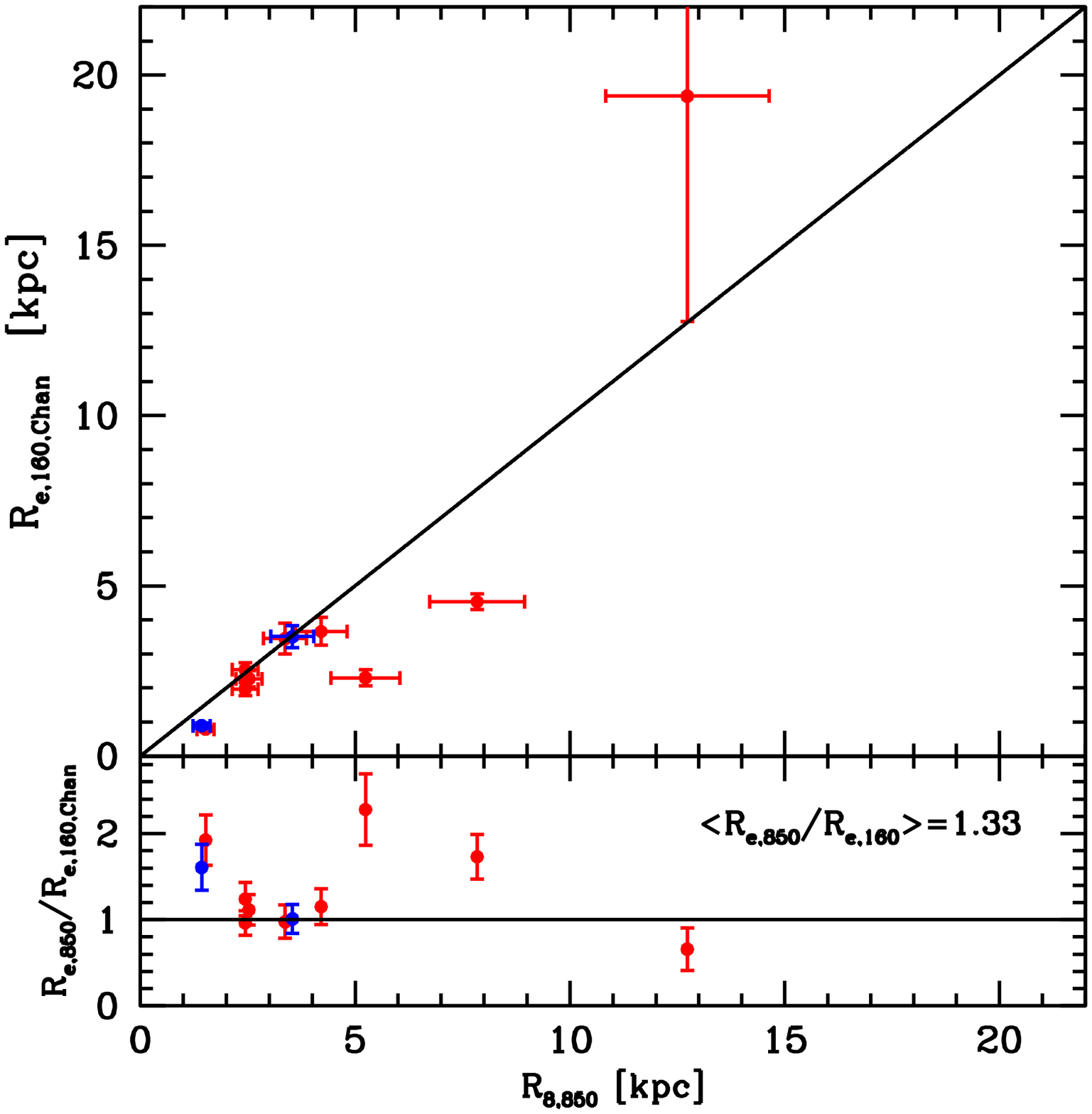}
\caption{\textit{Left panel}: comparison between R$_e$ we derived in the F850LP image and the R$_e$ derived in the F160W image (upper panel). The ratio R$_{e,850}$/R$_{e,160}$ is plotted as a function of R$_{e,850}$ in the lower panel.
\textit{Right panel}: comparison between R$_e$ we derived in the F850LP image and the R$_e$ Chan et al. derived in the F160W image (upper panel). In blue we plotted the two galaxies for which Chan et al. derived positive F850LP-F160W gradients. The ratio R$_{e,850}$/R$_{e,160,Chan}$ is plotted as a function of our R$_{e,850}$ in the lower panel.
}
\label{fig:comp_re}
\end{figure*}

For two galaxies (595 and 684) Chan et al. derived a positive gradient, 
whereas we measured a negative gradient. 
Actually, as can be seen, one of the two (684 in our sample) is 
affected by a large uncertainty that makes it consistent even also with a 
negative gradient. 
In Fig. \ref{fig:prof} we show the surface brightness profiles we derived 
for these galaxies in the three HST bands (third and fourth panels, respectively) 
and in Fig. \ref{fig:col2} we show their colour profiles, which clearly 
present a negative trend. 
Since the authors do not report the effective radii in the F850LP band, 
to investigate why they derived positive gradients for these two galaxies, 
we compare our R$_e$ in the F850LP with their R$_e$ in 
the F160W image. 
The results are shown in Fig. \ref{fig:comp_re} (right panel). 
We plotted in blue the effective radii of galaxies 595 and 684. 
These galaxies have a  ratio R$_{e,850}$/R$_{e,160}$  > 1, so they should 
present a negative colour gradient.
Indeed, if this ratio is $>1$, the light is more concentrated in the reddest 
band considered and the colour gradient is negative; 
the opposite happens for values $< 1$ of the ratio.
Since, instead, Chan et al. measured a positive gradient for these two galaxies,
it follows that the R$_e$ they derived in the F850LP image is smaller than 
the R$_e$ they derived in the F160W image.
However, as shown in Fig. \ref{fig:comp_re} (left panel), we find that systematically 
all the galaxies (with the exception of the central one) have the R$_e$ in 
the F160W smaller than the R$_e$  in the F850LP: 
$<R_{e.850}/R_{e,160}>\sim 1.5$. 
We are not able to explain the reason why the radii of 595 and 684 
should deviate from this behaviour.

%%%%%%%%%%%%%%%%%%%%%%%%%%%%%%%%%%%%%%%%%%%%%%%%%%

% Don't change these lines
\bsp	% typesetting comment
\label{lastpage}
\end{document}